\documentstyle[epsfig]{mn2e}
\begin{document}
\input BoxedEPS.tex
\newcommand{\aip}{{\small ${\cal AIPS}$}}
\newcommand{\gtsim}{\mbox{{\raisebox{-0.4ex}{$\stackrel{>}{{\scriptstyle\sim}}
$}}}}
\newcommand{\ltsim}{\mbox{{\raisebox{-0.4ex}{$\stackrel{<}{{\scriptstyle\sim}}
$}}}}
\newcommand{\s}{$\stackrel{\rm s}{.}$}
\newcommand{\h}{$^{\rm h}$}
\newcommand{\m}{$^{\rm m}$}
\newcommand{\pp}{$\stackrel{\prime\prime}{.}$}
\newcommand{\de}{$^{\circ}$}
\newcommand{\p}{$^{\prime}$}
\newcommand{\arc}{$^{\prime\prime}$}
\newcommand{\marc}{^{\prime\prime}}
\newcommand{\rs}{{\em $r_s$}}
\newcommand{\DPM}{{\em DPM}}
\newcommand{\alf}{{\displaystyle\biggl({\nu_{\rm h} \over \nu_{\rm l}}\biggr)^{\alpha}} }

\newcommand{\figstart}[1]
    { \begin{figure}[htb]
      \begin{picture}(0,#1) }
\newcommand{\figend}[4]
    { \end{picture}
      \special{#1}
      \caption[#2]{#3}
      \label{#4}
      \end{figure} }
\newcommand{\fig}[5]
    { \figstart{#1}
      \figend{#2}{#3}{#4}{#5} }
\newcommand{\bHS}{\beta_{\mbox{\scriptsize HS}}}
\newcommand{\bBF}{\beta_{\mbox{\scriptsize BF}}}
\newcommand{\nT}{\nu_{\mbox{\scriptsize T}}}
\newcommand{\et}{E_{\mbox{\scriptsize T}}}
\newcommand{\nTn}{\nu_{\mbox{\scriptsize Tn}}}
\newcommand{\nTf}{\nu_{\mbox{\scriptsize Tf}}}
\newcommand{\tn}{\tau_{x\mbox{\scriptsize n}}}
\newcommand{\tf}{\tau_{x\mbox{\scriptsize f}}}
\newcommand{\xn}{x_{\mbox{\scriptsize n}}}
\newcommand{\xf}{x_{\mbox{\scriptsize f}}}
\newcommand{\yn}{y_{\mbox{\scriptsize n}}}
\newcommand{\yf}{y_{\mbox{\scriptsize f}}}
\newcommand{\lln}{l_{\mbox{\scriptsize n}}}
\newcommand{\llf}{l_{\mbox{\scriptsize f}}}
\newcommand{\Dn}{f(\Delta_{\mbox{\scriptsize n}})}
\newcommand{\Df}{f(\Delta_{\mbox{\scriptsize f}})}
\newcommand{\B}{\mbox{$B$}}
\newcommand{\Bo}{\mbox{$B$}_{0}}

\SetEPSFDirectory{/scratch/sbgs/figures/hst/}
\SetRokickiEPSFSpecial
\HideDisplacementBoxes

\title[Photometric redshifts in the SWIRE Survey]{Photometric redshifts in the SWIRE Survey}
\author[Rowan-Robinson M.]{Michael Rowan-Robinson$^1$, Tom Babbedge$^1$, Seb Oliver$^2$,  
\newauthor
Markos Trichas$^1$, Stefano Berta$^{9,10}$, Carol Lonsdale$^3$, Gene Smith$^3$, 
\newauthor
David Shupe$^4$, Jason Surace$^4$,Stephane Arnouts$^5$, Olivier Ilbert$^5$,
\newauthor  
Olivier Le F\'{e}vre$^5$, Alejandro Afonso-Luis$^6$, Ismael Perez-Fournon$^6$,
\newauthor
Evanthia Hatziminaoglou$^7$, Mari Polletta$^3$, Duncan Farrah$^8$, Mattia Vaccari$^9$ \\
$^1$Astrophysics Group, Blackett Laboratory, Imperial College of Science 
Technology and Medicine, Prince Consort Road,\\ 
London SW7 2AZ; $^2$Astronomy Centre, Department of Physics and Astronomy, University of Sussex;\\
$^3$Center for Astrophysics and Space Science, University of California, San Diego, La Jolla, CA 92093-0424, USA;\\
$^4$Spitzer Science Center, California Institute of Technology, Pasadena, CA 91125, USA;\\
$^5$Laboratoire d'Astrophysique de Marseille, UMR 6110 CNRS-Universite de Provence, 
BP8, 13376 Marseille Cedex 12, France;\\
$^6$Instituto de Astrofisica de Canarias, C/ Via Lactea, 38200 La Laguna, S/C de Tenerife, Spain;\\
$^7$European Southern Observatory, Karl-Swartzschild-Str. 2, D85748 Garching bei Munchen, Germany;\\
 $^8$Department of Astronomy, Cornell University, 220 Space Science Building, Ithaca, NY 14853;\\
 $^9$Dipartimento di Astronomia, Universita di Padova, Vicolo Oservatorio 3, I-35122 Padua, Italy.\\
$^10$Max-Planck-Institut f\"{u}r Extraterrestrische Physik (MPE), Postfach 1312, 85741 Garching, Germany.
}
\maketitle
\begin{abstract}
We present the SWIRE Photometric Redshift Catalogue, 1025119 redshifts of unprecedented reliability and of accuracy
comparable with or better than previous work.
Our methodology 
is based on fixed galaxy and QSO templates applied to data at 0.36-4.5 $\mu$m, and on a set of 4
infrared emission templates fitted to infrared excess data at 3.6-170 $\mu$m.  The galaxy templates are initially empirical, but 
are given greater physical validity by fitting star-formation histories to them, which also allows us to
estimate stellar masses.  The code involves two passes through the
data, to try to optimize recognition of AGN dust tori.  A few carefully justified priors are used and are the key to supression of 
outliers.  Extinction, $A_V$, is allowed as a free parameter.  The full reduced $\chi^2_{\nu}(z)$ distribution 
is given for each source, so the full error distribution can be used, and aliases investigated.

We use a set of 5982 spectroscopic redshifts, taken from 
the literature and from our own spectroscopic surveys, to analyze the performance of our method as a function of the number 
of photometric bands used in the solution and the reduced $\chi^2_{\nu}$.  For 7 photometric bands (5 optical + 3.6, 4.5 
$\mu$m) the
rms value of $(z_{phot}-z_{spec})/(1+z_{spec})$ is 3.5$\%$, and the percentage of catastrophic outliers (defined as $>$ 
15$\%$ error in (1+z) ), is $\sim 1\%$.  These rms values are comparable with the best achieved in other studies, and the 
outlier fraction is significantly better.  The inclusion of the 3.6 and 4.5 $\mu$m IRAC bands is 
crucial in supression of outliers.

We discuss the redshift distributions at 3.6 and 24 $\mu$m.  In individual fields, structure in the redshift distribution
corresponds to clusters which can be seen in the spectroscopic redshift distribution, so the photometric redshifts are
a powerful tool for large-scale structure studies.  10$\%$ of sources in the SWIRE photometric redshift catalogue have 
z $>$2, and 5$\%$ have z$>$3, so this catalogue is a huge resource for high redshift galaxies.  

A key parameter for understanding the evolutionary status of infrared galaxies is $\L_{ir}/L_{opt}$.
For cirrus galaxies this is a measure of the mean extinction in the interstellar medium of the
galaxy.  There is a population of ultraluminous galaxies with cool dust and we have shown SEDs for some of the reliable examples.  
For starbursts, we estimate the specific star-formation rate, $\phi_*/M_*$.  Although the very highest
values of this ratio tend to be associated with Arp220 starbursts, by no means all ultraluminous galaxies are.  We discuss an interesting
population of galaxies with elliptical-like spectral energy distributions in the optical and  luminous starbursts in the infrared.

For dust tori around Type 1 AGN, $L_{tor}/L_{opt}$ is a measure of the torus covering factor and we deduce a mean covering
 factor of 40$\%$.

Our infrared templates also allow us to estimate dust masses for all galaxies with an infrared excess.

\end{abstract}


\section{Introduction}

The ideal data-set for studying the large-scale structure of the universe and the evolution
and star-formation history of galaxies is a large-area spectroscopic redshift survey.  
The advent of the Hubble Deep Field, the first of a series of deep extragalactic surveys with
deep multiband photometry but no possibility of determining spectroscopic redshifts
for all the objects in the survey, has led to an explosion of interest in photometric redshifts
(Lanzetta et al 1996, Mobasher et al 1996, Gwyn and Hartwick 1996, Sawicki et al 1997, 
Mobasher et al and Mazzei 1998, Arnouts et al 1999, Fernando-Soto et al 1999, 2002, 
Benitez 2000, Fontana et al 2000, Bolzonella et al 2000, Thompson et al
2001, Teplitz et al 2001, Lanzetta et al 2002, Le Borgne and Rocca-Volmerange 2002, 
Firth et al 2002, 2003,  Chen and Lanzetta 2003, 
Rowan-Robinson 2003, Wolf et al 2004, Rowan-Robinson et al 2004, 2005, Babbedge et al 2004, 
Mobasher et al 2004, Vanzella et al 2004, Benitez et al 2004, Collister and Lahav 2004,
Gabasch et al 2004, Hsieh et al 2005, Ilbert et al 2006, Brodwin et al 2006, Polletta et al 2007).  Large planned photometric surveys,
both space- and ground-based, make these methods of even greater interest.  

Here we apply the template-fitting method of Rowan-Robinson (2003), Rowan-Robinson et al
(2004, 2005), Babbedge et al (2004), to the data from the SPITZER-SWIRE Legacy Survey 
(Lonsdale et al 2003).  We have been
able to make significant improvements to our method.  Our goal is to derive a robust method which 
can cope with a range of optical photometric bands, take advantage of the SPITZER IRAC bands, 
give good discrimination between galaxies and QSOs, cope with extinguished objects, and deliver 
accurate redshifts out to z = 3 and beyond.  With this paper we supply photometric redshifts for 
1,025,119 infrared-selected galaxies and quasars which we believe are of unparalleled reliability, and of accuracy
at least comparable with the best of previous work.

The areas from the SWIRE Survey in which we have optical photometry and are able to derive 
photometric redshifts are:
(1)  8.72 sq deg of ELAIS-N1, in which we have 5-band (U'g'r'i'Z') photometry from the Wide Field Survey (WFS, 
McMahon et al 2001, Irwin and Lewis 2001), (2) 4.84 sq deg of ELAIS-N2, in which we have 5-band 
(U'g'r'i'Z') photometry from the WFS, 
(3) 7.53 sq deg of the Lockman Hole, in which we have 3-band photometry (g'r'i') from the SWIRE photometry 
programme, with U-band photometry in 1.24 sq deg, (4) 4.56 sq deg in Chandra Deep Field South 
(CDFS), in which we have 3-band (g'r'i') photometry from the SWIRE photometry programme, (5)6.97 sq 
deg of XMM-LSS, in which we have 5-band (UgriZ) photometry from Pierre et al (2007). 
In addition within XMM 
we have 10-band photometry (ugrizUBVRI) from the VVDS programme of McCracken et al (2003), Le F\'evre et al (2004) (0.79 sq deg), 
and very deep 5-band photometry (BVRi'z') in 1.12 sq deg of the Subaru XMM Deep Survey 
(SXDS, Sekiguchi et al 2005, Furusawa et al 2008).   The SWIRE data are described in Surace et al (2004, 2008 in preparation).

Apart from the use of flags denoting objects that are morphologically stellar at some points in the code (see section 2), we 
make no other use of morphological information in determining galaxy types.  So reference to 'ellipticals' etc
refers to classification by the optical spectral energy distribution.

A cosmological model with $\Lambda$ = 0.7, $h_0$=0.72 has been used throughout.

\section{Photometric redshift method}

In this paper we have analysed the catalogues of SWIRE sources with optical associations 
with the I{\scriptsize MP}z code (Rowan-Robinson 2003 (RR03), Rowan-Robinson et al 2004, 2005, Babbedge et al 
2004 (B04)) which uses a small set of optical templates for galaxies and AGN.  
The galaxy templates were based originally on empirical spectral energy distributions (SEDs) by Yoshii and Takahara (1988) and 
Calzetti and Kinney (1992) but have been subsequently modified in Rowan-Robinson (2003), 
Rowan-Robinson et al (2004, 2005) and Babbedge et al (2004).   Here we have modified the templates further
(section 3), taking account of the large VVDS sample with spectroscopic redshifts and good
 multiband photometry.

The other main changes we have made to the code described by Rowan-Robinson (2003), Babbedge et al (2004) are:

\begin{itemize}
\item{The code involves two passes through the data.  In the first pass we use 6 galaxy templates and 3 AGN 
templates (see section 3) and each source selects the best solution from these 9 templates. The redshift 
resolution is 0.01 in $log_{10}(1+z)$.  Quasar templates are permitted only if the source is flagged as morphologically stellar. 
3.6 and 4.5 $\mu$m data are only used in the solution if 
$log_{10}(S(3.6)/S_r) < 0.5$: this is to avoid distortion of the solution by contributions from AGN dust tori 
in the 3.6 and 4.5 $\mu$m bands.  Longer wavelength Spitzer bands are not used in the optical template fitting.}
\item{Before commencing the solution we create a colour table for all the galaxy and QSO templates, for each 
of the photometric bands required, for 0.002 bins in $log_{10}(1+z)$, from 0-0.85 (z=0-6).  This  involves a full integration over 
the profile of each filter, and calculation of the effects of intergalactic absorption, for each template and redshift bin,
as described in Babbedge et al (2004).  Prior creation of a colour table saves computational time.}
\item{ After pass one we fit the Spitzer bands for which there is an excess relative to the starlight or QSO solution 
with one of four infrared templates: cirrus, M82 starburst, Arp 220 starburst and AGN dust torus, provided 
there is an excess at either 8 or 24 $\mu$m (cf Rowan-Robinson et al 2004).
  3.6 and 4.5 $\mu$m data are not used in the second pass redshift solution
if the emisision at 8 $\mu$m is dominated by an AGN dust torus.}
\item{In the second pass we use a redshift resolution of 0.002 in $log_{10}(1+z)$.  Galaxies and quasars are 
treated separately and for galaxies the limit on $log_{10}(S(3.6)/S_r) $ for use of 3.6 and 4.5 $\mu$m data 
in the solution is raised to 2.5, which comfortably includes all the galaxy templates for z $< $6.}   
\item{In pass 2 we use two elliptical galaxy templates, one corresponding to an old (12 Gyr) stellar population 
(which is also used in the first pass)
and one corresponding to a much more recent (1 Gyr) starburst (Maraston et al 2005).  For z $>$ 2.5 we 
permit only the latter template.}
\item{In pass 2 we interpolate between the 5 spiral and starburst templates to yield a total of 11 such templates.
Finer interpolation between the templates did not improve the solution.}
\item{Extinction up to $A_V$ = 1.0 is permitted for galaxies of type Sbc and later, ie no extinction for Sab 
galaxies or ellipticals.  For quasars the limit on the permitted extinction is $A_V$ = 0.3, unless the condition 
S(5.8) $>$ 1.2 S(3.6) is satisfied, in which case we allow extinction up to $A_V$ = 1.0.  This condition is a good 
selector of objects with strong AGN dust tori (Lacy et al 2004).  If we allow $A_V$ up to 1.0 for all QSO fits then we find serious
aliasing with low redshift galaxies.  The assumptions made here allow the identification of a small class of heavily
reddened QSOs which is clearly of some interest. A few galaxies and quasars might find a better fit to their SEDs with
$A_V > 1$ but we found that allowing this possibility resulted in a significant increase in aliasing and a
degradation of the overall quality of the redshift fits.  A separate search will be needed to identify such objects.

As in Rowan-Robinson et al (2004) a prior assumption of a power-law distribution function of $A_V$
is introduced by adding $1.5 A_V$ to $\chi^2_{\nu}$.  This avoids an excessive number of large values of $A_V$ being selected.

For galaxies we used a standard Galactic extinction.  For QSOs we used an LMC-type extinction, following Richards et al (2003). }
\item{As in our previous work there is an important prior on the range of absolute B magnitude, essentially corresponding,
for galaxies, to a range in stellar mass.  Because there is strong evolution in the mass-to-light ratio between z = 0
and z = 2.5, the $M_B$ limit needs to evolve with redshift.  We have assumed an upper luminosity limit
for galaxies of  $M_B$ = -22.5-z for z $<$2.5, =-25.0 for z $>$ 2.5, and a lower limit of
 $M_B$ = -17.0-z for z $<$2.5, =-19.5 for z $>$ 2.5.  For quasars we assume the upper and lower limits on
luminosity correspond to $M_B$ = -26.7 and -18.7 (see section 4 and Figs 2 for a justification of these 
assumptions).  We do not make any prior assumption about the shape of the luminosity function.  Strictly speaking 
it would be better to limit the near infrared luminosity to constrain the range of stellar mass, but we found that
by specifying a limited range of $M_B$ we were also eliminating unreasonably large star-formation rates
in late type galaxies. }
\item{We have accepted the argument of Ilbert et al (2006) that it is necessary to apply a multiplicative in-band correction
factor to each band to take account of incorrect photometry and calibration factors.  Table 1 gives these
factors for each of our areas, determined from samples with spectroscopic redshifts.}
\item{In calculating the reduced $\chi_{\nu}^2$ for the redshift solution we use the quoted photometric uncertainties,
but we set a floor to the error in each band, typically 0.03 magnitudes for g,r,i, 0.05 magnitudes for u,z, 1.5 $\mu$Jy for
3.6 $\mu$m, 2.0 $\mu$Jy for 4.5 $\mu$m.

This also implies a maximum weight given to any band in the solution.  Without this assumption, there would be cases 
where a band attracted unreasonably high weight because of a spuriously low estimate of the photometric error.}
\end{itemize}

\begin{table*}
\caption{Correction factors for fluxes in each band, by area}
\begin{tabular}{lllllllllll}
area & U' & g' & r' & i' & Z' &  J & H & K & 3.6 $\mu$m & 4.5 $\mu$m\\
&&&&&&&&&\\
EN1,EN2 & 1.114 & 0.988 & 0.951 & 0.994 & 1.004 &  1.238 & 1.171 & 1.130 & 0.954 & 0.993\\
Lockman & 1.057 & 1.031 & 0.901 & 0.910 & &&&& 1.050 & 1.002\\
VVDS & 1.046 & 0.969 & 0.928 & 0.960 & 0.972  & 1.173 & & 1.029 & 1.085 & 1.015\\
XMM-LSS & 1.046 & 0.969 & 0.928 & 0.960 & 0.972 & &&& 1.085 & 1.015\\
CDFS & 1.046 & 0.969 & 0.916 & 0.991 & 0.972 & &&& 1.045 & 1.037\\
\end{tabular}
\end{table*}

We have not used some of the other priors used by Ilbert et al (2006), for example use of prior information about
the redshift distribution, nor their very detailed 
interpolation between templates.  We have found that while carefully chosen priors are essential to
eliminate aliases, which are the main cause of outliers, it is also crucial to keep the code as simple as
possible since unnecessary complexity creates new outliers.

Aperture matching between bands is a crucial issue for photometric redshifts and we have proceeded as follows.  At optical wavelengths we use colours determined in a point-source aperture, but then apply an aperture correction to all bands defined in the r-band by the difference between the Sextractor mag-auto and the point-source magnitude.  For the IRAC bands we use
'ap-2' magnitudes, corresponding to a 1.9"  radius aperture, except that if the the source is clearly extended, defined by Area(3.6)$>$250, 
we use Kron magnitudes.
For the 2MASS J,H,K magnitudes we use the extended source catalogue magnitude in a 10" aperture, if available, and the point-source catalogue magnitude otherwise.  For 80$\%$ of sources with 2MASS associations the JHK magnitudes improve the fit slightly, but for 20$\%$ the addition of the JHK magnitudes makes the $\chi^2$ for the solution significantly worse.  We found the same problem for associations  with the UKIDSS DXS survey in EN1, with the corresponding proportions being 70:30.  The issue is almost certainly one of aperture matching.  We have not included JHK data in the analysis of the performance of the photometric redshift method given here.

The requirement for entry to the code is that a source be detected in at least one SWIRE band and in at least
one optical band out of gri.  We are then able to determine redshifts for $\sim 95 \%$ of sources, except in EN1, where the figure drops to 85$\%$.  The additional failure rate n EN1 is primarily due to photometry in one of the optical bands being out of line with the other bands.  Other reasons for failure can be that after imposing 3-$\sigma$ limits on all bands, there are no longer 2 bands available out of UgriZ, 3.6,4.5 $\mu$m.  Of course redshifts determined from only two photometric bands are of low reliability (see section 4.3).

\section{Templates}
\subsection{Galaxy Templates}
\label{subsec:gseds}
The choice of the number and types of template to use is very important.   
If there are too few templates, populations
of real sources will not be represented, whilst if there are too many there will be too much opportunity for aliases and degeneracy.

The I{\scriptsize MP}Z code uses six galaxy templates (RR03, B04 and Babbedge et al 2006); E, Sab, Sbc, Scd, Sdm and starburst 
galaxies.  These six templates (or similar versions) have been found to provide a good general representation of 
observed galaxy SEDs.  The original empirical templates used in RR03 were adapted from Yoshii and Tokahari (1988), apart from the starburst template, which was  adapted from observations by Calzetti and Kinney (1992).  These templates have been
subsequently modified, as described below, to improve the photometric redshift results.

In B04 those empirical templates were regenerated to higher resolution using Simple Stellar Populations (SSPs), 
each weighted by a different SFR and extinguished by a different amount of dust, $A_V$.  This procedure,
 based on the synthesis code of Poggianti (2001), gave the templates a physical validity.  
Minimization was based on the Adaptive Simulated Annealing algorithm.  Details on this algorithm and on the fitting 
technique are given in Berta et al (2004).  These templates were used by I{\scriptsize MP}Z in Babbedge et al (2006) in 
order to obtain photometric redshifts and calculate luminosity functions for IRAC and MIPS sources in the ELAIS N1 field of the SWIRE survey.

We have now improved these templates further in a two stage process using the rich multi-wavelength photometry 
and spectroscopic redshifts in the CFH12K-VIRMOS survey (LeFevre et al 2004): a deep BVRI 
imaging survey conducted with the CFH-12K camera in four fields.  Additionally, there are U band data from the 2.2 m 
telescope in La Silla, J and K data from the NTT and IRAC and MIPS photometry from the SWIRE survey.  The spectroscopic 
redshifts come from the follow-up VIRMOS-VLT Deep Survey (VVDS: LeFevre 2003), a large 
spectroscopic follow-up survey which overlaps with the SWIRE data in the XMM-LSS field. 

The spectroscopic sample is sufficiently large (5976 redshifts) and wide-ranging in redshift (0.01$<$z$<$4) to allow a detailed
 comparison between the template SEDs and observed SEDs.  We first find the best-fitting template SED for each VVDS 
source with the photometric redshift set to the spectroscopic redshift (for sources with i$<$22.5, nzflag=3 or 4
(Le F\'evre et al 2005, Gavignaud et al 2006), ie reliable
galaxy redshifts).  
We typically find $\sim$100-400 sources for each of the six galaxy templates (after discarding those which 
do not obtain a fit with reduced $\chi_{\nu}^{2}<$5).  Comparison of the renormalised, extinction-corrected rest-frame fluxes from 
each set of sources to their best-fit galaxy template then highlights potential wavelength ranges of the template that fail to 
reproduce well the observed fluxes (typically from $\sim$1200$\AA$ to $>$5$\mu$m).  
This comparison is shown in Fig. 1.  We then adapt those regions of the templates 
to follow the average track shown by the observed datapoints.  These adapted templates are 
once again empirical and in order to recover their physical basis we then 
reproduce them via the same SSP method as used in B04.  These final templates are 
then both physically-based and provide a very good representation of the observed 
sources from the VVDS survey.  For a comparison see Fig 1 and for details of the 
SSPs and their contribution to each template see Table \ref{table:seds}.  

\begin{figure*}
\epsfig{file=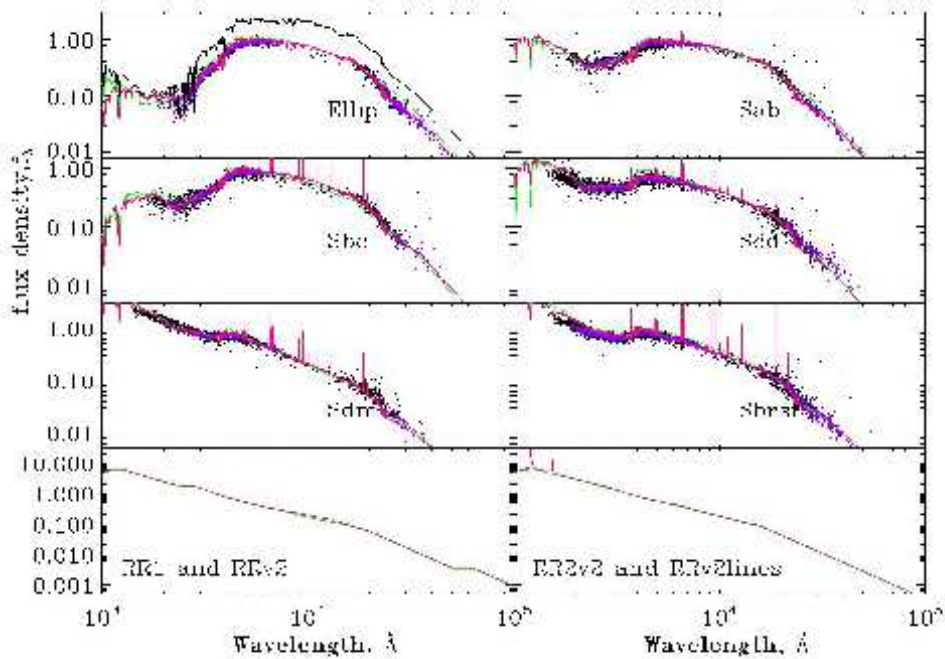,angle=90,width=14cm}
\caption{The new SED templates (red) compared with the old ones (green)
and data from VVDS photometry (purple: r$<$21.5, black: r $>$ 21.5).  The
 black SED (displaced) is the modified Maraston (2005) proto elliptical.
}
\end{figure*}

\subsubsection{Ellipticals}
\label{subsubsec:maraston}
In addition to our standard elliptical template we now include (in the second pass) a young elliptical template.  
This young elliptical is based on a 1Gyr old SSP provided by Claudia Maraston (2005).  A significant improvement 
for high redshift galaxy studies is the treatment in this template of thermally 
pulsing the asymptotic giant branch (TP-AGB) phase of stellar evolution, particularly in the near-IR.  
Maraston (2006) has demonstrated that this phase has a strong contribution to the infrared SED of galaxies 
in the high redshift Universe (z$\sim$2). Because of the limited time for evolution available at higher redshifts, we
restrict ellipticals to just this template for z $>$ 2.5. 

The Maraston elliptical template did not have a `UV bump' but since we have found that our higher redshift 
ellipticals did exhibit some UV emission, we have replaced the template shortward of 2100$\AA$ with the UV 
behaviour of our old (12Gyr) elliptical.  We are not able to say whether this UV upturn is due to a recent burst of
star formation (but see section 8 for some hint that this may be relevant), to horizontal branch stars, or to
binary star evolution.  Additionally, as the TP-AGB stellar templates incorporated into the evolutionary 
population synthesis of Maraston (2006) are empirical, they do not extend longward of 3$\mu$m 
so we have followed a similar procedure as above to derive the behaviour at $\lambda>3\mu$m.

An important benefit of these stellar synthesis fits to our templates is that we can estimate stellar masses
for all SWIRE galaxies (see section 8).
\begin{table*}
\caption{\scriptsize{The six galaxy templates and the SSPs that were used to
create them.  SFR is scaled to give a total mass of 1E11M$_{\odot}$.}}
\centering
\scriptsize
\begin{tabular}{c r r | c*{12}{c}}
\hline
\hline
\multicolumn{3}{c|}{SSP} 	&\multicolumn{12}{c}{SED Template}\\
\hline
\#& \multicolumn{1}{c}{Age (yr)} & \multicolumn{1}{c|}{$\Delta t$ (yr)}
&&\multicolumn{2}{c}{Elliptical(E1)}&\multicolumn{2}{c}{Sab}&
\multicolumn{2}{c}{Sbc}&\multicolumn{2}{c}{Scd}&\multicolumn{2}{c}
{Sdm}&\multicolumn{2}{c}{Sbrst}\\
\cline{5-16}
    	&		&	       && SFR  & E(B-V) &SFR  & E(B-V)  &
SFR     & E(B-V) & SFR   & E(B-V) &SFR  & E(B-V) & SFR &E(B-V) \\
1	&$1\, 10^6$		&$2\, 10^6$     && -- &   --  & -- &  -- &
--   & -- &  --  & -- & -- & -- & 181.5 & 1.4   \\
2	&$3\, 10^6$		&$4\, 10^6$     && -- &   --  & -- &  -- &
--   & -- &  --  & -- & -- & -- & 8.74 & 4.8E-6  \\
3	&$7\, 10^6$		&$1.3\, 10^7$     && -- &   --  & 1.71 &
0.11  & 8.7 & 0.17 & 53.4   & 0.77 &  47.6  & 0.079 & -- & --\\
4	&$8\, 10^6$		&$3\, 10^6$     && -- &   --  & -- &  -- &
--   & -- & --  & -- &  --  & -- & 1.2E-3 & 4.8E-6  \\
5	&$1\, 10^7$		&$4\, 10^6$     && -- &   --  & -- &  -- &
--   & -- & -- & -- &  --  & -- & 3.2E-4 & 4.8E-6 \\
6	&$5\, 10^7$		&$6.2\, 10^7$     && -- &   --  & -- &  -- &
--   & -- & --  & -- &  --  & -- & 2.9E-4 & 4.8E-6 \\
7	&$9\, 10^7$		&$1.87\, 10^8$     && -- &   --  & 3.1E-3 &
0.11 & 2.5E-3 & 0.17  & 1.12   & 0.047 &  4.6E-2  & 0.079 & -- & --  \\
8	&$1\, 10^8$		&$1.25\, 10^8$     && 0.53 &   4.3E-7  & --
& -- & -- &  -- & --   & -- &  --  & -- & 3.4E-4 & 4.8E-6  \\
9	&$3\, 10^8$		&$2\, 10^8$     && 1.5E-4 &   4.3E-7 &  &
-- & -- & -- & --   & -- &  --  & -- & 45.4 & 7.7E-10  \\
10	&$4.5\, 10^8$		&$5.5\, 10^8$     && -- &   --  & 2.27 &
1.8E-4  & 6.1 & 1.9E-4 & 7.15   & 0.007 &  45.5  & 8.1E-9 & -- & --  \\
11	&$5\, 10^8$		&$3.5\, 10^8$     && 2.7E-3 &   4.3E-7 & --
&  --  & -- & -- & --   & -- &  --  & -- & 1.0E-3 & 7.7E-10 \\
12	&$1\, 10^9$		&$1.25\, 10^9$     && 3.38 &   3.3E-8  &
1.3E-3 &  1.8E-4  & 2.7E-5 & 1.9E-4 & 7.9E-3   & 0.0071 &  --  & -- & 2.6E-5
& 7.7E-10 \\
13	&$2\, 10^9$		&$1\, 10^9$     && 1.5E-3 &   3.3E-8  &
4.6E-3 &  1.8E-4  & 1.2E-4 & 1.9E-4 & 8.0E-2   & 0.0071 &  2.3E-2  & 8.1E-9
& 2.0E-5 & 7.7E-10  \\
14	&$4\, 10^9$		&$1.5\, 10^9$     && 1.1E-3 &  3.3E-8  &
6.3E-4 &  1.8E-4  & -- & -- & 4.0E-2  & 0.0071 &  2.2E-3  & 8.1E-9 & 9.6E-6
& 7.7E-10 \\
15	&$7\, 10^9$		&$3\, 10^9$     && 3.0E-4 &  3.3E-8  &
5.2E-3 &  1.8E-4  & 8.9E-5 & 1.9E-4 & 6.4E-2   & 0.0071 &  9.7E-3  & 8.1E-9
& 3.8E4 & 7.7E-10 \\
16	&$1\, 10^{10}$		&$3\, 10^9$     && 3.3E-3 &  3.3E-8  &
1.9E-4 &  1.8E-4  & 3.3E-3 & 1.9E-4 & 26.8  & 0.0071 &  2.5E-2  & 8.1E-9 &
2.9E-4 & 7.7E-10\\
17	&$1.2\, 10^{10}$		&$2\, 10^9$     && 47.8 &  3.3E-8  &
49.4 &  1.1E-6  & 48.3 & 2.3E-5 & 23.5   & 2.1E-4 &  44.5  & 8.1E-9 & 45.3 &
7.7E-10  \\
\hline
\hline
\end{tabular}
\footnotesize
\label{table:seds}
\normalsize
\end{table*}
\subsection{AGN Templates}
\label{subsec:agn}
As well as galaxy templates, the inclusion of a number of different AGN templates has 
been considered to allow the I{\scriptsize MP}Z code to identify quasar-type objects 
as well as normal galaxies.  B04 tested the success of including the the SDSS median 
composite quasar spectrum (VandenBerk 2001) and $red$ AGN templates such as the $z=2.216$ 
FIRST J013435.7--093102 source from Gregg (2002) but found that two simpler AGN templates 
were most useful.  These were based on the mean optical quasar  spectrum of Rowan-Robinson (1995),
 spanning 400${\rm \AA}$ to 25$\mu$m.  For wavelengths longer than Ly$\alpha$ the templates 
are essentially $\alpha_{\lambda}\approx-1.5$ power-laws, with slight variations included 
to take account of observed SEDs of ELAIS AGN (Rowan-Robinson et al 2004).  Following a number 
of further tests and template alterations, we now make use of updated versions of these 
two AGN templates:  AGN1 has been improved (now called RR1v2) in a similar procedure as 
applied in section 3.1 using the spectroscopic dataset to indicate regions of poor agreement between 
the template and photometry; AGN2 (now RR2v2) has also been modified to create a third 
template, RR2v2lines, by adding Ly$\alpha$ and CIV emission lines and the whole template then 
adapted via comparison to photometry (as with the other templates).  This means 
I{\scriptsize MP}Z makes use of three AGN templates, whose main difference is the 
presence$/$lack of Ly$\alpha$ and CIV emission, as well as the amount of flux longward 
of 1$\mu$m.  The numbers of quasars selecting the 3 templates are 6568 (0.65$\%$) (RR1v2), 
5253 (0.52$\%$) (RR2v2) and 5853 (0.58$\%$) (RR2v2lines).  Use of a template like that derived by 
1Richards et al (2003)
from SDSS quasars, which has very strong emission lines, generated many more aliases.

\section{Results from the IMPz2 photometric redshift code}

\subsection{Justification of priors on $M_B$}

Figures 2 show the absolute B magnitude versus spectroscopic redshift
for galaxies and quasars.  The photometric redshift code was run with z 
forced to be equal to the spectroscopic redshift, but other parameters
(template type, extinction) allowed to vary, and $M_B$ then calculated 
for the best solution.  The assumed limits on $M_B$ are shown and look
reasonable.  The limit at the low end does exclude some very low
luminosity galaxies and for these the photometric redshift would be biassed to
slightly higher redshift.  However lowering the lower boundary results in low-redshift 
aliases for galaxies whose true redshift is much higher.  

The increase of maximum luminosity with redshift reflects
the strong evolution in mass-to-light ratio with redshift since z$\sim$2.
Van der Wel et al (2005) find $d ln(M_*/L_K) \sim -1.2 z$ and $d ln(M_*/L_B) \sim -1.5 z$ for $0 < z < 1.2$
for early-type galaxies.  We needed to continue this trend to redshift 2.5 to accomodate some of the
$z \sim 3$ galaxies found by Berta et al (2007), based on the IRAC 'bump' technique of Lonsdale  et al (2008, in preparation).
At higher redshifts the maximum luminosity should decline, reflecting the accumulation of galaxies,
but we have not tried to model that in detail here.  The apparent increase in the maximum luminosity
with redshift may also be in part due to the increasing volume being sampled as redshift increases.

\begin{figure*}
\epsfig{file=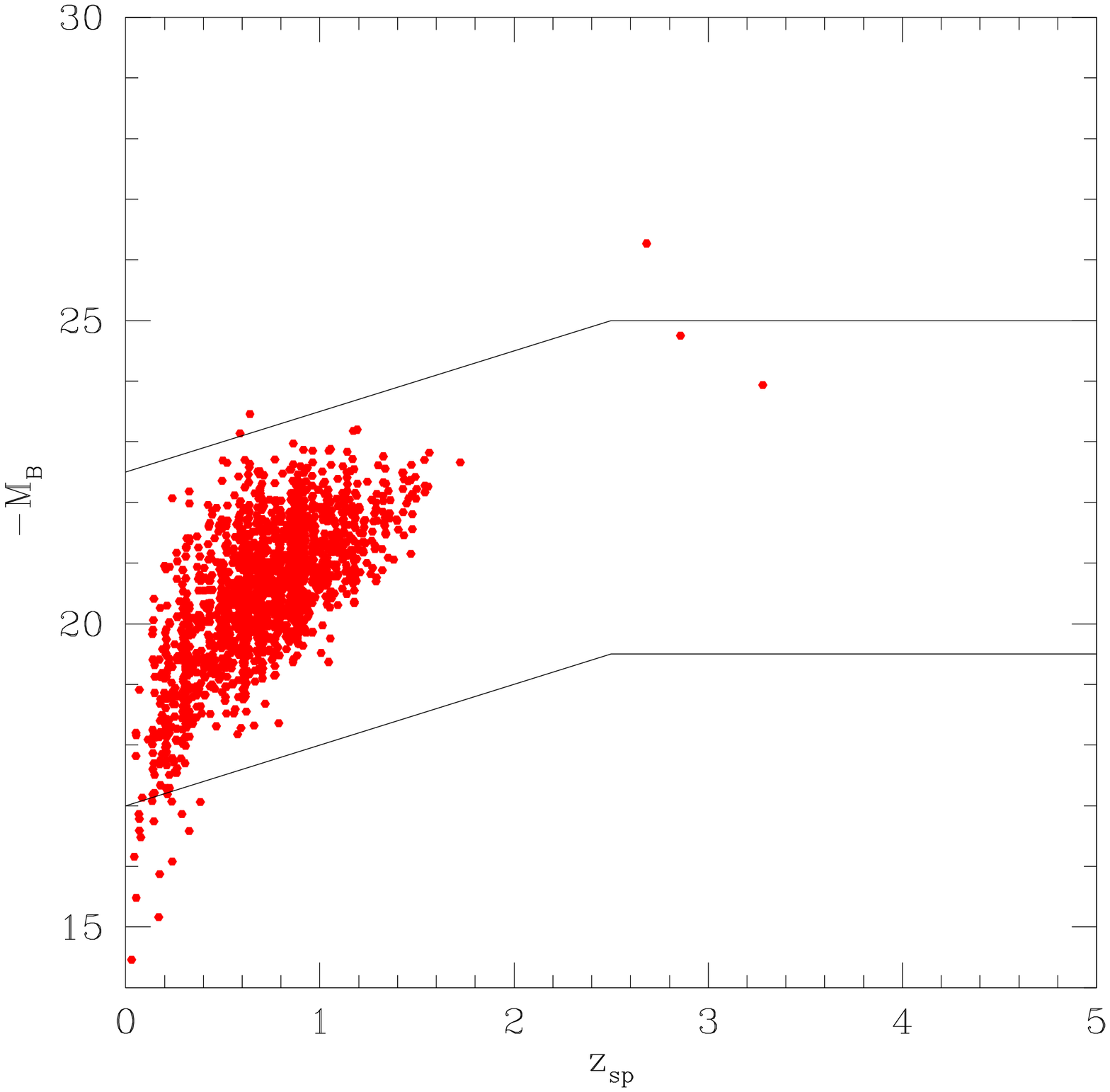,angle=0,width=7cm}
\epsfig{file=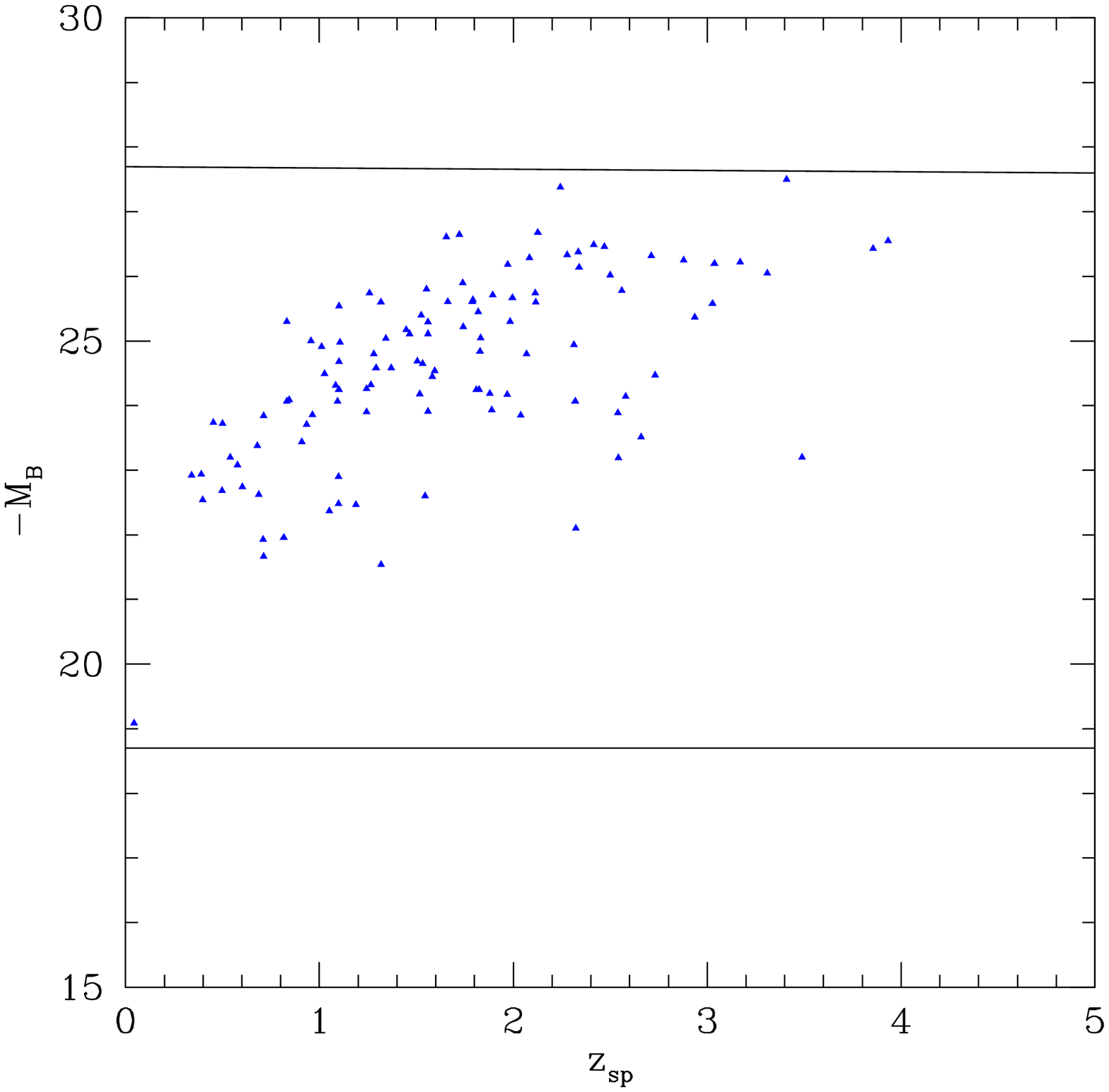,angle=0,width=7cm}
\caption{LH: Absolute blue magnitude, $M_{B}$, versus spectroscopic redshift for galaxies in SWIRE-VVDS (L) and QSOs in SWIRE-Lockman (R).}
\end{figure*}

\subsection{Impact of number of photometric bands available}

To analyze the impact of the number of optical photometric bands available, and of what improvement is
afforded by being able to use the SPITZER 3.6 and 4.5 $\mu$m data, we have carried out a detailed
analysis of the spectroscopic sample from the VVDS survey (Lefevre et al 2004).  Figures 3-7 show
results from our code, paired so that the left plot in each case is  the result for optical data only, while
the right plots show the impact of including 3.6 and 4.5 $\mu$m data in the solution.  Note that there
are some additional objects on the left-hand plots which are undetected at 3.6 or 4.5 $\mu$m.
Figures 3L, 4L, 5L, 6L, 7L show
a comparison of $log_{10}(1+z_{phot})$ versus $log_{10}(1+z_{spec})$ for VVDS using 10 (ugrizUBVRI), 5 (ugriz), 4 (ugri or griz) and 3
(gri) optical bands without using 3.6 and 4.5 $\mu$m data.  Figs 3R, 4R, 5R, 6R show corresponding plots when
3.6 and 4.5 $\mu$m data are used.  The latter show a dramatic reduction in the number of outliers.
Inclusion of 3.6 and 4.5 $mu$m bands has a more significant effect than Increasing the number of optical bands from 5 to 10. 
 Absence of the U band does significantly worsen perfomance, especially at z $<$ 1, but with gri + 3.6, 4.5 there is 
still an acceptable performance. 

Fig 7R shows $log_{10}(1+z_{phot})$ versus $log_{10}(1+z_{spec})$ for the whole SWIRE survey requiring at least 4 bands in total, S(3.6) $> 7.5 \mu Jy$, r $<$23.5,  4 photometric bands (which could be, say, gri+3.6 $\mu$m) is the minimum number for
reliable photometric redshifts.  Figs 8 show the corresponding plots in the SWIRE-EN1 and SWIRE-Lockman areas, where we have carried out programs of spectroscopy (Trichas et al 2007, Berta et al 2007, Smith et al 2007, Owen and Morrison 2008, in preparation), for a minimum of 6 photometric bands in total, and Fig 9 shows the same plot for the whole of SWIRE, 
requiring a minimum of 6 photometric bands.  We see that our method gives an excellent performance for galaxies out to z = 1 (and beyond).  Demanding more photometric bands discriminates
against high redshift objects, which will start to drop out in short wavelength bands, and against quasars because
they tend to have dust tori and therefore the 3.6 and 4.5 $\mu$m bands are not used in the solution.

The performance worsens for z $>$ 1.5, though we do not really have enough spectroscopic data to fully characterize this.
For 31 galaxies with $z_{ph}$, detected in at least 5 bands, with r $<$ 23.5, S(3.6) $> 7.5 \mu$Jy, we found that 9 (29$\%$)
had $|log_{10}(1+z_{phot})-log_{10}(1+z_{spec}| > 0.15$, and the rms of $(z_{phot}-z_{spec})/(1+z_{spec})$ for the remainder was
11 $\%$. 

In EN1 there appears to be a slight systematic overestimation of the redshift around z $\sim$1, by about 0.1.  This is not seen in the VVDS or Lockman samples, or in the overall plot for the whole SWIRE Catalogue (Fig 9).  The most likely explanation is some bias in the WFS photometry at fainter magnitudes. 


\begin{figure*}
\epsfig{file=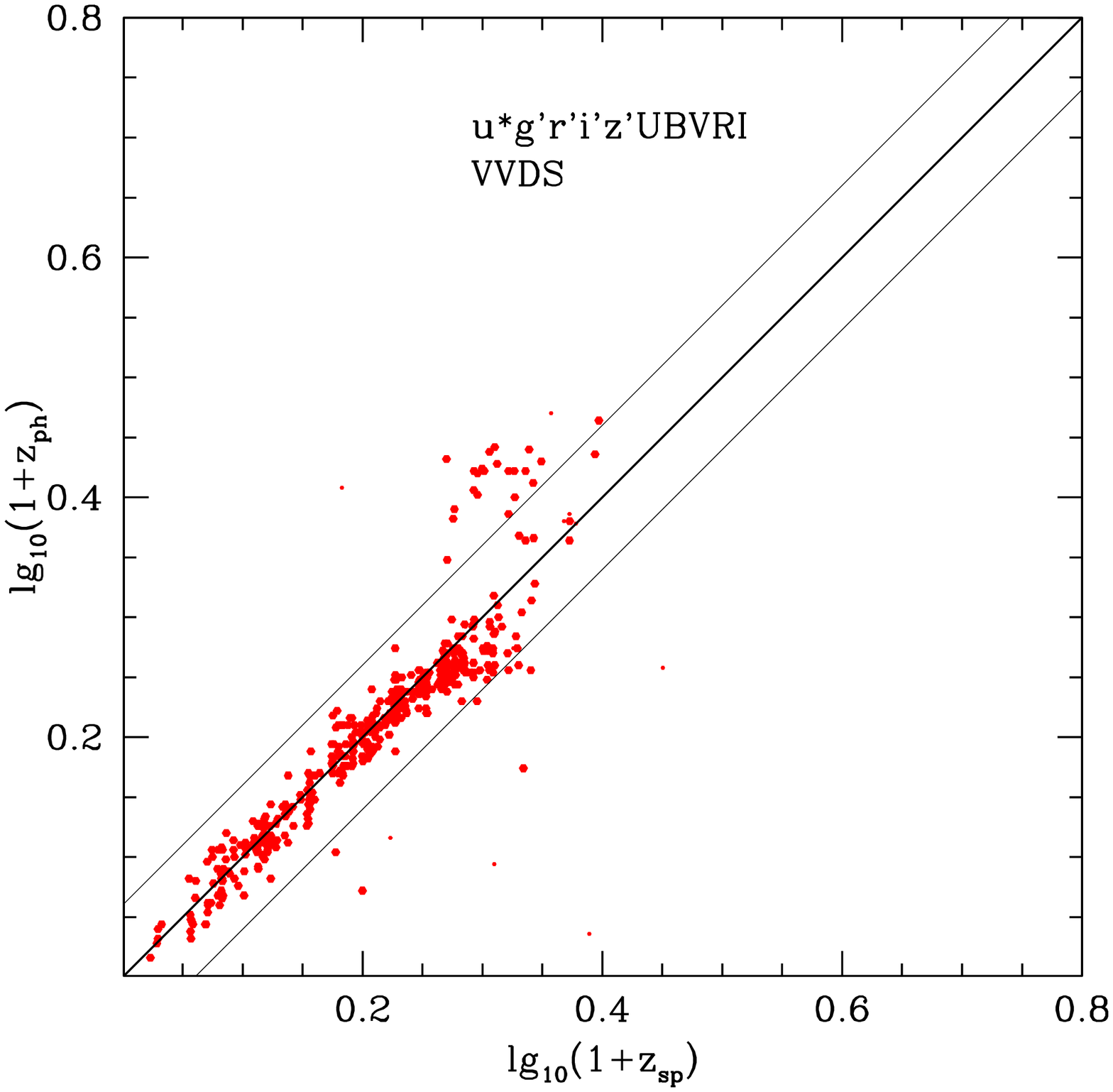,angle=0,width=7cm}
\epsfig{file=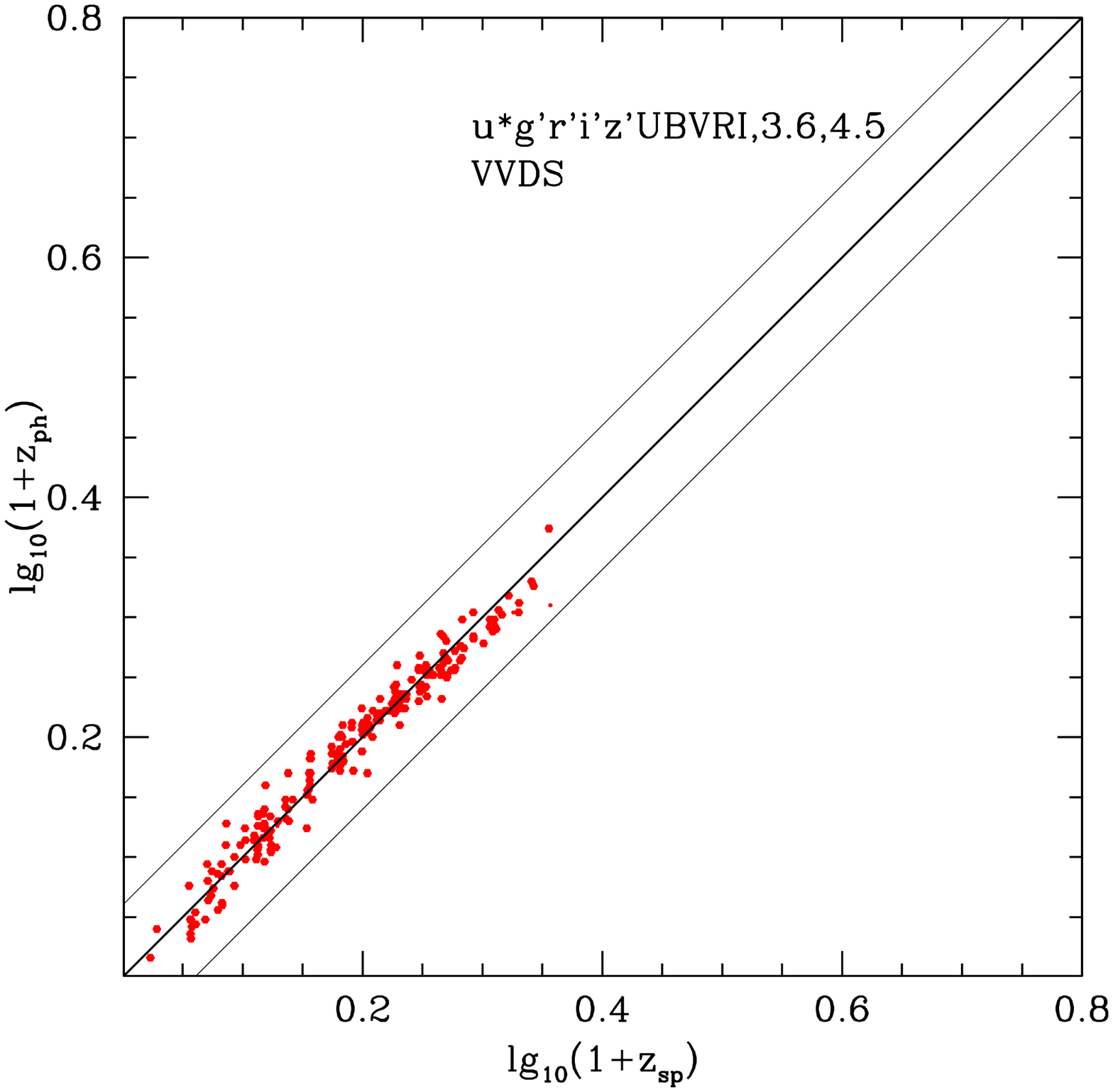,angle=0,width=7cm}
\caption{LH: Photometric versus spectroscopic redshift for SWIRE-VVDS sources, using 10 optical bands
(ugrizBVRI) and requiring r$<$23.5. Larger dots are those classified spectroscopically as galaxies, while
small denotes are classified spectroscopically as quasars (but our code selects a galaxy template).
RH: Photometric versus spectroscopic redshift for SWIRE-VVDS sources, using 10 optical bands
(ugrizBVRI) + 3.6, 4.5 $\mu$m, and requiring r$<$23.5, S(3.6)$> 7.5 \mu Jy$.
The tram-lines in this and subsequent plots correspond to 
$\Delta log_{10}(1+z) = \pm 0.06$, ie $\pm 15\%$, our boundary for catastrophic outliers.}
\end{figure*}

\begin{figure*}
\epsfig{file=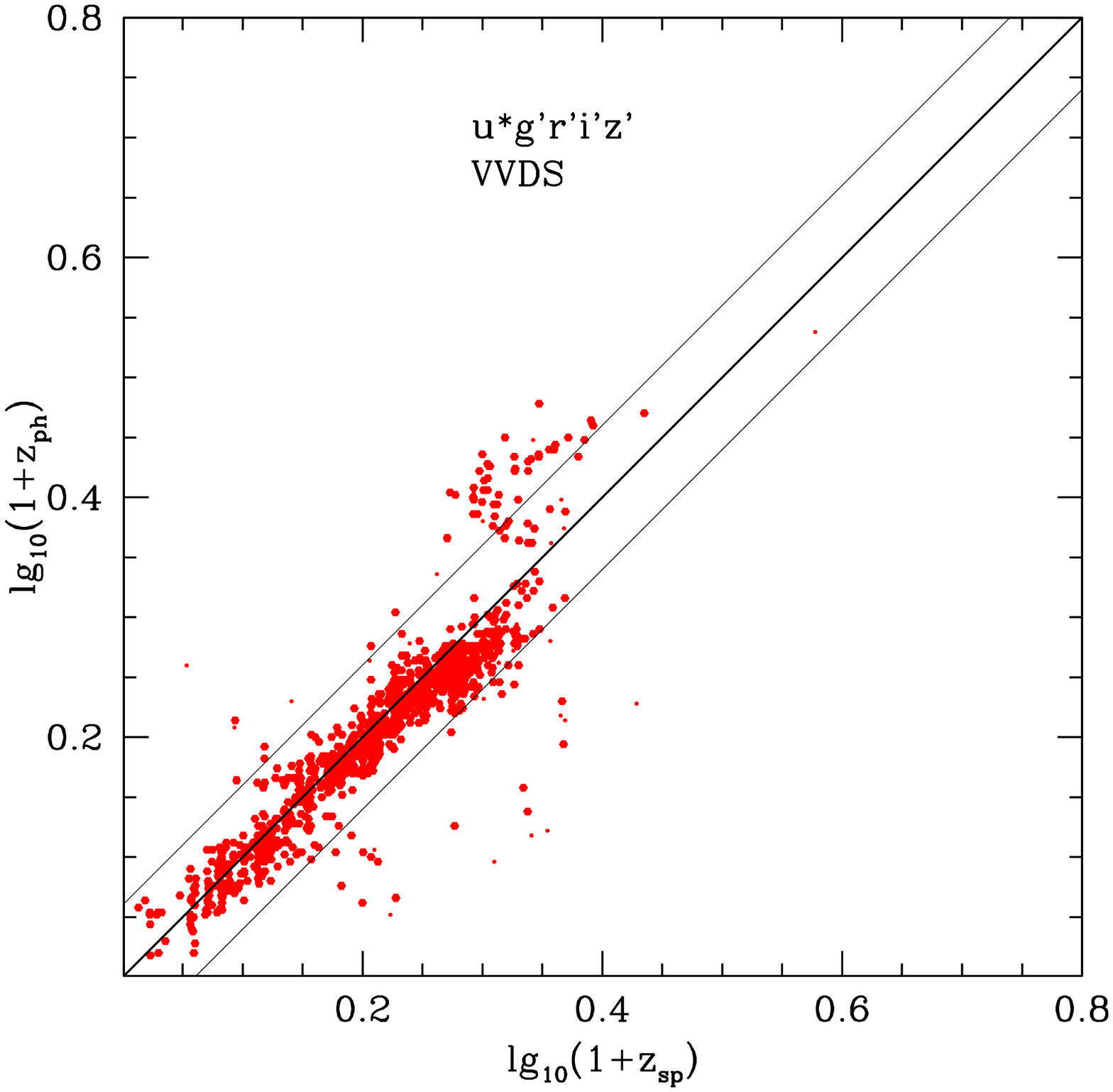,angle=0,width=7cm}
\epsfig{file=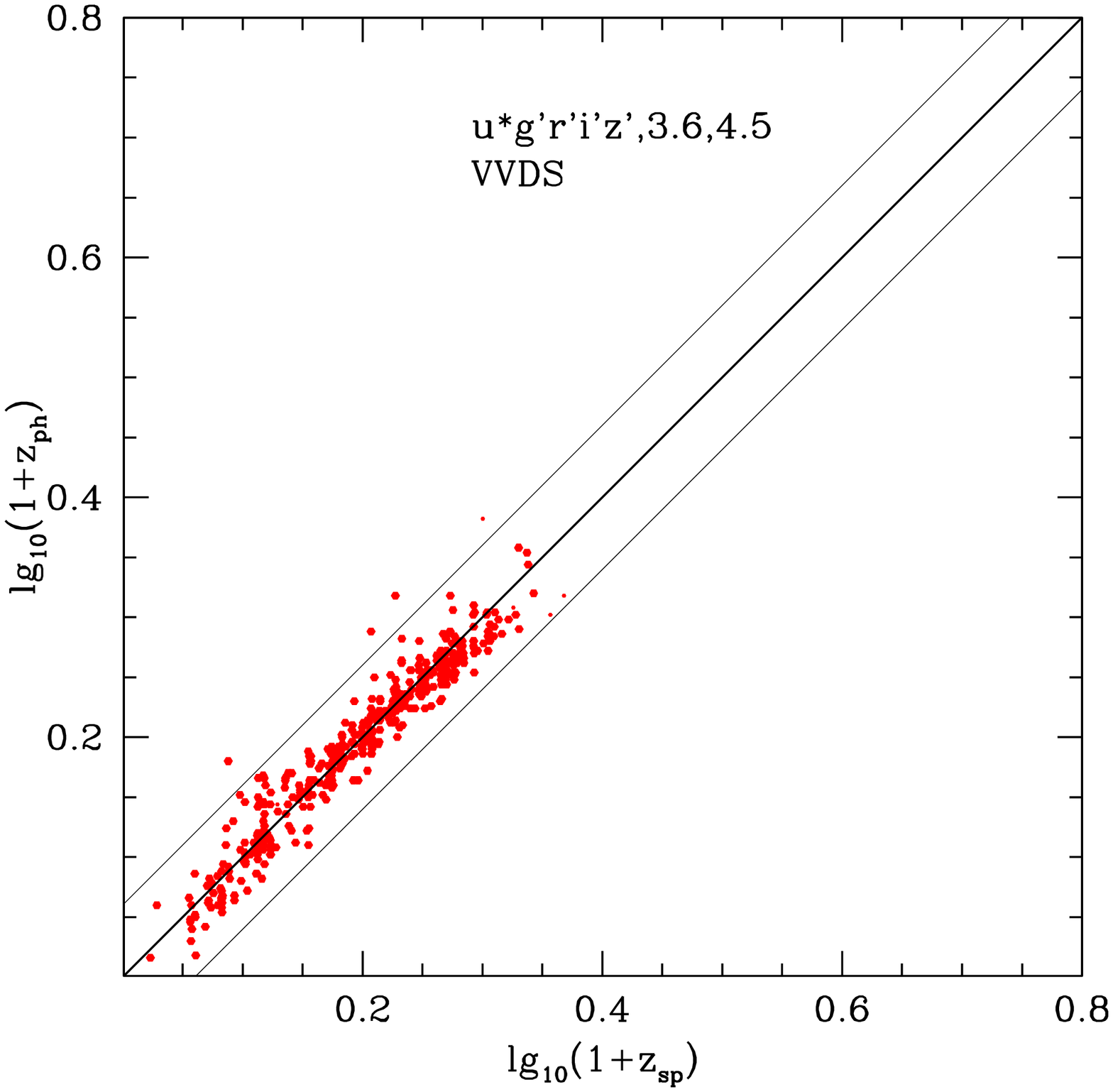,angle=0,width=7cm}
\caption{LH: Photometric versus spectroscopic redshift for SWIRE-VVDS sources, using 5 optical bands only (ugriz)
and requiring r$<$23.5. 
RH: Photometric versus spectroscopic redshift for SWIRE-VVDS sources, using 5 optical bands (ugriz)
+ 3.6, 4.5 $\mu$m, and requiring r$<$23.5, S(3.6)$> 7.5 \mu Jy$.
 }
\end{figure*}

\begin{figure*}
\epsfig{file=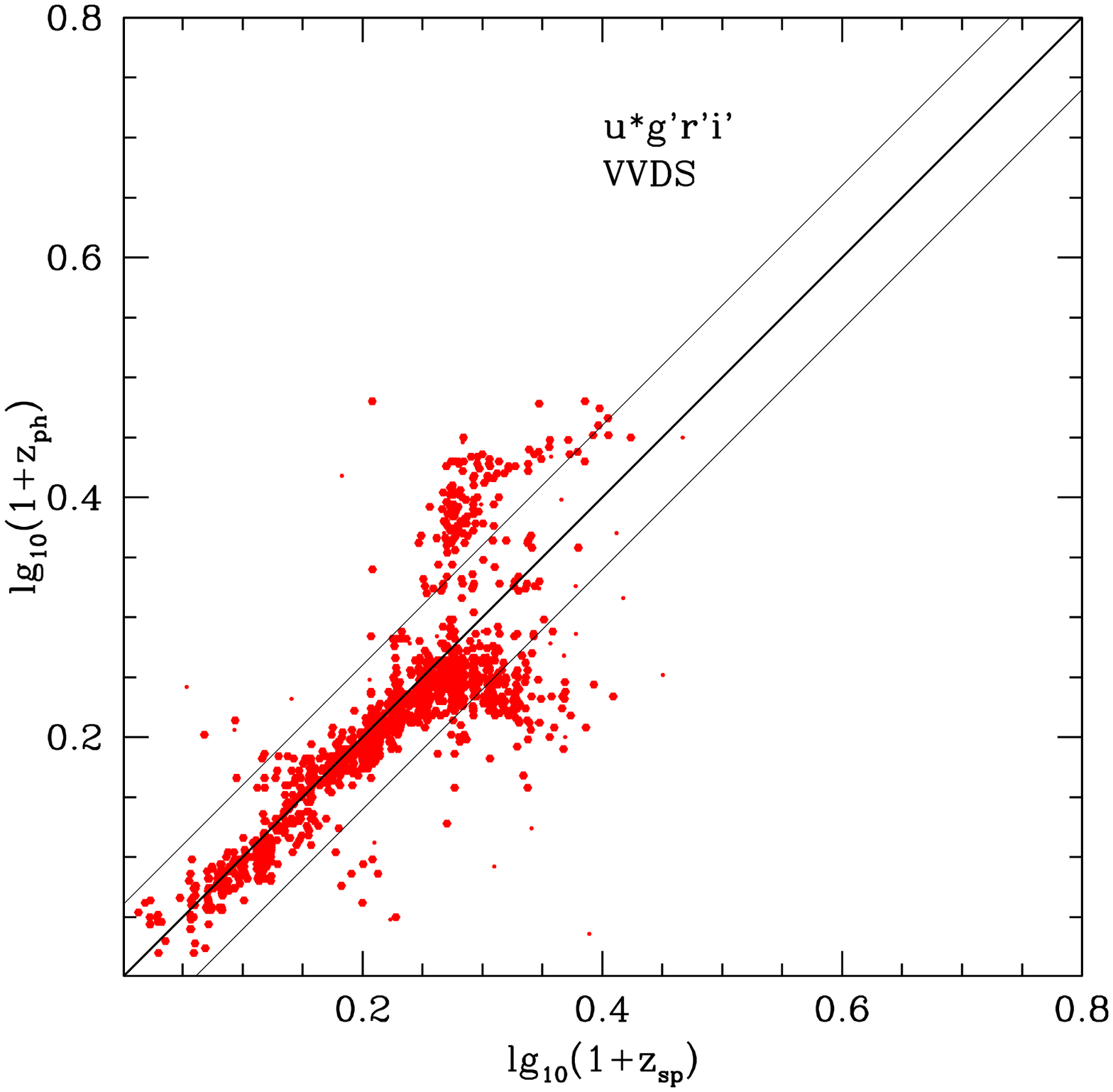,angle=0,width=7cm}
\epsfig{file=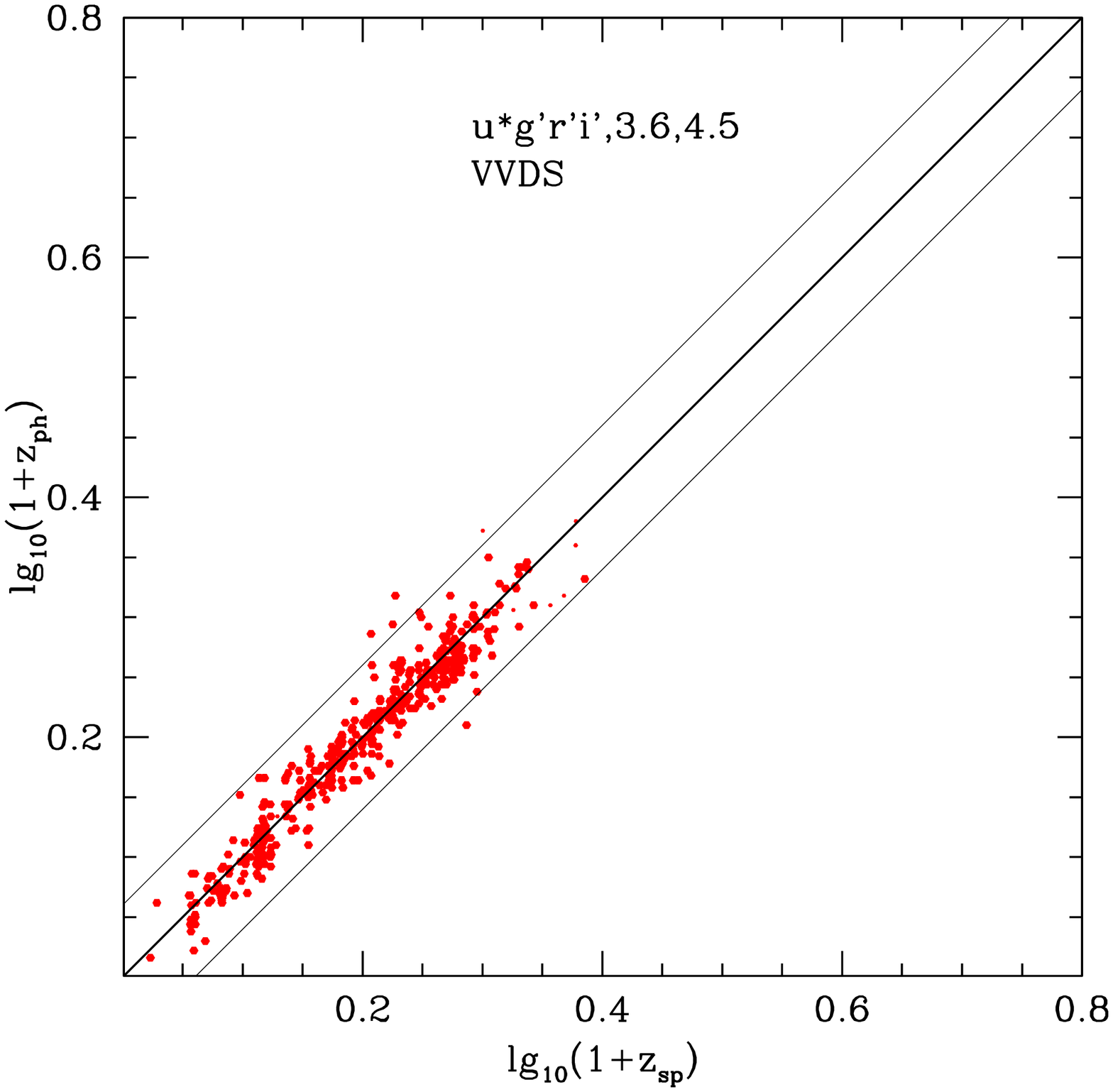,angle=0,width=7cm}
\caption{LH: Photometric versus spectroscopic redshift for SWIRE-VVDS sources, using 4 optical bands only
(ugri) and requiring r$<$23.5.
RH: Photometric versus spectroscopic redshift for SWIRE-VVDS sources, using 4 optical bands
(ugri)+ 3.6, 4.5 $\mu$m, and requiring r$<$23.5, S(3.6)$> 7.5 \mu Jy$.
}
\end{figure*}

\begin{figure*}
\epsfig{file=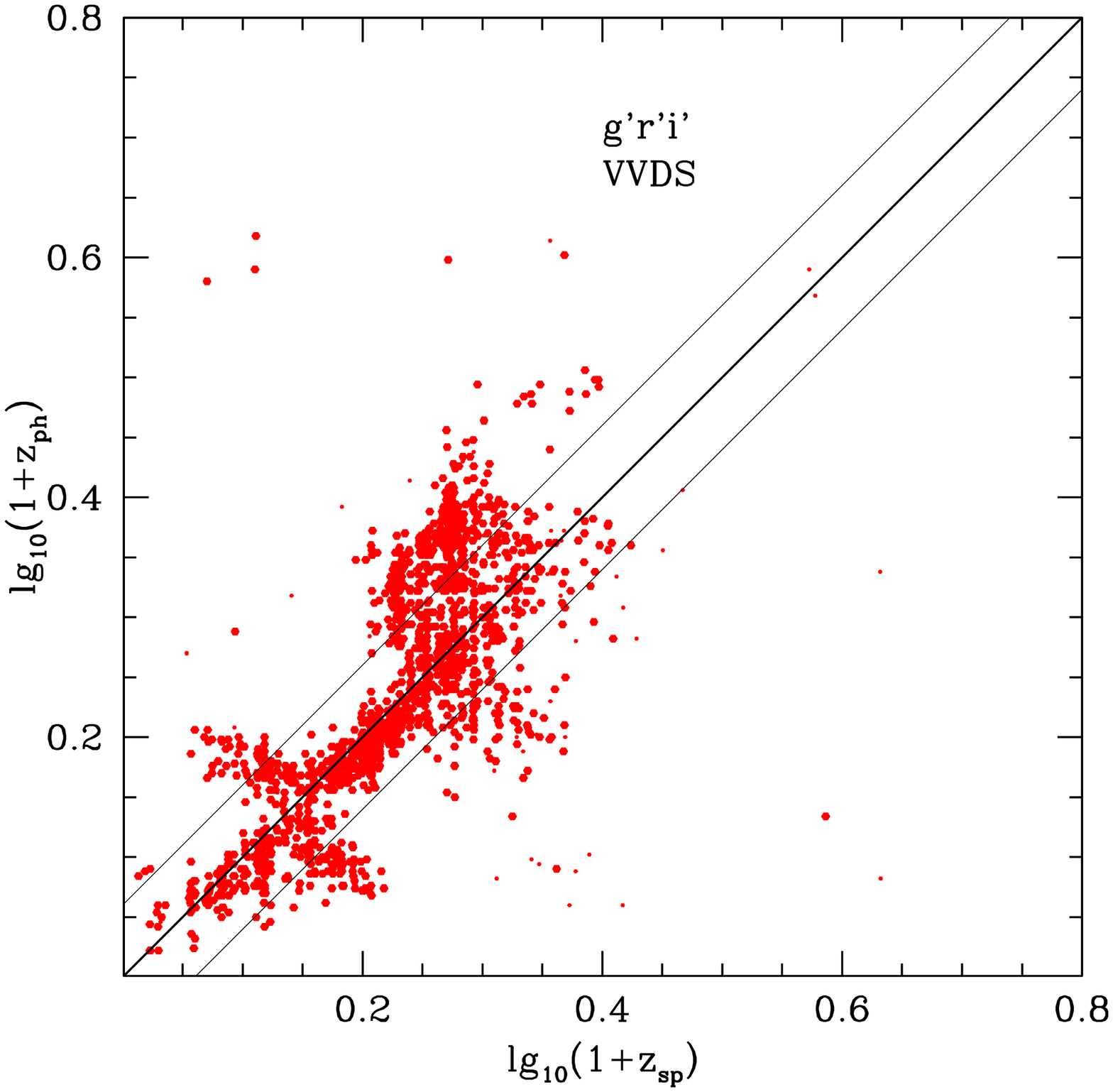,angle=0,width=7cm}
\epsfig{file=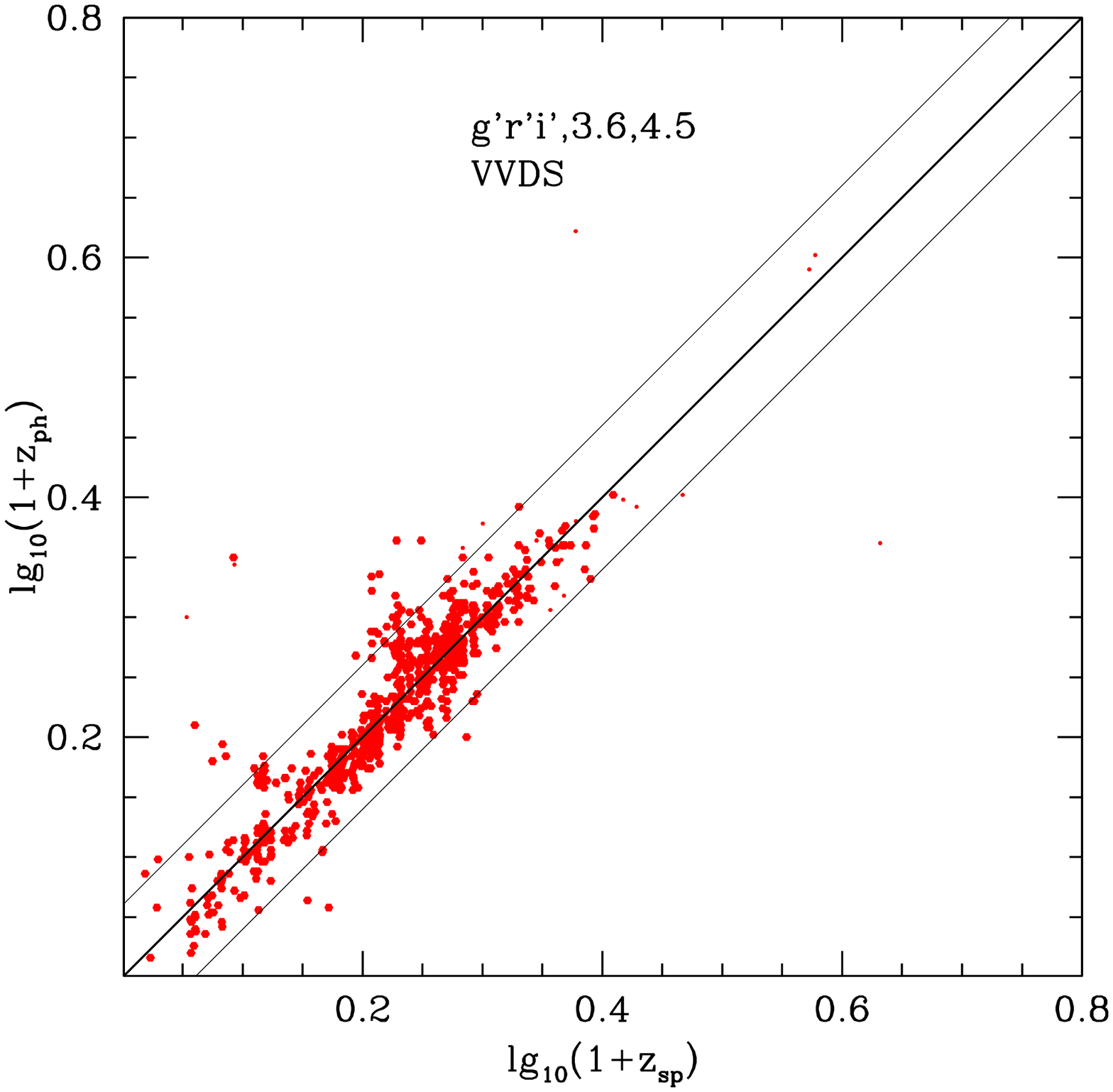,angle=0,width=7cm}
\caption{LH: Photometric versus spectroscopic redshift for SWIRE-VVDS sources, using 3 optical bands only (gri)
and requiring r$<$23.5.
RH: Photometric versus spectroscopic redshift for SWIRE-VVDS sources, using 3 optical bands (gri)
+ 3.6, 4.5 $\mu$m, and requiring r$<$23.5, S(3.6)$> 7.5 \mu Jy$.
}
\end{figure*}

\begin{figure*}
\epsfig{file=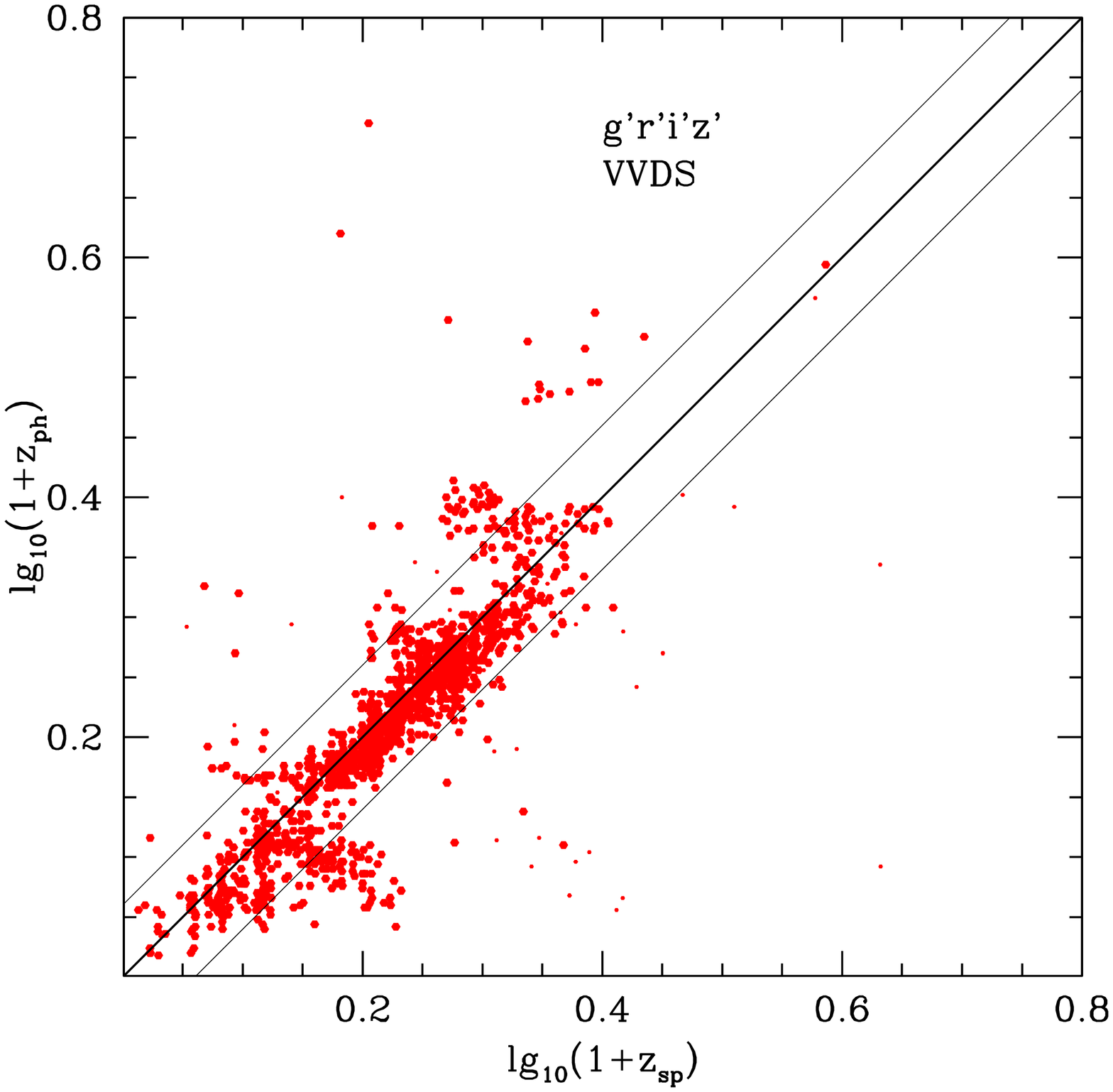,angle=0,width=7cm}
\epsfig{file=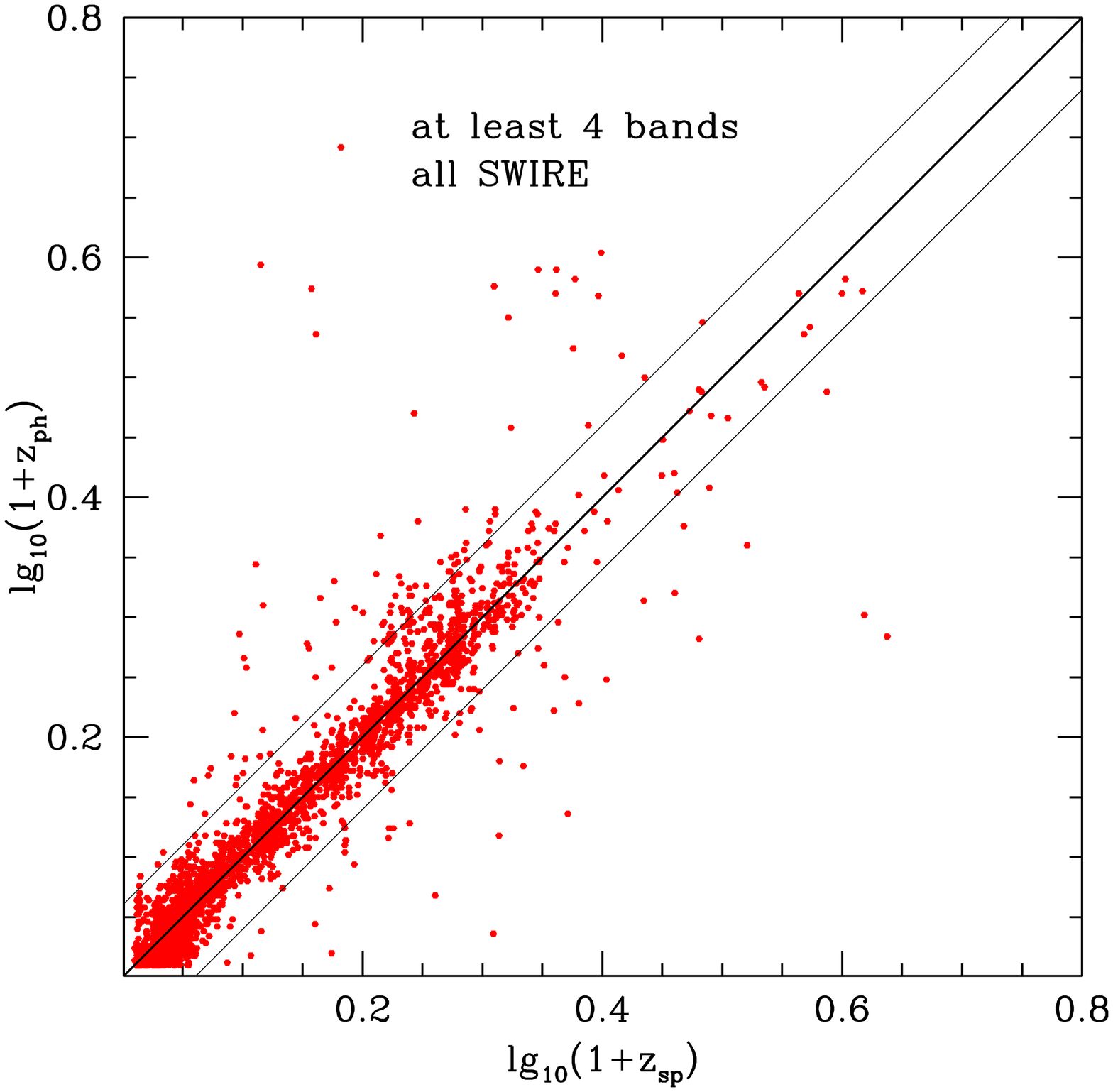,angle=0,width=7cm}
\caption{LH: Photometric versus spectroscopic redshift for SWIRE-VVDS sources, using 4 optical bands only
(griz) and requiring r$<$23.5.
RH: Photometric versus spectroscopic redshift for all SWIRE sources with spectroscopic redshift, using 
at least 4 bands in total and requiring r$<$23.5, S(3.6)$> 7.5 \mu Jy$}
\end{figure*}

\begin{figure*}
\epsfig{file=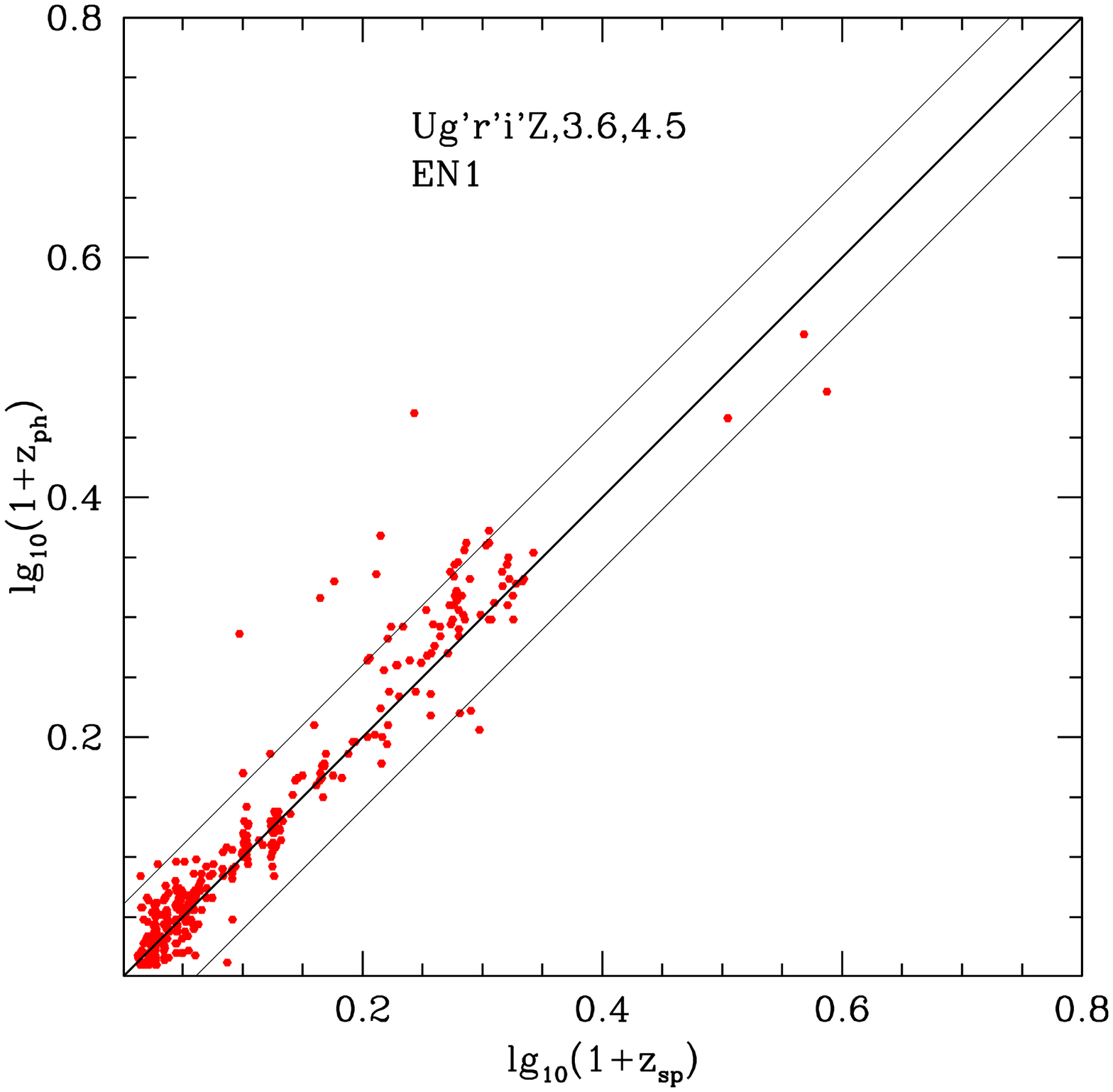,angle=0,width=7cm}
\epsfig{file=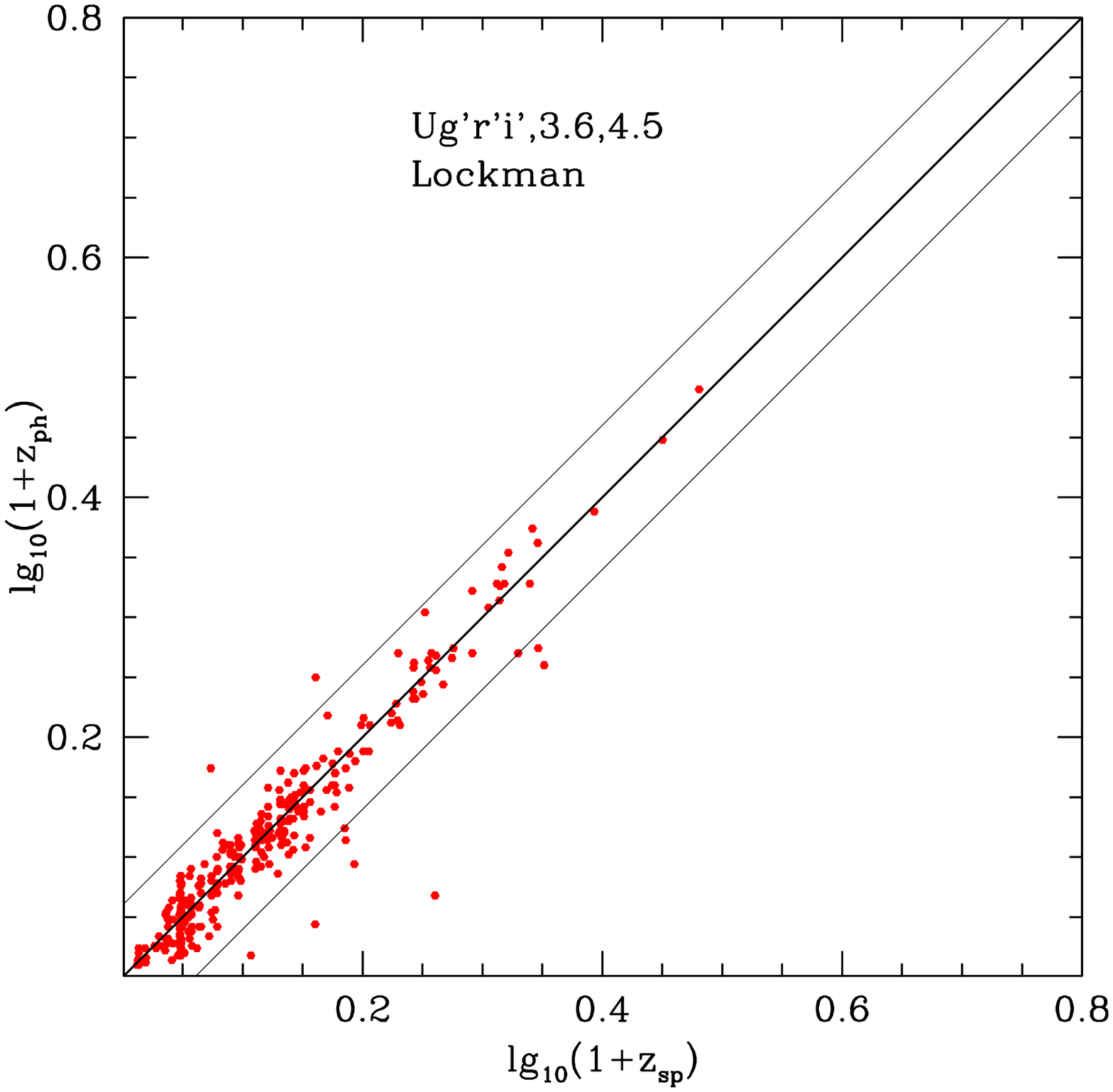,angle=0,width=7cm}
\caption{
LH: Photometric versus spectroscopic redshift for SWIRE-EN1 sources (UgriZ
+ 3.6, 4.5 $\mu$m), using at least 6 bands and requiring r$<$23.5, S(3.6)$> 7.5 \mu Jy$. Most of the spectroscopic redshifts
at z $>$ 0.5 are from the programme of Berta et al (2007), Trichas et al (2007).
RH: Photometric versus spectroscopic redshift for SWIRE-Lockman sources 
(Ugri+ 3.6, 4.5 $\mu$m), using 6 bands
and requiring r$<$23.5, S(3.6)$> 7.5 \mu Jy$.
Most of the spectroscopic redshifts
at z $>$ 0.5 are from the programme of Berta et al (2007), Smith et al (2007).}
\end{figure*}

\begin{figure*}
\epsfig{file=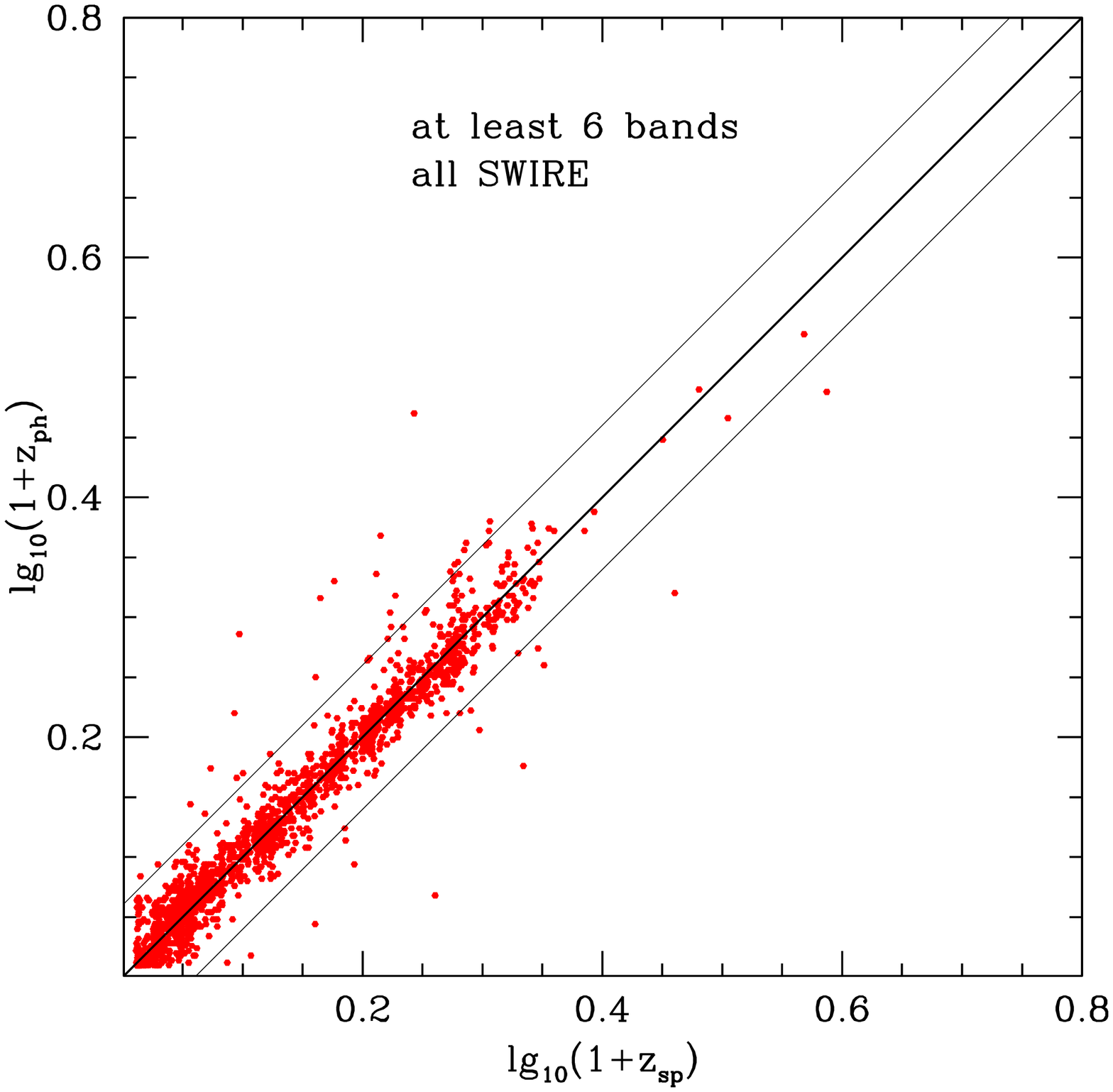,angle=0,width=14cm}
\caption{
Photometric versus spectroscopic redshift for all SWIRE galaxies with spectroscopic redshifts, using
at least 6 bands and requiring r$<$23.5, S(3.6)$> 7.5 \mu Jy$.
}
\end{figure*}

\subsection{Dependence of performance on r and $\chi^2_{\nu}$}

The rms deviation of $(z_{phot}-z_{spec})/(1+z_{spec})$ depends on the number of photometric
bands, the limiting optical magnitude, and the limiting value of $\chi^2$ (see Figs 11 and 12), 
but a typical value for galaxies is 4 $\%$, a significant improvement on our earlier work.  
For comparison typical rms values found by Rowan-Robinson (2003) and Rowan-Robinson et al (2004)
from UgriZ, JHK data alone were 9.6 and 7 $\%$, respectively. 

Figs 10 show the dependence of $log_{10}(1+z_{phot})/(1+z_{spec})$ on the r-magnitude and
on the value of the reduced $\chi^2_{\nu}$.  The percentage of outliers stars to increase for r $>$ 22.  Although
we treat $\chi^2 _{nu}>$ 10 as a failure of the photometric method (due for example to poor photometry
or confusion with a nearby optical object), there is still a good correlation of photometric and
spectroscopic redshift.  

Figs 11 show how the rms value of $log_{10}(1+z_{phot})/(1+z_{spec})$, $\sigma$ and the
percentage of outliers, $\eta$, defined as objects with $|log_{10}(1+z_{phot})/(1+z_{spec})| > 0.06$,
as a function of the total number of photometric bands.  The rms values are comparable or slightly
better than those of Ilbert et al (2006) derived from the VVDS optical data using the LE PHARE code.  For 7 or more bands 
(eg UgriZ + 3.6, 4.5 $\mu$m)
the rms performance is comparable to the 17-band estimates of COMBO-17 (Wolf et al 2004),
significantly better than those of Mobasher et al (2004) with GOODS data, and comparable to Mobasher 
et al (2007).  The outlier performance is
significantly better than Wolf et al (2004), Ilbert et al (2006), Mobasher et al (2007) , which we attribute to the use of the
SWIRE 3.6 and 4.5 $\mu$m bands.  Fig 12L show the same quantities as
a function of the limit on the reduced $\chi^2_{\nu}$, together with a histogram of $\chi^2_{\nu}$ values, for SWIRE 
sources with at least 7 photometric bands.
The figures for COMBO-17
data were derived from their published catalogue using exactly the same procedure as for the SWIRE data.
Those for GOODS were taken from Table 1 of Mobasher et al (2004).  Of course, where the number of optical
bands is greater than 5, there is generally considerable overlap between the bands (typically U'griZ' plus UBVRI) , 
so the number of bands is not a totally fair measure of the amount of independent data being used (this applies to
our analysis of the VVDS data, and to the GOODS and COMBO17 analyses).  

While the rms values found in several different studies are comparable, the dramatic improvement here is in the
reduction of the number of outliers, when sufficient photometric bands are available.  The remaining outliers tend to be 
due either to bad photometry in one band or to aliases.

\begin{figure*}
\epsfig{file=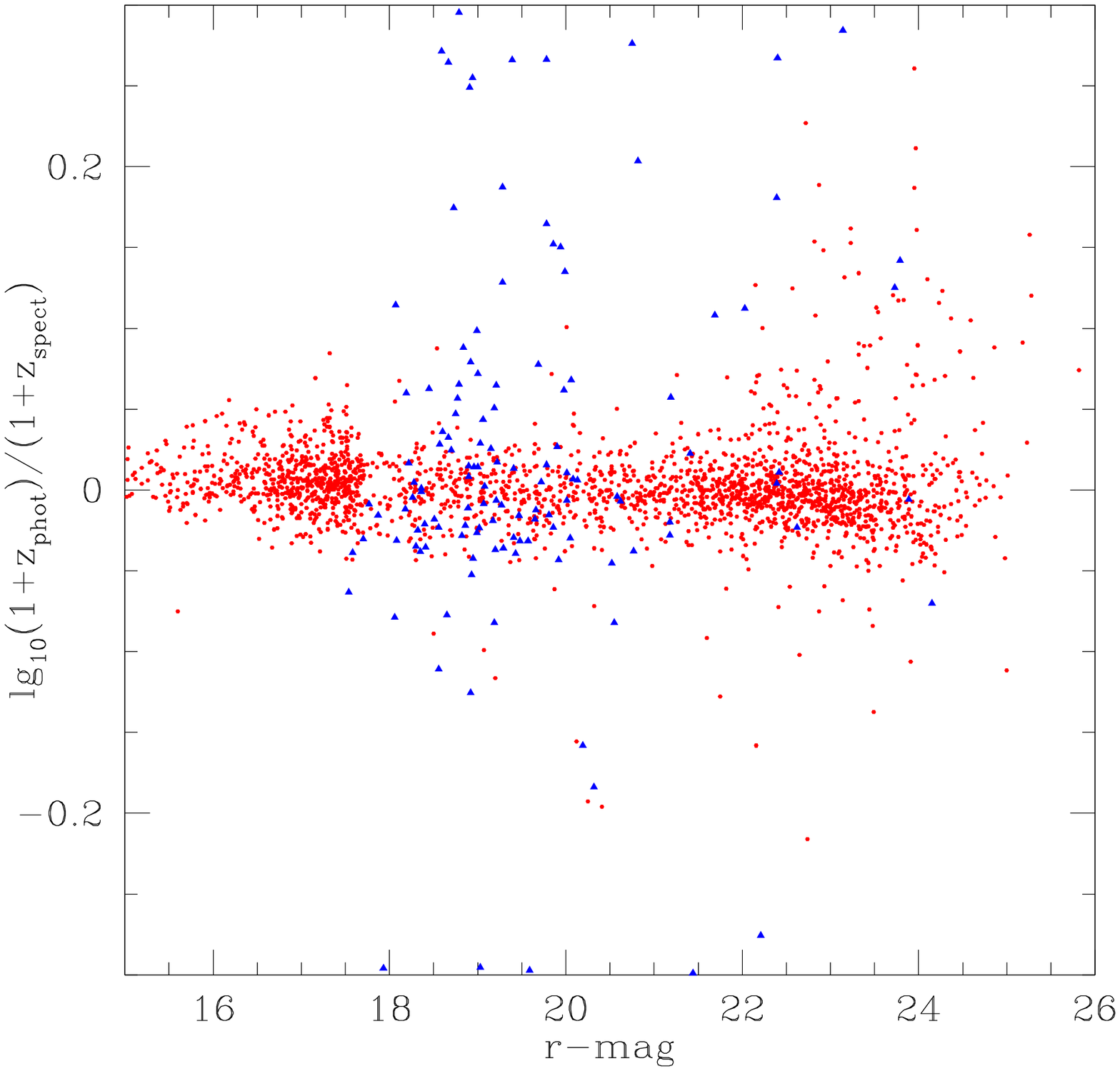,angle=0,width=7cm}
\epsfig{file=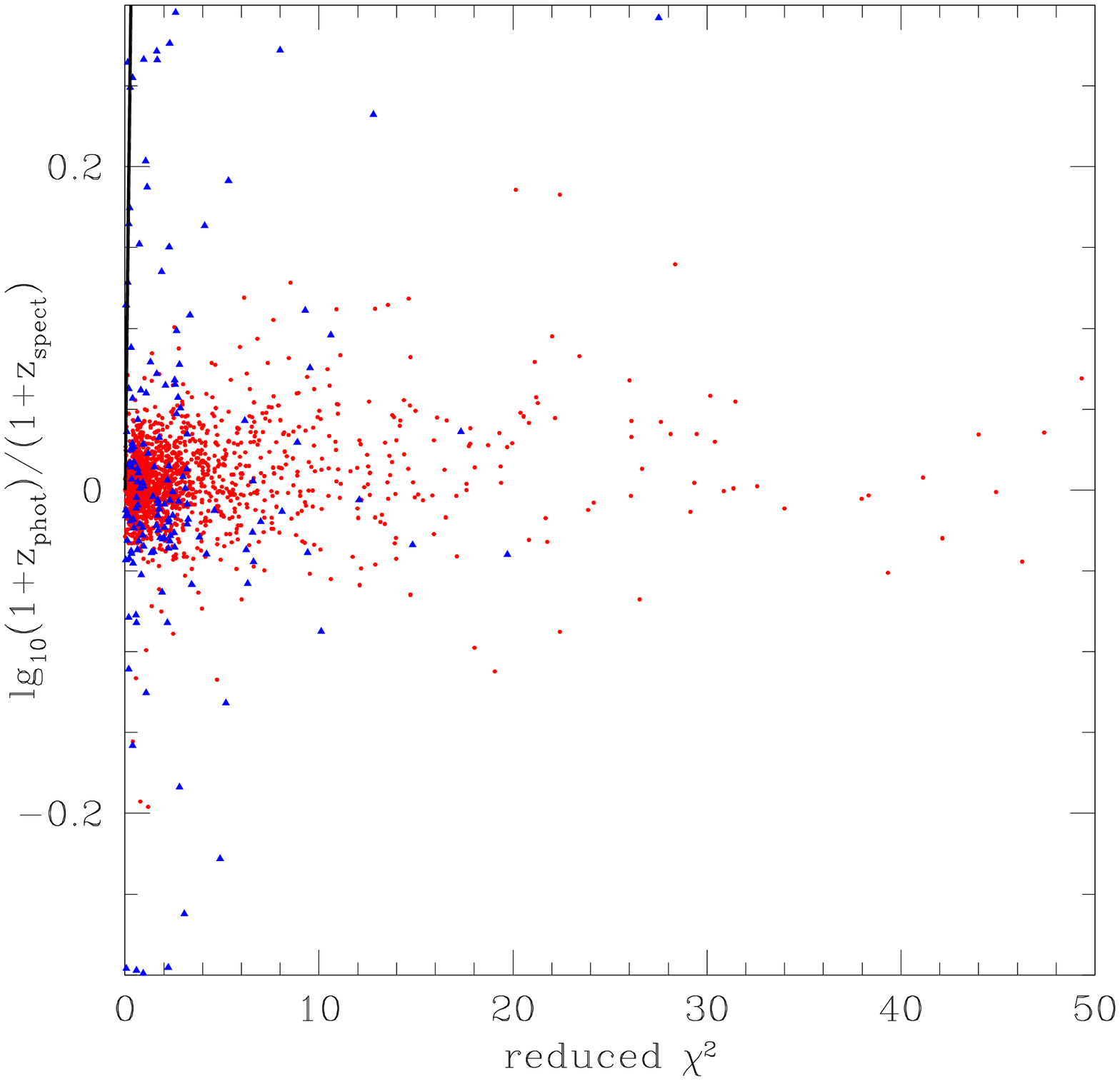,angle=0,width=7cm}
\caption{
LH: $log_{10}(1+z_{phot}/(1+z_{spec})$ versus r for all SWIRE sources.  Red = galaxies, blue = quasars.
RH: $log_{10}(1+z_{phot}/(1+z_{spec})$ versus $\chi^2$ for all SWIRE sources.
}
\end{figure*}

\begin{figure*}
\epsfig{file=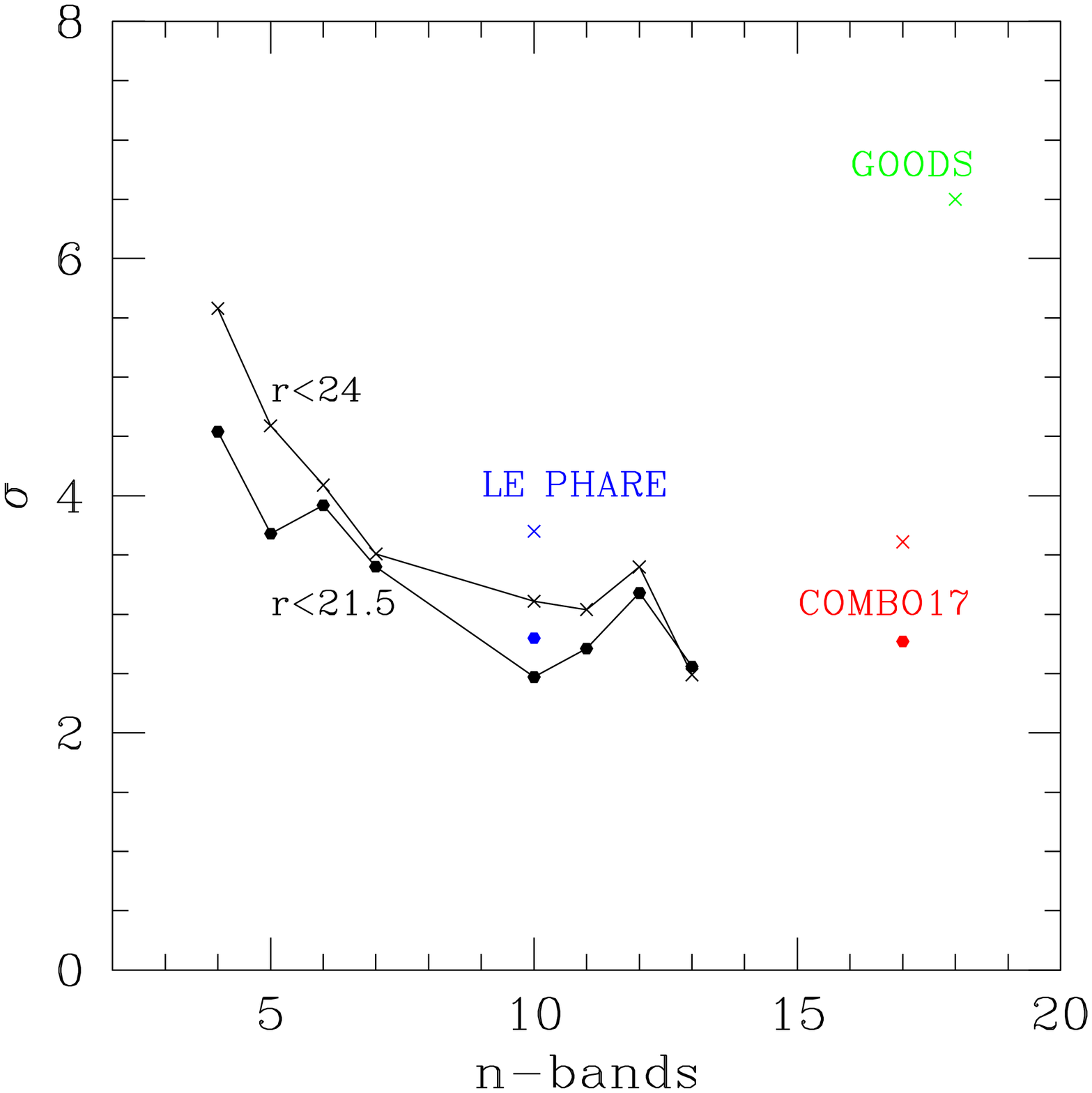,angle=0,width=7cm}
\epsfig{file=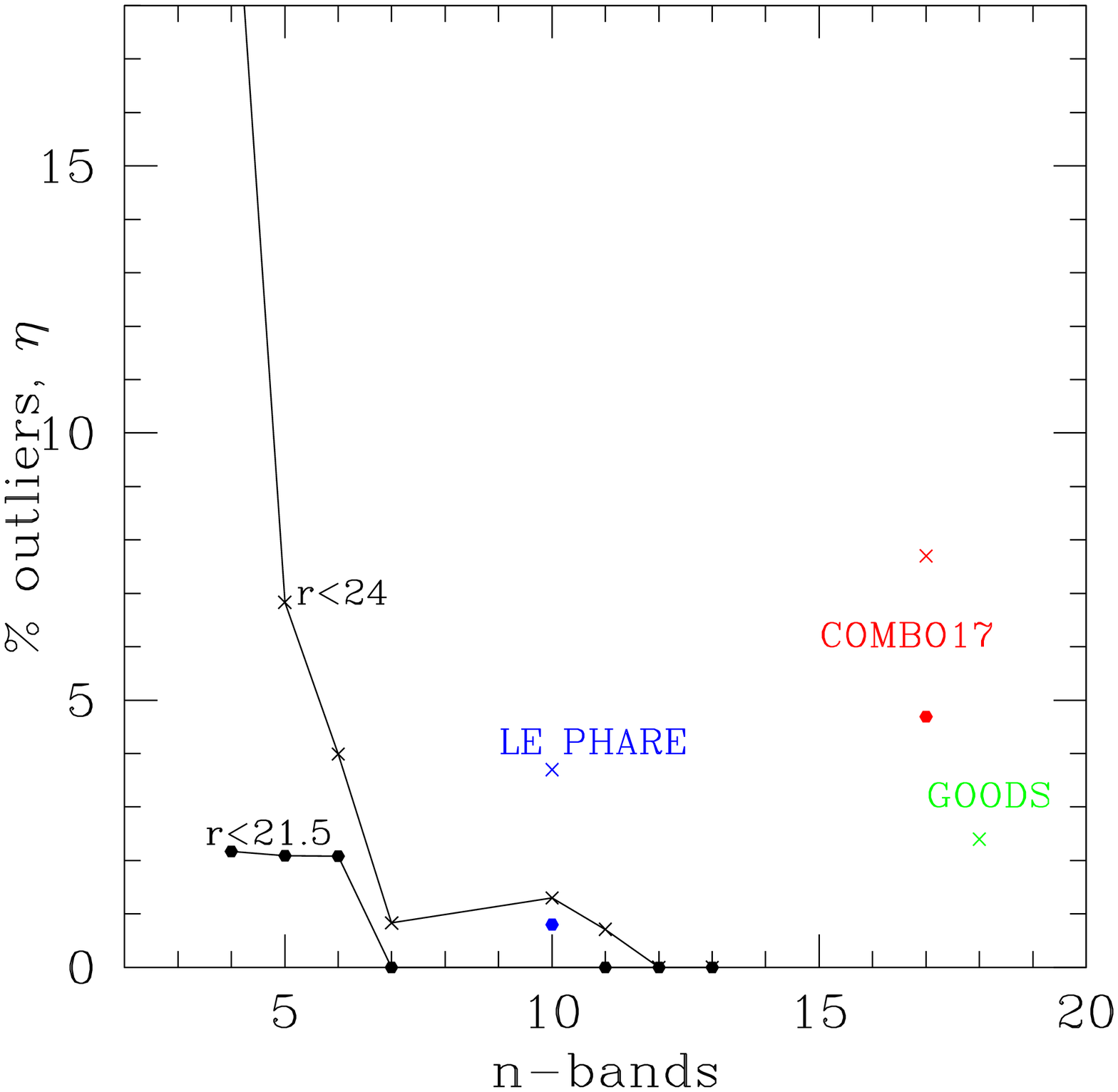,angle=0,width=7cm}
\caption{
LH: $\%$ rms of $(z_{phot}-z_{spec})/(1+z_{spec})$, excluding outliers, versus total number of photometric bands 
for SWIRE sources.  Figures for 10 or more bands are from the SWIRE-VVDS survey.  Figures for
COMBO-17 (Wolf et al 2004) and GOODS (Mobasher et al 2004) are also shown.
RH: $\eta, \%$ outliers versus total number of photometric bands for SWIRE sources. 
Filled circles are values for $r<21.5$, crosses for $r<24$.}
\end{figure*}

\begin{figure*}
\epsfig{file=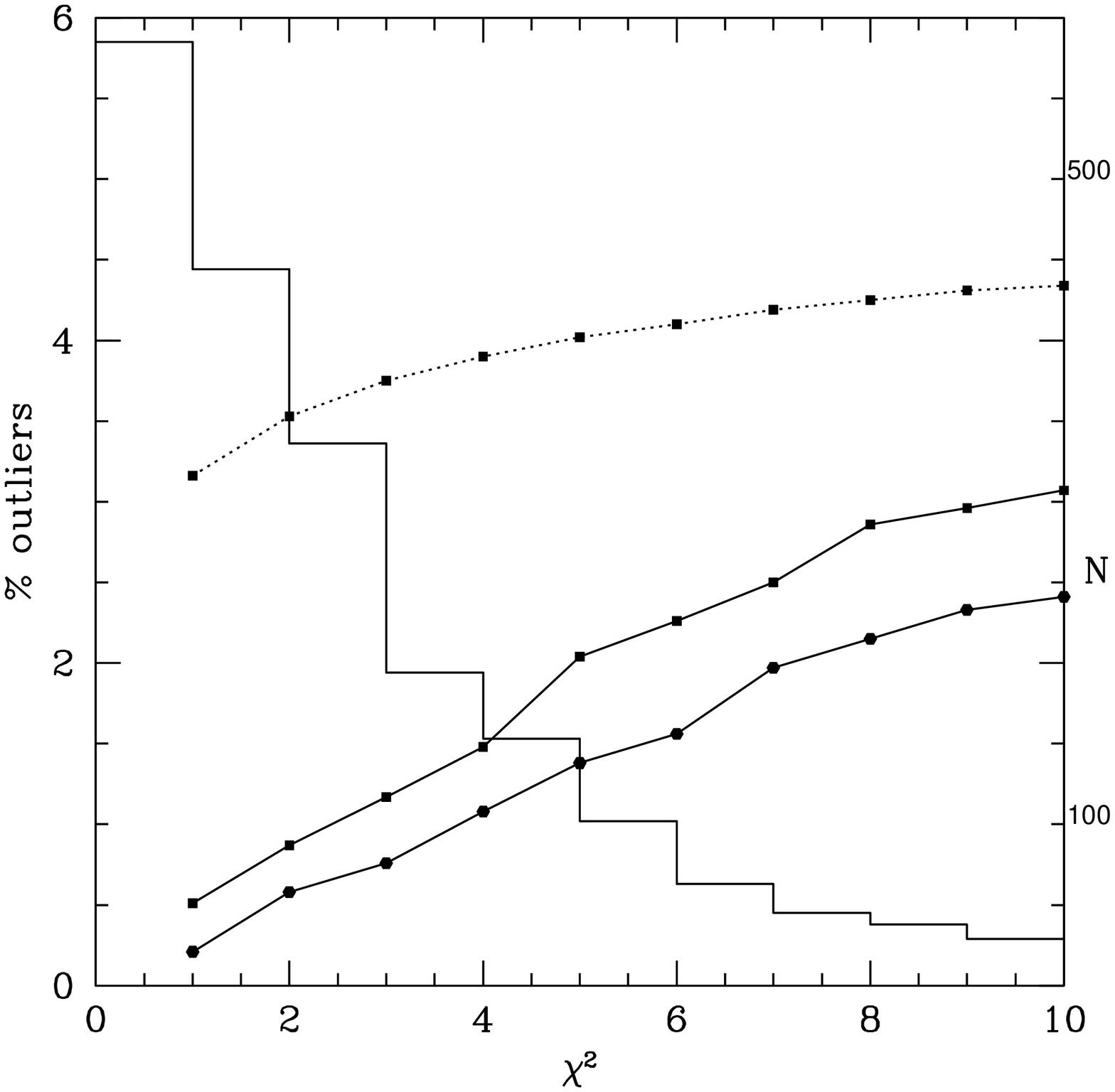,angle=0,width=7cm}
\epsfig{file=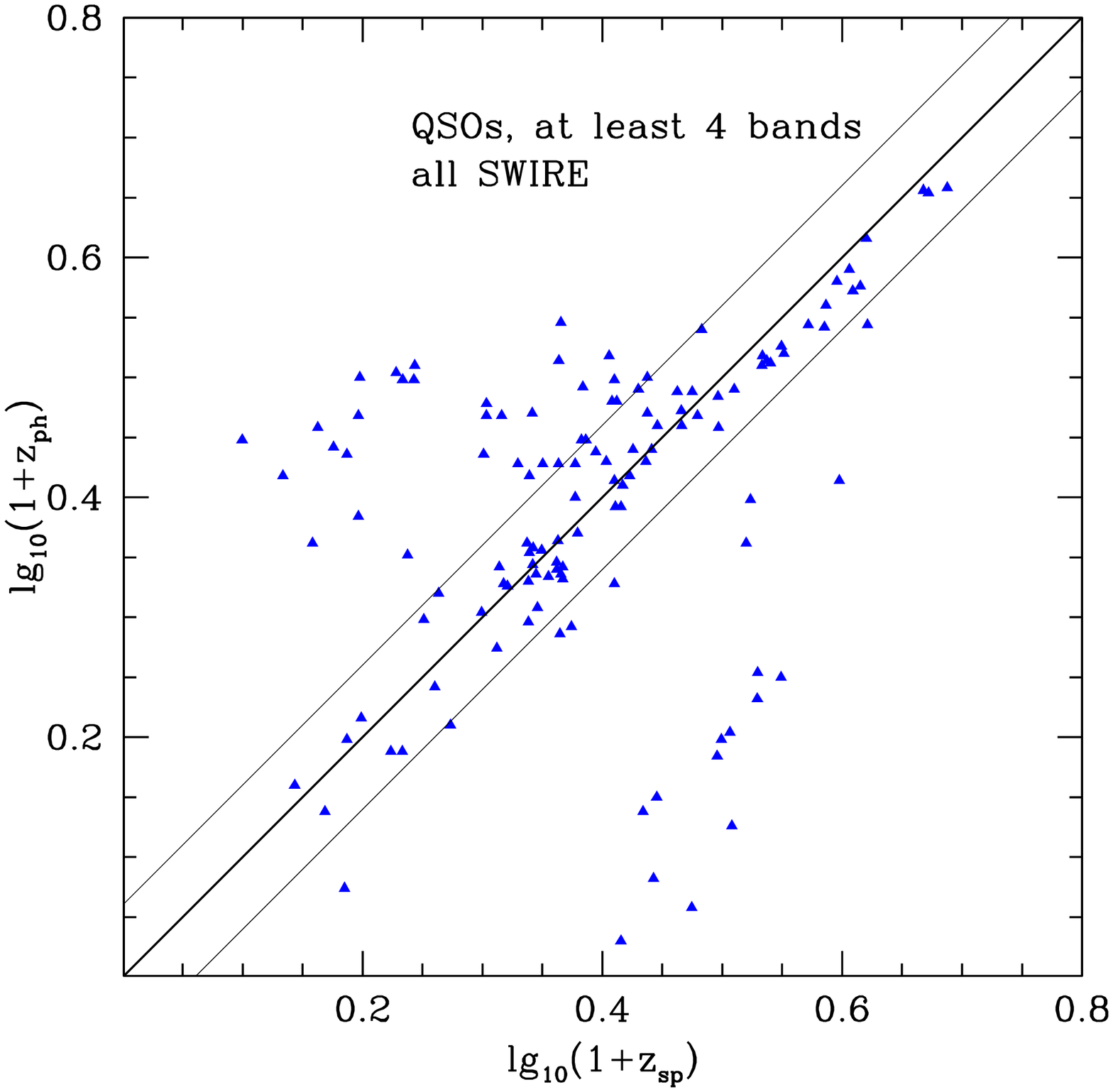,angle=0,width=7cm}
\caption{
LH: $\%$ rms of $(z_{phot}-z_{spec})/(1+z_{spec})$ versus $\chi^2$ limit (dotted curve, LH scale)
for SWIRE sources with at least 7 photometric bands.  $\%$ outliers versus $\chi^2_{\nu}$ limit 
(solid curves, LH scale, upper curve r$<$24, lower curve r$<$21.5). Also shown is a histogram of $\chi^2_{\nu}$ values (RH scale).
RH: Photometric versus spectroscopic redshift for SWIRE quasars detected in
4 photometric bands. 
}
\end{figure*}

\subsection{Performance for QSOs}

The photometric estimates of redshift for AGN are more uncertain than those for galaxies, due to aliassing problems,
but the code is effective at identifying Type 1 AGN from the optical and near ir data.  
For some quasars there is significant torus dust emission in the 3.6 and 4.5 $\mu$m bands,
and since the strength of this component varies from object to object,
inclusion of these bands in photometric redshift determination with a fixed template can make the fit worse rather than better.
We have therefore omitted the 3.6 and 4.5 $\mu$m bands if S(3.6)/S(r) $>$ 3.
Note that only 1.75 $\%$ of SWIRE sources are identified by the photometric redshift code as Type 1 AGN, and
of these only 5$\%$ are found to have $A_V > 0.5$. 

Figure 12R shows photometric versus spectroscopic redshift for SWIRE quasars
detected in at least 4 photometric bands.  One third of QSOs (53/158) have
$|log_{10}(1+z_{phot})/log_{10}(1+z_{spec})|>0.10$ and the rms deviation for
 the remainder is 9.3$\%$.   Many of the outliers are cases where because an
almost power-law SED is being fitted, the redshift uncertainty is very wide indeed.

Redshift estimates for AGN can be affected by optical variability since optical photometry for
different bands may have been taken at different epochs, as in the INT-WFS photometry in
EN1 and EN2 (Afonso-Luis et al 2004).

Richards et al (2001) and Ball et al (2007) have given similar estimates of rms and outlier rate
for photometric redshifts for quasars.  Because we are rarely able to use 3.6 and 4.5 $\mu$m data 
in the photometric redshift solution, we do not expect to have any advantage over fits using 
purely optical data.  But it is worth emphasizing that our code appears to be effective in selecting
the Type 1 AGN without any spectroscopic information.

\section{Catalogue description}

The SWIRE Photometric Redshift Catalogue consists of 1025119 sources, split 
between the SWIRE fields as follows: EN1 (218117), EN2 (125364), Lockman
(229238), SWIRE-VVDS (34630), SWIRE-SXDS (55432), XMM (excluding VVDS and SXDS) (212572), Chandra DFS
(149766).  It is available via IRSA and also at
$http://astro.ic.ac.uk/\sim mrr/swirephotzcat$

The columns listed are: 

\noindent nidir(i8), infrared identifier

\noindent nidopt(i8), optical identifier

\noindent ra(f11.6), RA in degrees

\noindent dec(f11.6), dec in degrees

\noindent s36(f10.2), S(3.6$\mu$m) in $\mu$Jy

\noindent s24(f10.2), S(24$\mu$m) in $\mu$Jy

\noindent s70(f10.2), S(70$\mu$m) in $\mu$Jy

\noindent am3(f8.2), r-band mangnitude (AB magnitudes for XMM-LSS, Vega for all the other regions)

\noindent j1(i3), optical template type (1-11 galaxies (1 E, 2 E', 3 Sab, 5 Sbc, 7 Scd, 9 Sdm, 11 sb), 13-15 QSOs) for $A_V$ = 0 solution

\noindent alz(f7.3), $log_{10}(1+z_{phot})$ for $A_V$ = 0 solution

\noindent err0(f10.3), reduced $\chi^2$ for $A_V$ = 0 solution

\noindent j2(i3), optical template type (1-11 galaxies, 13-15 QSOs) for free $A_V$  solution

\noindent alz2(f7.3), $log_{10}(1+z_{phot})$ for free $A_V$ solution

\noindent av1(f6.2), $A_V$

\noindent err1(f10.3), reduced $\chi^2$ for free $A_V$ solution

\noindent n91(i3), total number of photometric bands in solution

\noindent nbopt(i3), number of optical bands in solution

\noindent amb2(f8.2), $M_B$ (corrected for extinction) for free $A_V$ solution

\noindent alb(f8.2), $log_{10} L_B$, in solar units

\noindent spectz0(f10.5), spectroscopic redshift

\noindent nzref0(i4), reference for spectroscopic redshift [N1,N2: nzref0 = 1-9, as in Rowan-Robinson et al 2003;
= 11 Keck redshifts (Berta et al 2007), = 20 WIYN redshifts (Trichas et al 2007), = 25 GMOS redshifts (Trichas et al 
2007), = 30 (Swinbank et al 2007);
VVDS: nzref0 = 3-14 VVDS quality flag; Lockman: nzref0 = -1, 1 to 4 (Steffen et al 2004), = 5 NED, = 10 WIYN, Keck, 
Gemini redshifts (Smith et al 2007), = 11 Keck redshifts (Berta et al 2007); XMM, CDFS: nzref0 = 5 NED]

\noindent nir(i3), number of bands with infrared excess

\noindent alp1(f6.2), relative amplitude of cirrus component at 8 $\mu$m (s.t. alp1+alp2+alp3+alp4=1)

\noindent alp2(f6.2), relative amplitude of M82 starburst component at 8 $\mu$m

\noindent alp3(f6.2), relative amplitude of AGN dust torus component at 8 $\mu$m

\noindent alp4(f6.2), relative amplitude of A220 component at 8 $\mu$m

\noindent errir3(f10.3), reduced $\chi_{\nu}^2$ of infrared template fit

\noindent alcirr(f8.2), $log_{10} L_{cirr}$ in solar units (1-1000 $\mu$m), upper limits shown -ve, assuming alp1 $\le$0.05

\noindent alsb(f8.2), $log_{10} L_{M82}$ in solar units, upper limits shown -ve, assuming alp2 $\le$0.05

\noindent alagn(f8.2), $log_{10} L_{tor}$ in solar units, upper limits shown -ve, assuming alp3 $\le$0.05

\noindent al220(f8.2), $log_{10} L_{A220}$ in solar units, upper limits shown -ve, assuming alp4 $\le$0.05

\noindent alir(f8.2), $log_{10} L_{ir}$ in solar units

\noindent nirtem(i3), ir template type (dominated by cirrus(1), M82 starburst(2),
A220 starburst(3), AGN dust torus(4), single band excess(5), no excess(6)

\noindent als70(f6.2), $log_{10} S70$, predicted from ir template fit, in mJy

\noindent als160(f6.2), $log_{10} S160$, predicted from ir template fit, in mJy

\noindent als350(f6.2), $log_{10} S350$, predicted from ir template fit, in mJy

\noindent als450(f6.2), $log_{10} S450$, predicted from ir template fit, in mJy

\noindent als850(f6.2), $log_{10} S850$, predicted from ir template fit, in mJy

\noindent als1250(f6.2), $log_{10} S1250$, predicted from ir template fit, in mJy

\noindent al36(f8.2), $log_{10} L(3.6\mu)$ in units of $L_{\odot}$

\noindent alm(f8.2) $log_{10} M_*/M_{\odot}$, stellar mass in solar units

\noindent alsfr(f8.2) $log_{10} sfr$, star formation rate in $M_{\odot} yr^{-1}$

\noindent almdust(f8.2) $log_{10} M_{dust}/M_{\odot}$, dust mass in solar units

\noindent chi2(85f6.2), array of reduced $\chi_{\nu}^2$ as function of alz2, minimized over all templates, in bins of 0.01 in $log_{10} (1+z_{ph})$, from 0.01-0.85.

\section{Redshift distributions}
Figure 13L shows the redshift distributions derived in this way for SWIRE-SXDS sources with S(3.6) $>$
5 $\mu$Jy, above which flux the optical identifications (to r $\sim$ 27.5) are relatively complete,
with a breakdown into elliptical, spiral + starburst and quasar SEDs based on the photometric redshift fits.  
Fig 13R shows the corresponding histograms for the whole SWIRE catalog.  Here we have subdivided this
into E, Sab, Sbc, Scd, starbursts and quasars.  The latter distribution starts to cut off at slightly lower redshift, $\sim$1.5,
because the typical depth of the optical data is r $\sim$24-25 instead of 27.5.  

A small secondary redshift peak appears
for galaxies at z $\sim$3.  About two-thirds of these have aliases at lower redshifts and some of these z $\sim$3
sources are spurious (see Fig 7R).  However there is a real effect favouring detection of galaxies at z $\sim$3,
that the starlight peak at $\sim 1 \mu$m entering the IRAC bands
generates a negative K-correction at 3.6 and 4.5 $\mu$m.  Spectroscopy is needed to ascertain the reality of this
peak.

Ellipticals cut off fairly sharply at z $\sim$1.  At z$>$2 sources tend to be starbursts or quasars.  The structure seen
in the redshift distribution may be partly a result of redshift aliasing, but bearing in mind the estimated accuracy
of these redshifts (typically 4$\%$ in (1+z)), some of the grosser features may indicate large-scale structure 
within the SWIRE fields.  Fig 14L shows the
redshift distribution in EN1, in which peaks appear at z $\sim$ 0.3, 0.5, 0.9, 1.1.  From Fig 8L we see that the
spectroscopic redshifts show clear clusters at z = 0.31, 0.35, 0.47, 0.9 and 1.1, so the peaks in Fig 14L
do seem to correspond to real structure.  The peak at z $\sim$ 0.9 corresponds to the supercluster seen by
Swinbank et al (2007).

Figure 14R shows the redshift distribution for SWIRE 24 $\mu$m sources with S(24) $> 200 \mu$Jy.  Here the broad
peak at redshift $\sim$1 is partly due to the shifting of the 11.7 $\mu$m PAH feature through the 24 $\mu$m band. 

10$\%$ of sources in the SWIRE photometric redshift catalogue have z $>$2, and 4$\%$ have z$>$3, so this catalogue 
is a huge resource for high redshift galaxies.

\begin{figure*}
\epsfig{file=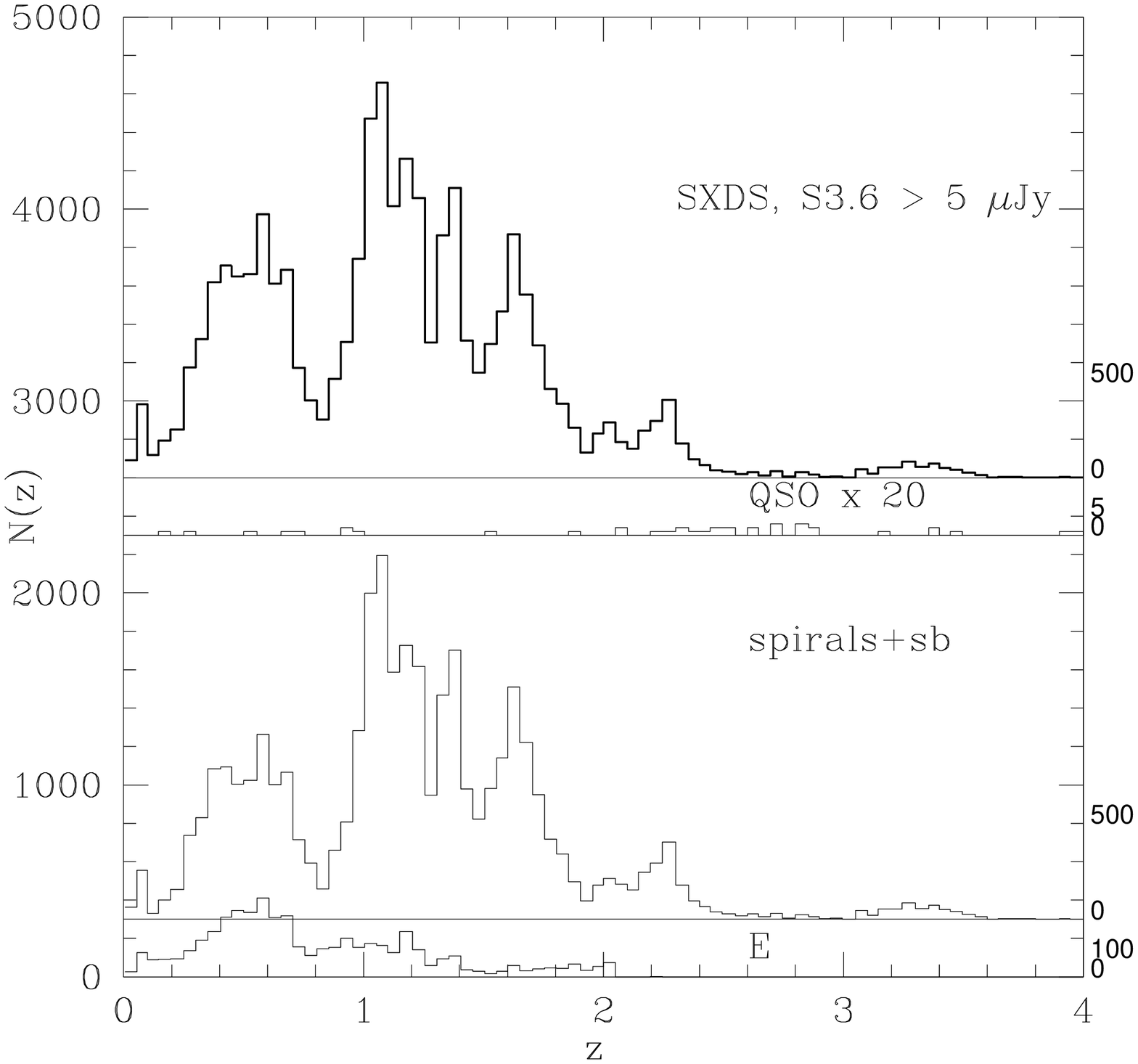,angle=0,width=7cm}
\epsfig{file=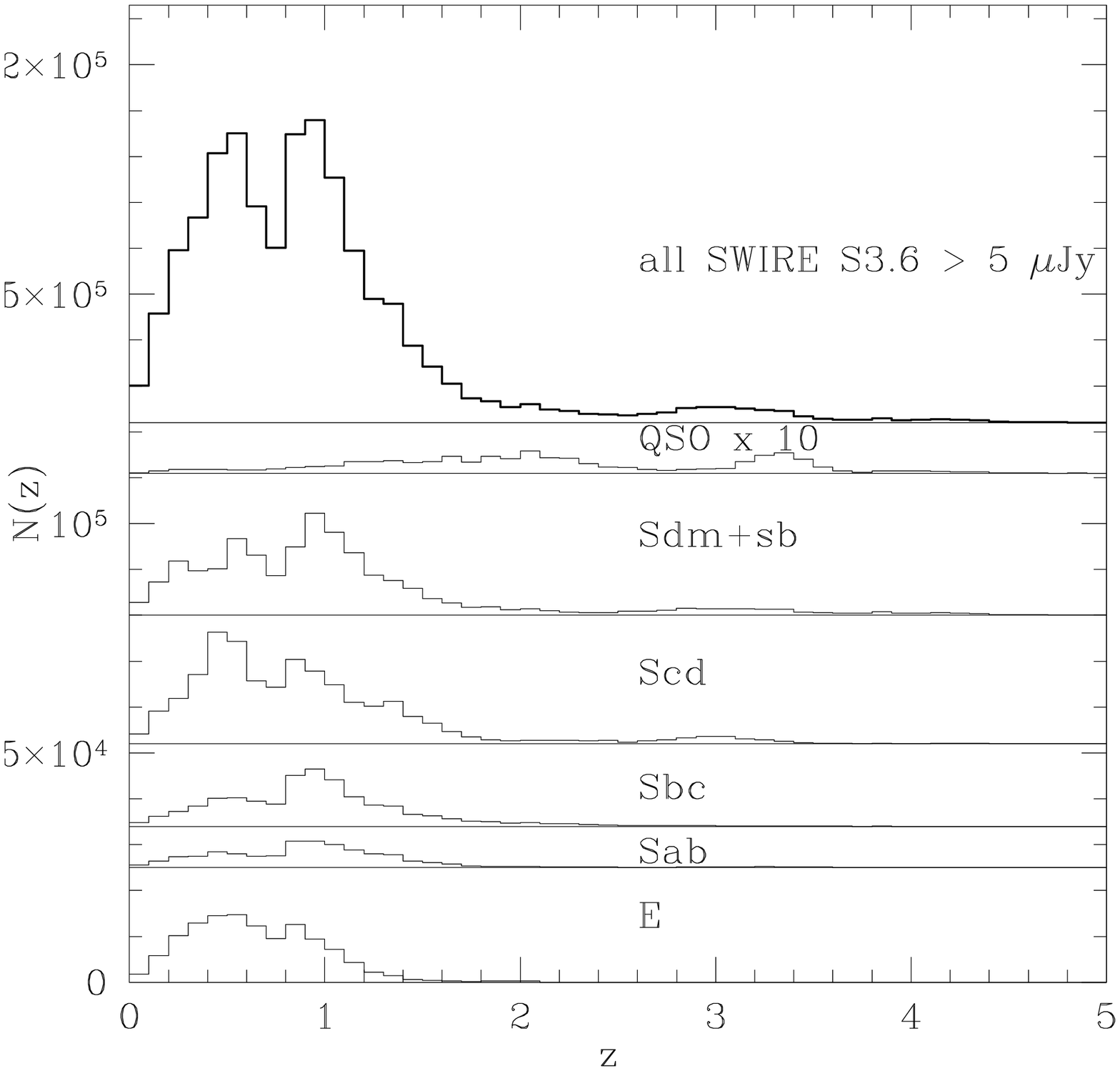,angle=0,width=7cm}
\caption{LH: Photometric redshift histogram for SWIRE-SXDS sources with 6 good photometric bands, and S(3.6) $>$ 5 $\mu$Jy, r $<$ 27.5. 
Top panel: all sources;
lower panels: separate breakdown of contributions of ellipticals, spirals
(Sab+Sbc+Scd)+ starbursts (Sdm+sb), and quasars.  The histogram for quasars has
been multiplied by 20.
RH: Photometric redshift histogram for all SWIRE sources with 4 good photometric bands, and S(3.6) $>$ 5 $\mu$Jy. 
Top panel: all sources;
lower panels: separate breakdown of contributions of ellipticals, Sab, Sbc, Scd, starbursts (Sdm+sb), and quasars. 
 The histogram for quasars has been multiplied by 10.
}
\end{figure*}

\begin{figure*}
\epsfig{file=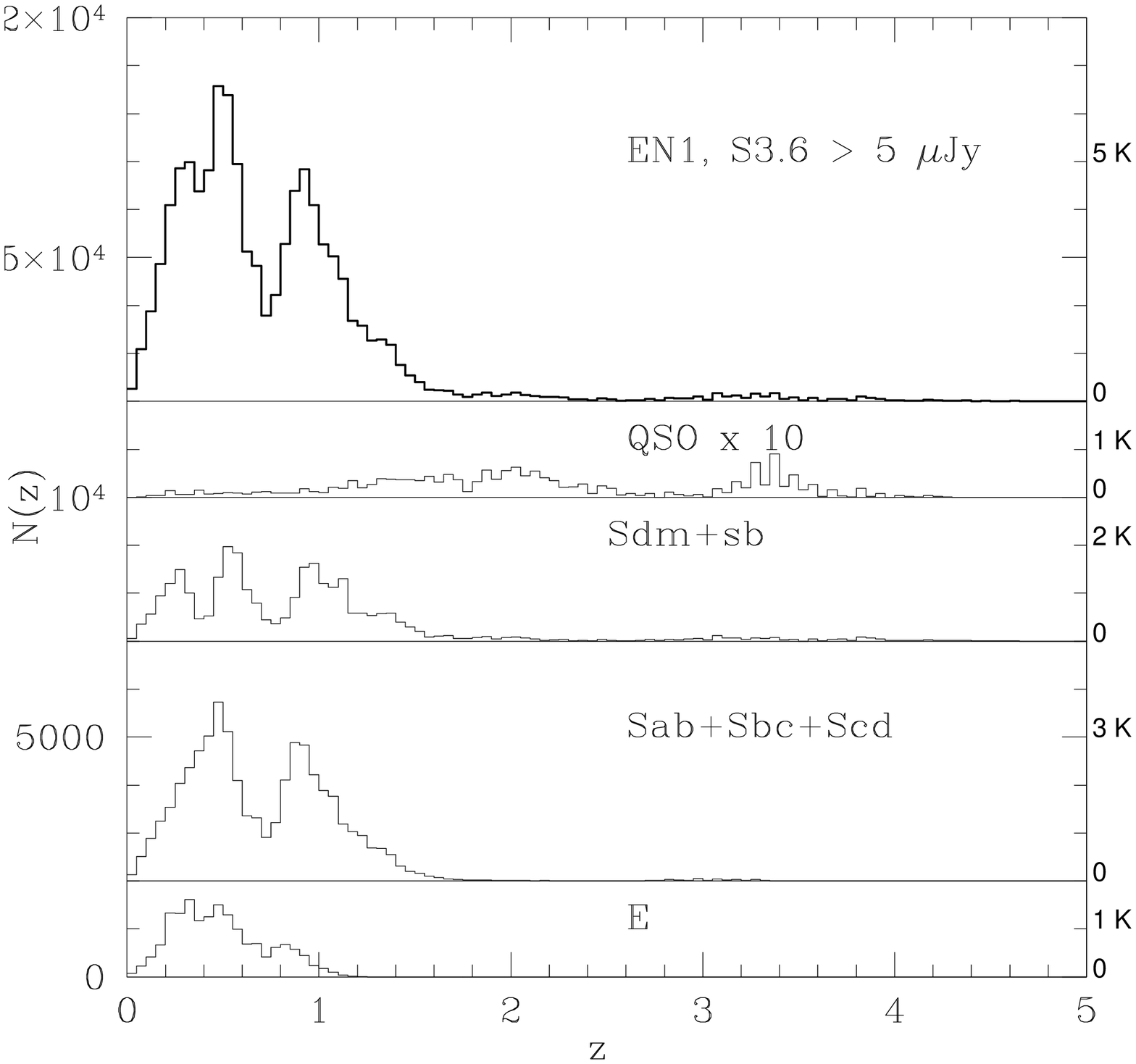,angle=0,width=7cm}
\epsfig{file=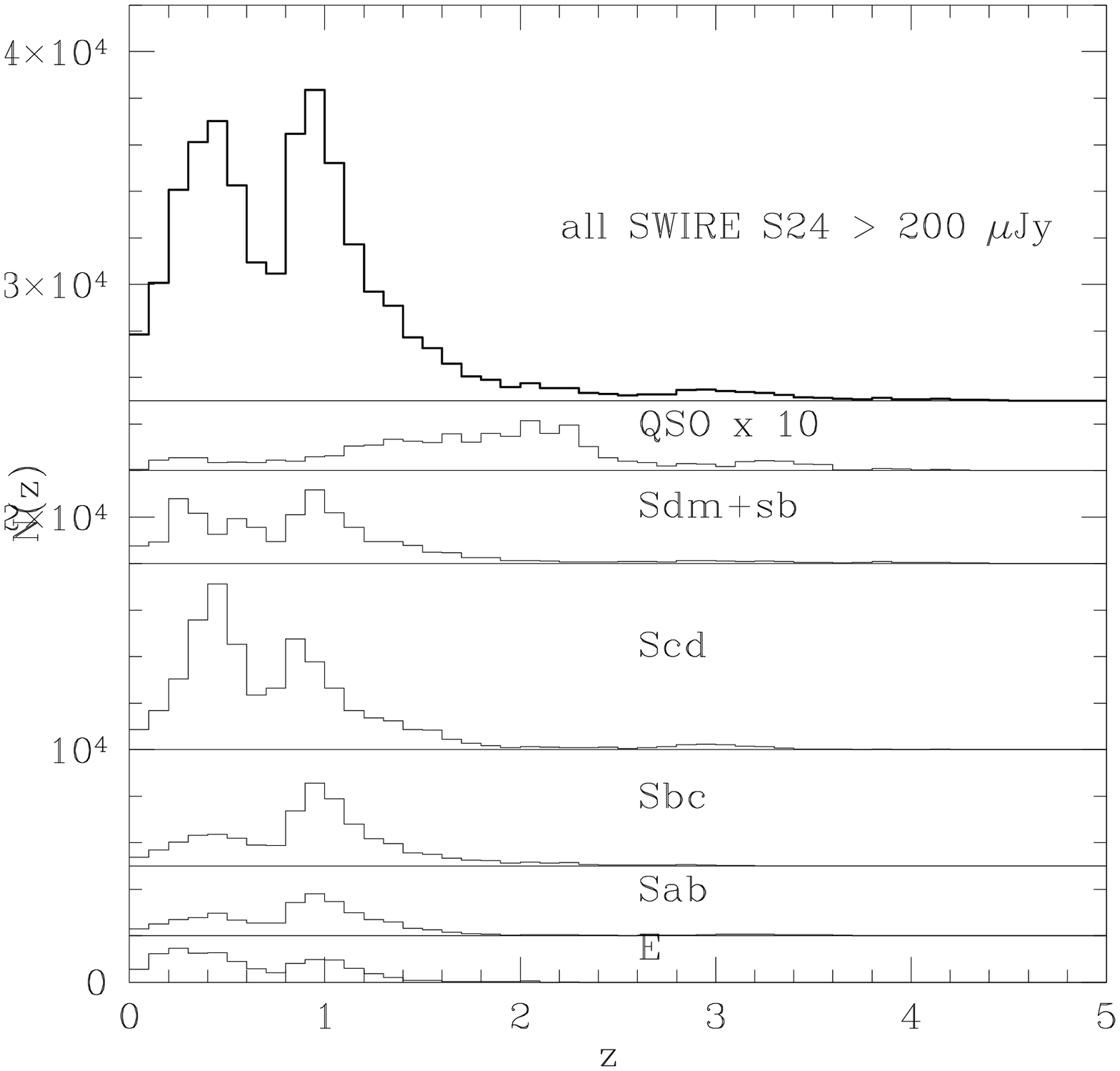,angle=0,width=7cm}
\caption{LH: Photometric redshift histogram for SWIRE-EN1 sources with at least 5 photometric bands, 
and S(3.6) $>$ 5 $\mu$Jy. 
Top panel: all sources;
lower panels: separate breakdown of contributions of ellipticals, spirals
(Sab+Sbc+Scd)+ starbursts (Sdm+sb), and quasars.  The histogram for quasars has
been multiplied by 20.
RH: Photometric redshift histogram for all SWIRE sources with 3-band optical IDs, and S(324) $>$ 200 $\mu$Jy. 
Top panel: all sources;
lower panels: separate breakdown of contributions of ellipticals, Sab, Sbc, Scd, starbursts (Sdm+sb), and quasars. 
 The histogram for quasars has been multiplied by 10.
}
\end{figure*}

\section{Extinction estimates}
We have estimated the extinction for 456470 spiral galaxies (excluding E, Sab, Sbc) and 11538 quasars, 
those sources which have at least 3 optical bands
available.  The profile of the mean extinction for all galaxies, and for quasars, with redshift is shown in 
Fig 15 and follows a similar pattern to that found
by Rowan-Robinson (2003).  This can be understood in terms of simple star formation histories, where the
gas-mass in galaxies, and star formation rate, decline sharply from $z \sim 1$ to the present.  At higher redshift
the dust extinction in galaxies is expected to decline with redshift, reflecting the build-up of heavy elements with time.  

There is some tendency for aliasing between extinction and galaxy template type, and this is probably
responsible for the peaks and troughs in the $A_V$ distribution.

The extinctions measured here represents an excess extinction over the standard templates used, which already have
some extinction present.  The average extinction in local spiral galaxies is $A_V \sim 0.3$ (Rowan-Robinson 2003b).

We find that 9 $\%$ of SWIRE galaxies and 6 $\%$ of quasars have $A_V \ge 0.5$.  A few galaxies and quasars
would have found a better fit with $A_V > 1$, but these represent $< 1\%$ of the population.

The redshift solutions with extinction appear to be better than those with extinction set to zero.

\begin{figure*}
\epsfig{file=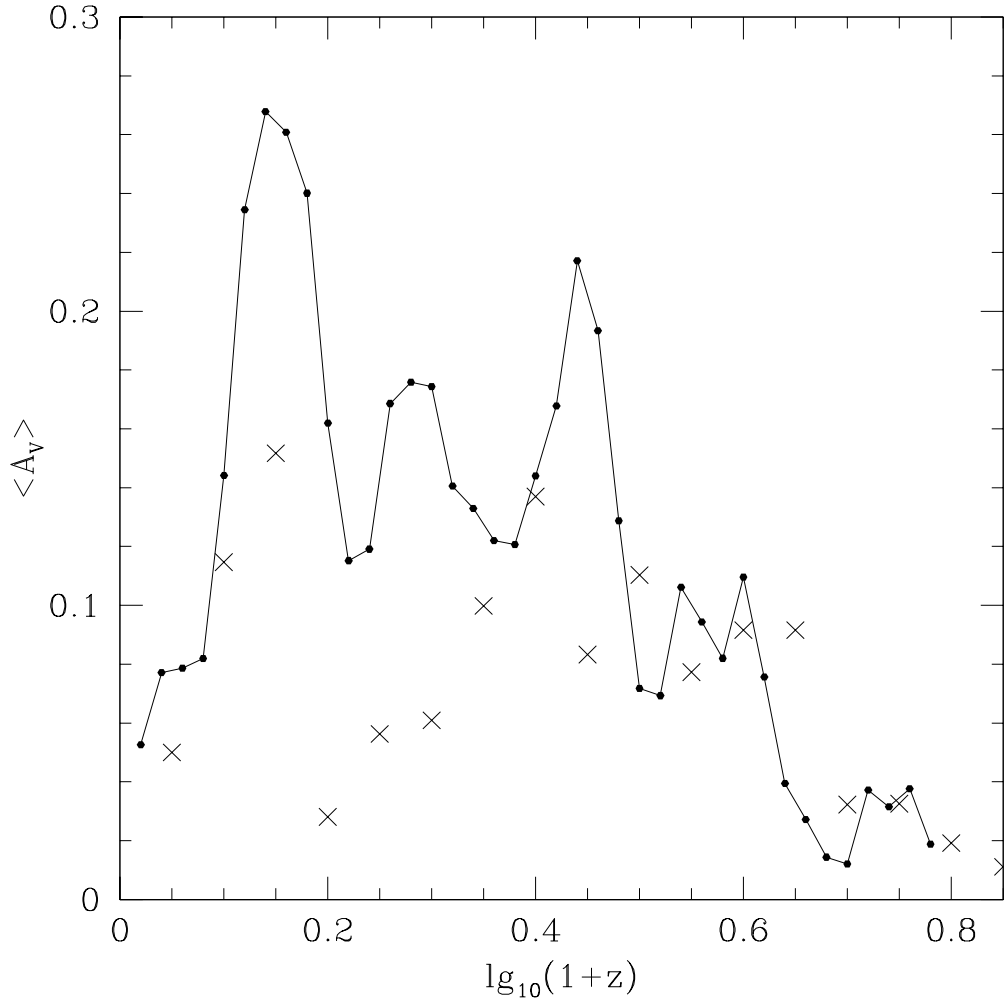,angle=0,width=14cm}
\caption{Mean value of $A_V$ as a function of redshift, for galaxies (filled circles and solid line) and quasars (crosses).}
\end{figure*}

\section{Infrared galaxy populations}

Our infrared template fits, for sources with infrared excess at $\lambda \ge 8 \mu$m, allow us to estimate the bolometric
infrared luminosity, $L_{ir}$, which can be a measure of the total star-formation rate in a galaxy (if there is no 
contribution from an AGN dust torus).   Since the optical
bolometric luminosity of a galaxy, corrected for extinction, is a measure of the stellar mass , the ratio
$L_{ir}/L_{opt}$ is a measure of the specific star-formation rate in the galaxy, ie the rate of star-formation per unit mass in stars.
However there are some caveats which should be borne in mind throughout this discussion.  Firstly, uncertainties in photometric
redshifts, and the possibility of catastrophic outliers, will affect the accuracy of the luminosities.  Because these depend on 
the number of photometric bands available, they are best evaluated by reference 
to Figs 11.  For redshifts determined from at least five bands, the rms uncertainty in $log_{10}(1+z_{phot}) \le$  0.017  and the 
corresponding uncertainty in luminosity at z = 0.2, 0.5, 1, 1.5 is
0.20, 0.10, 0.07 and 0.06 dex, respectively, with a few $\%$ being catastrophically wrong.  Secondly, if we only
have data out to 24 $\mu$m we need to bear in mind that the estimate of the infrared bolometric luminosity is  
uncertain by a factor of $\sim$2 due to uncertainties in the correct template fit (Rowan-Robinson et al 2005, 
Siebenmorgen and Krugel 2007; Reddy et al (2006) quote 2-3), and this applies 
also to derived quantities like the star-formation rate.   
Finally there is the issue of whether the 8-160 $\mu$m sources have
been associated with the correct optical counterpart in cases of confusion.  The process of band-merging of Spitzer data in the
SWIRE survey and of optical association has been described by Surace et al (2004, 2007 in preparation) and Rowan-Robinson et al 
(2005).  The probability of incorrect association of 3.6-24 $\mu$m sources is very low for this survey.  Similarly the incidence
of multiple possible associations between 3.6 and optical sources within our chosen search radius of 1.5 arcsec is very low
($<1 \%$).  Association of 70 and 160 $\mu$m sources with SWIRE 3.6-24 $\mu$m sources is made only if there is a 24 $\mu$m
detection and with the brightest 24 $\mu$m source if there are multiple associations.  A more sophisticated analysis would involve
distributing the far infrared flux between the different candidates.  The impact of confusion on the subsequent discussion is expected
to be small, though it may affect individual objects.  It does not affect any of the more unusual classes of galaxy discussed below.

\begin{figure*}
\epsfig{file=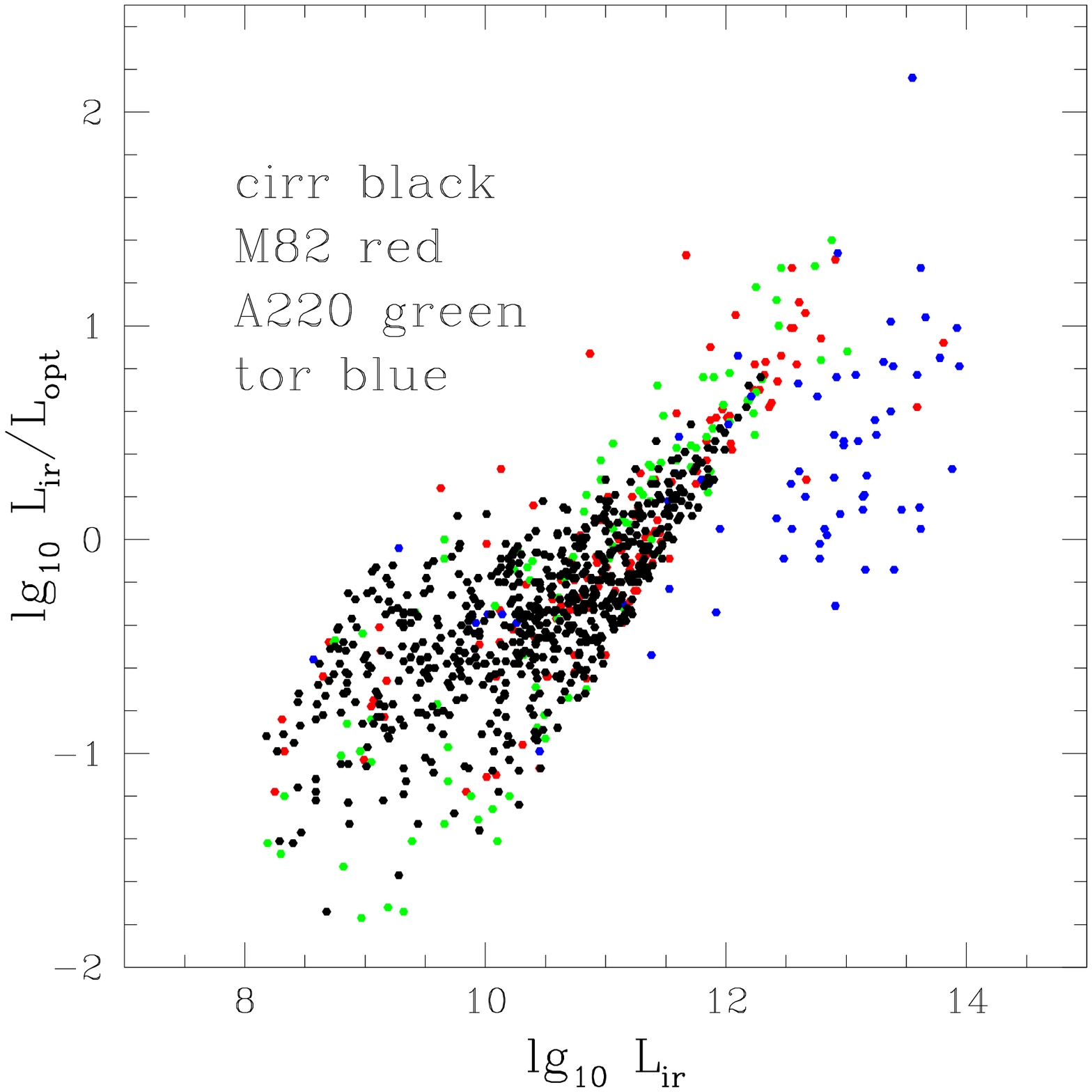,angle=0,width=7cm}
\epsfig{file=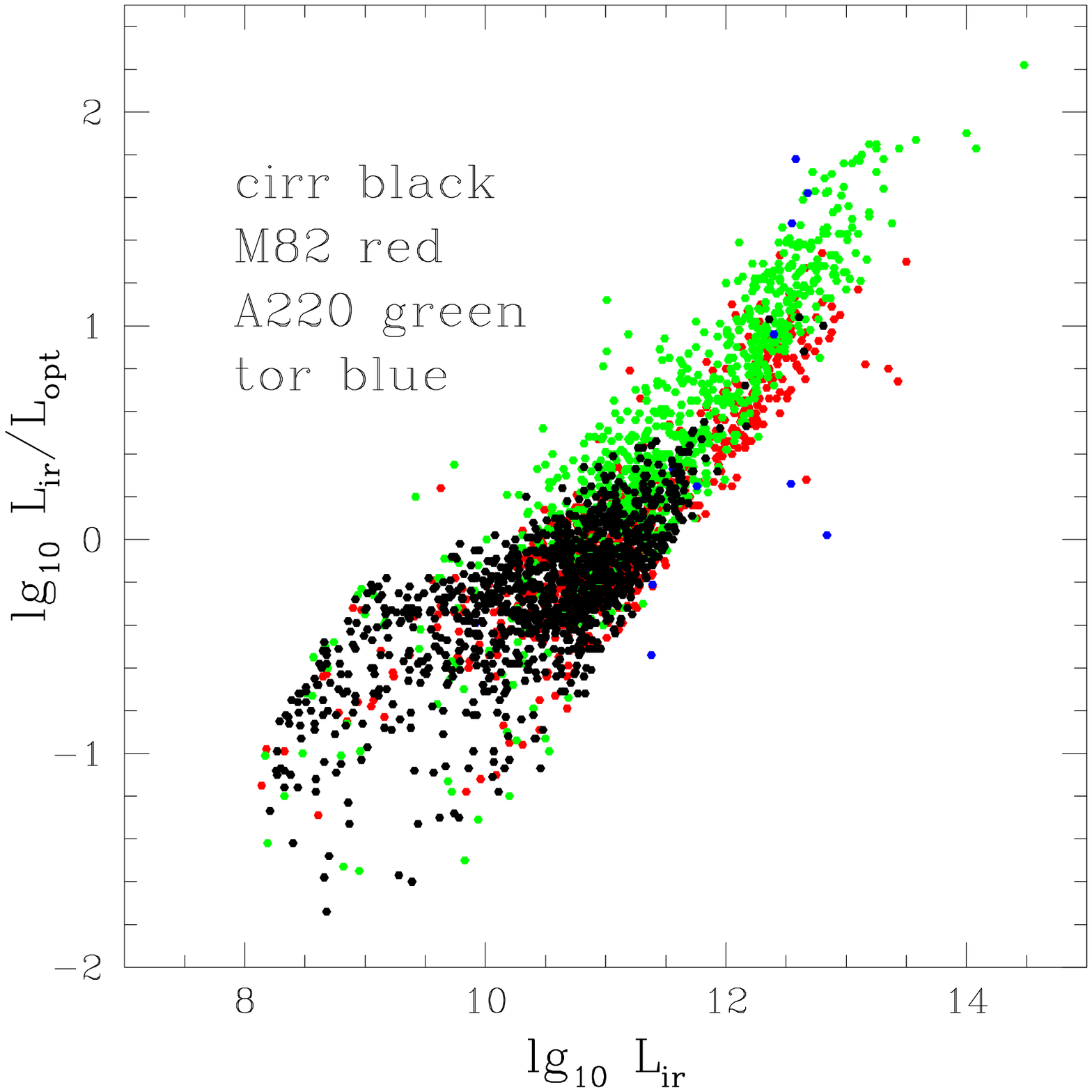,angle=0,width=7cm}
\caption{
LH: Ratio of infrared to optical bolometric luminosity, $log_{10} (L_{ir}/L_{opt})$, versus 1-1000 $\mu$m infrared luminosity,
$L_{ir}$ for SWIRE galaxies and quasars with spectroscopic redshifts and 70 $\mu$m detections.
RH: Ratio of infrared to optical bolometric luminosity, $log_{10} (L_{ir}/L_{opt})$, versus 1-1000 $\mu$m infrared luminosity,
$L_{ir}$ for SWIRE galaxies and quasars with photometric redshifts determined from at least 6 photometric bands and with 
reduced $\chi^2 <$5 and with 24 and 70 $\mu$m detections.
}
\end{figure*}

Fig 16L shows the ratio $L_{ir}/L_{opt}$ versus $L_{ir}$ for our most reliable sub-sample, galaxies with spectroscopic
redshift and 70 $\mu$m detections, with different colours coding sources dominated by cirrus, M82 or A220 starbursts or
`AGN dust tori.  Unfortunately this sample is heavily biassed to low redshifts and does not contain many examples of
ultraluminous galaxies ($L_{ir}>10^{12} L_{\odot}$), or of galaxies with $L_{ir}/L_{opt}>1$.

In Fig 16R we show the same plot but we have dropped the requirement of spectroscopic redshift to that of highly reliable 
photometric redshifts (at least 6 photometric bands, reduced $\chi^2 < 5$).  We now see large numbers of ultraluminous
galaxies, especially with A220 templates. We also see a small number of cool luminous galaxies, sources
with $L_{ir}>10^{12} L_{\odot}$ and  $L_{ir}/L_{opt}>1$ whose far infared spectra are fitted with a cirrus template,
as was seen in the ISO ELAIS survey (Rowan-Robinson et al 2004).  

We now look more closely at each infrared template type in turn.  To make things more precise we have estimated the 
stellar mass for each galaxy, based on our stellar synthesis templates (section 3, table 2).  For each galaxy we estimate 
the rest-frame 3.6 $\mu$m luminosity, $\nu L_{\nu}(3.6)$, in units of $L_{\odot}$, and from our stellar synthesis models
estimated the ratio $(M_*/M_{\odot})/(\nu L_{\nu}(3.6)/L_{\odot})$, which we find to be 38.4, 40.8, 27.6, 35.3, 18.7,
26.7, for types E, Sab, Sbc, Scd, Sdm, sb, respectively.  
[Note: we are measuring the 3.6 $\mu$m monochromatic luminosity in total solar units, not in units of the sun's
monochromatic 3.6 $\mu$m luminosity.]  We find values of $M_*$ agreeing with these within 10-20$\%$
if we base estimates on the B-band luminosity.  Estimates based on 3.6 $\mu$m should be more reliable, since there is a
better sampling of lower mass stars and less susceptibility to recently formed massive stars.
These mass estimates would be strictly valid only for low redshift.  For higher redshifts the mass-to-light estimates will be lower
since for the oldest stellar populations, $M/L$ varies strongly with age (Bruzual and Charlot 1993, see their Fig 3).
This can be approximately modeled using the Berta et al (2004) synthesis fits described in section 3.1 above, with an 
accuracy of 10$\%$, as 

$(M_*/M_{\odot})/(\nu L_{\nu}(3.6)/L_{\odot})(t) = 50/[a+1.17(t/t_0)^{-0.6})$

where $t_0$ is the present epoch and a = 0.15, 0.08, 0.61, 0.26, 1.44, 0.70 for SED types E, Sab, Sbc, Scd, Sdm and sb, 
respectively.  The dust masses in the photometric redshift catalogue (section 5) have been corrected for this evolution.

We have also estimated the star formation rate, using the conversion from 60 $\mu$m luminosity of Rowan-Robinson
et al (1997), Rowan-Robinson (2001):

$\phi_*(M_{\odot} yr^{-1}$) = 2.2 $\epsilon^{-1} 10^{-10} (L_{60}/L_{\odot})$

where $\epsilon$ is the fraction of uv light absorbed by dust, taken as $2/3$
[but note the discussion of far infrared emission arising from illumination by older stars by Rowan-Robinson(2003b),
Bell (2003), which can result in overestimation of the star-formation rate].
The bolometric corrections at 60 $\mu$m, needed to convert $L_{ir}$ to $L_{60}$, are 3.48, 1.67 and 1.43 for cirrus,
M82 and A220 templates, respectively.

Figure 17L shows $L_{ir}/L_{opt}$ versus $M_*$ 
for cirrus galaxies, colour-coded by optical SED type.  Since the emission is due to interstellar dust heated by the general
stellar radiation field, $L_{ir}/L_{opt}$ is a measure of the dust opacity of the interstellar medium.  Although many galaxies with elliptical SEDs have 
$L_{ir}/L_{opt} <<1$, consistent with low dust opacity, significant numbers of galaxies with elliptical SEDs in the optical seem to have
values of $L_{ir}/L_{opt}$
comparable with spirals.  There is a population of high mass spirals with dust opacity $\ge$1.

Figures 18 show the specific star-formation rate, $\phi_*/M_*$, versus $L_{ir}$ for M82 (LH) and A220 (RH) starbursts.  
The specific star formation rate 
ranges from 0.003 to 2 $Gyr^{-1}$ for M82 starbursts and from 0.03-10 for A220 starbursts.  It is by no means the case that
ultraluminous starbursts are predominantly of the A220 type, as is often assumed in the literature.  There is an
interesting population of galaxies with elliptical SEDs in the optical, with associated starbursts in the ir  luminosity range $10^9 - 10^{11} L_{\odot}$, similar to those found by Davoodi et al (2006).

\begin{figure*}
\epsfig{file=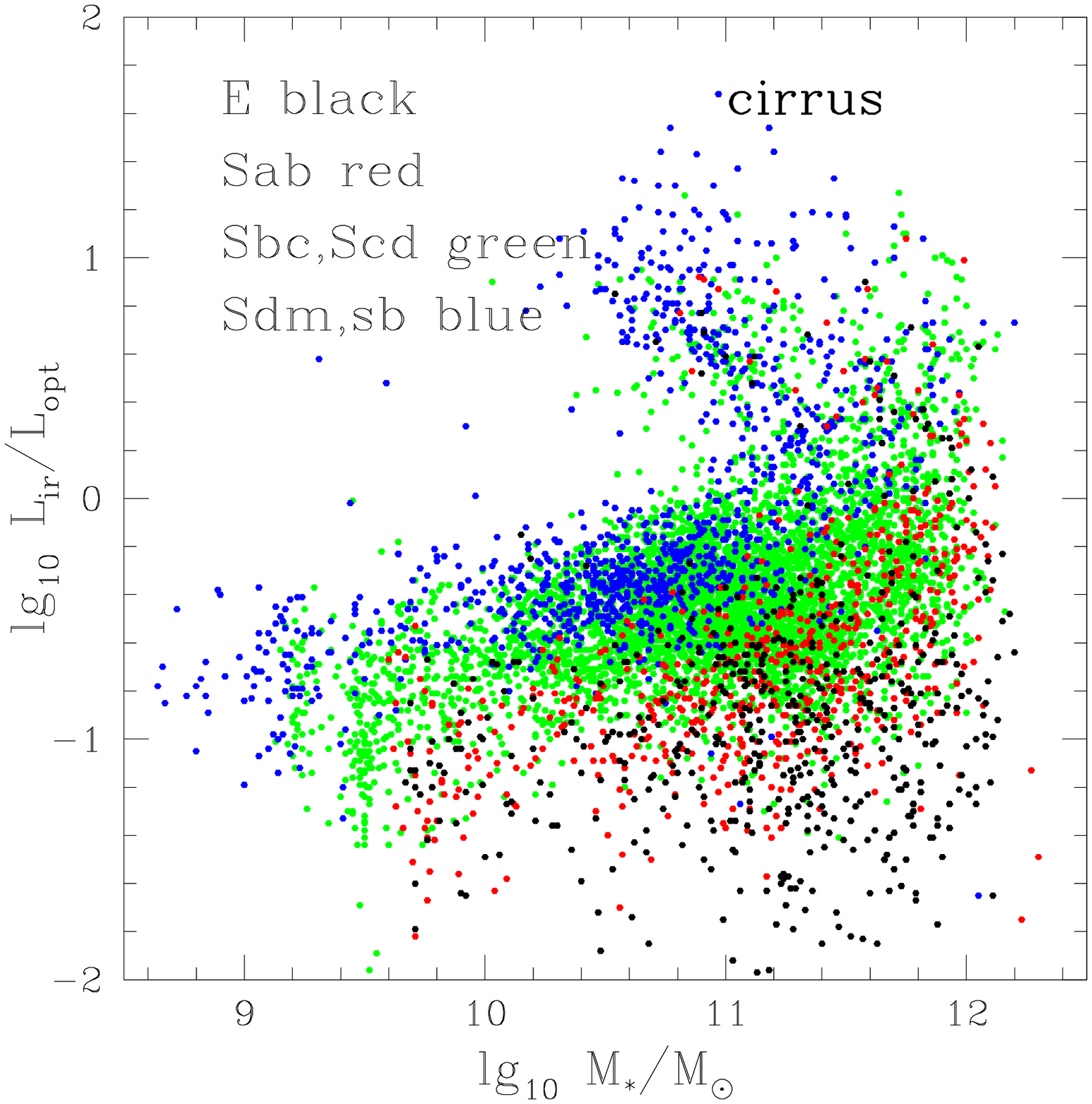,angle=0,width=7cm}
\epsfig{file=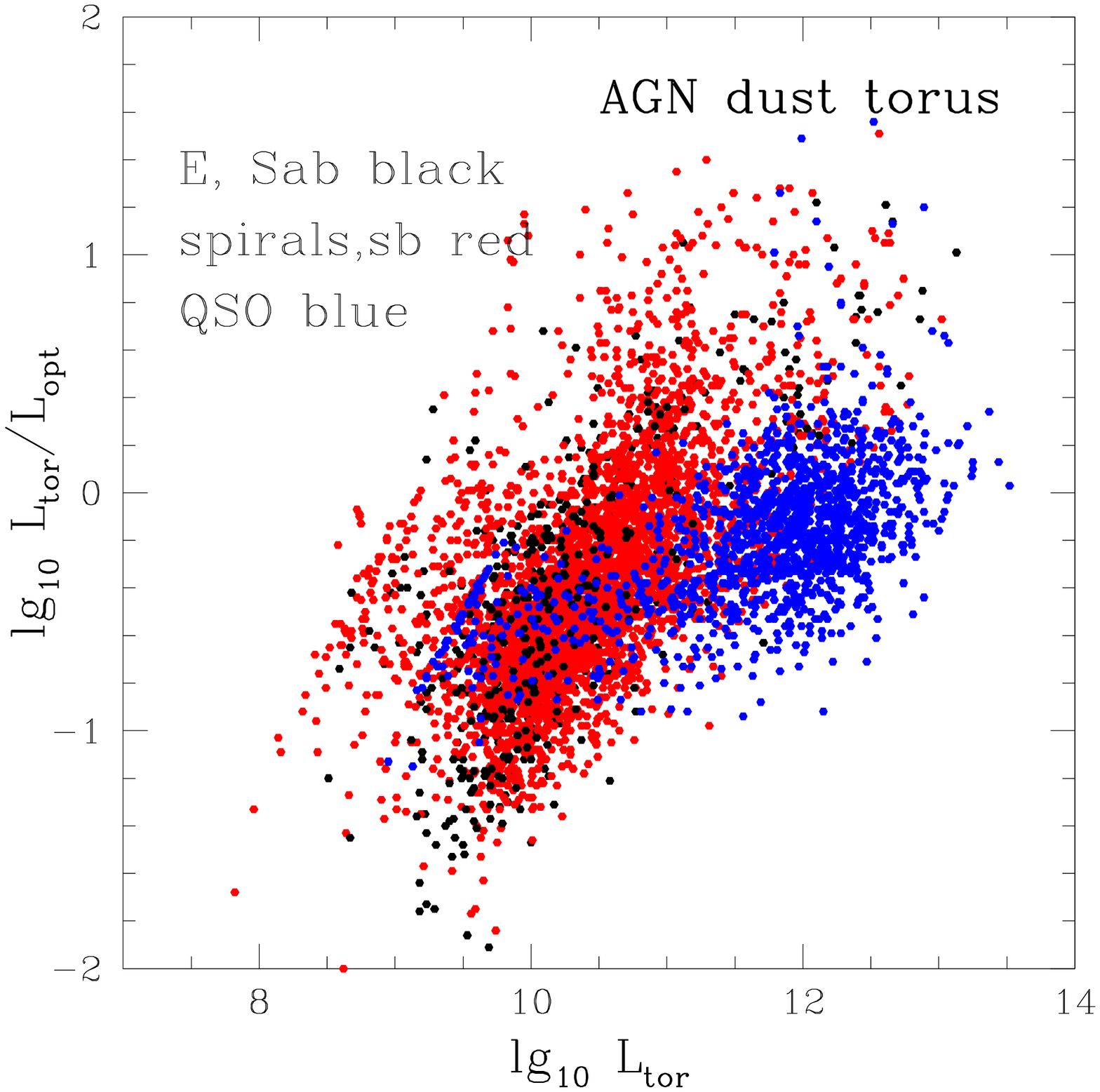,angle=0,width=7cm}
\caption{
LH: Ratio of infrared to optical bolometric luminosity, $log_{10} (L_{ir}/L_{opt})$, versus stellar mass,
$M_*$ for SWIRE galaxies with photometric redshifts determined from at least 6 photometric bands and with 
reduced $\chi^2 <$5 and with 24  $\mu$m detections, with infrared excess fitted by cirrus template.
RH: Ratio of dust torus luminosity to optical bolometric luminosity, $log_{10} (L_{tor}/L_{opt})$, versus dust torus luminosity,
$L_{tor}$, for SWIRE galaxies with photometric redshifts determined from at least 6 photometric bands and with reduced $\chi^2 <$5 and with 24  $\mu$m detections, with infrared excess fitted by AGN dust torus template.
}
\end{figure*}

\begin{figure*}

\epsfig{file=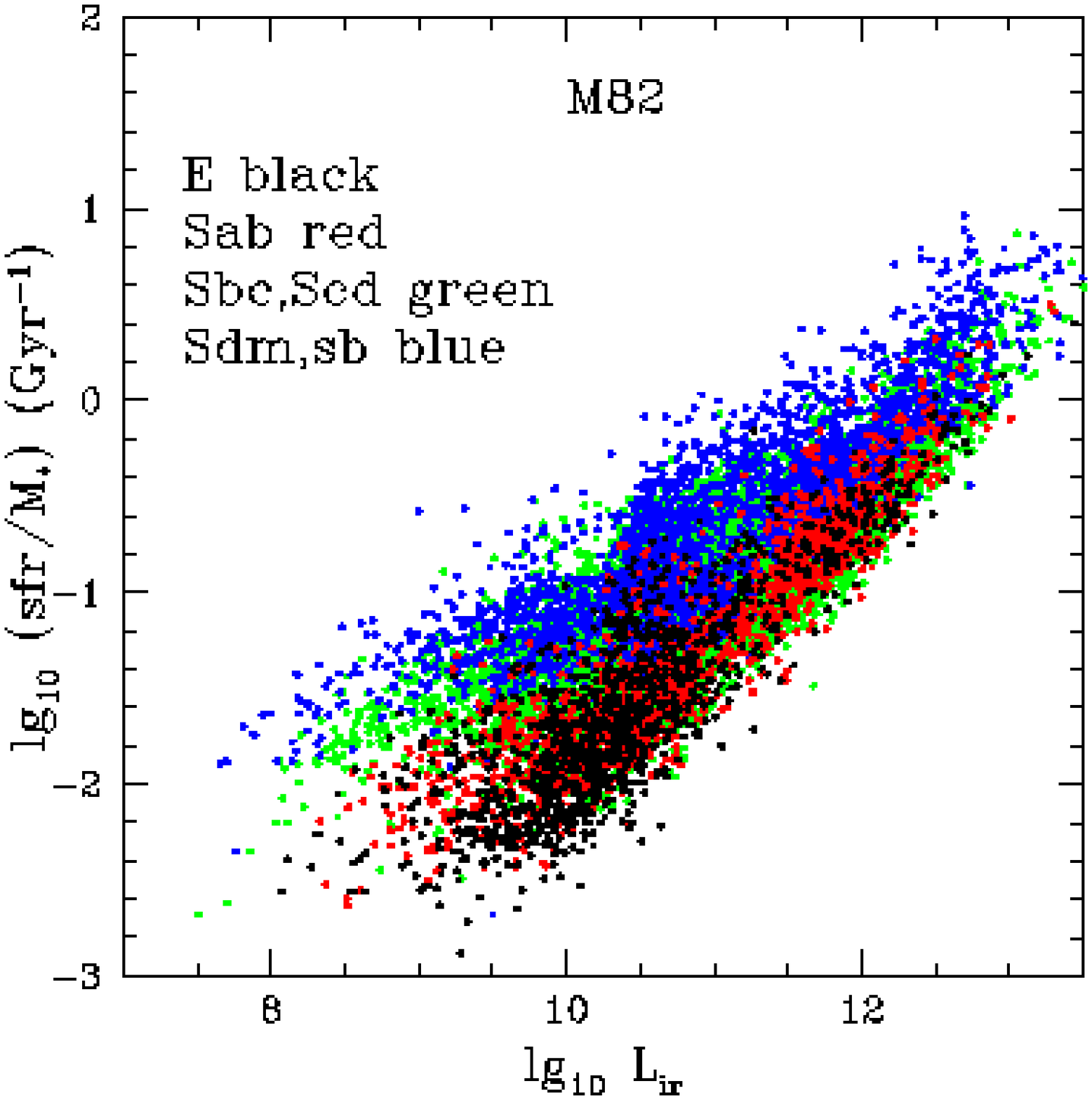,angle=0,width=7cm}
\epsfig{file=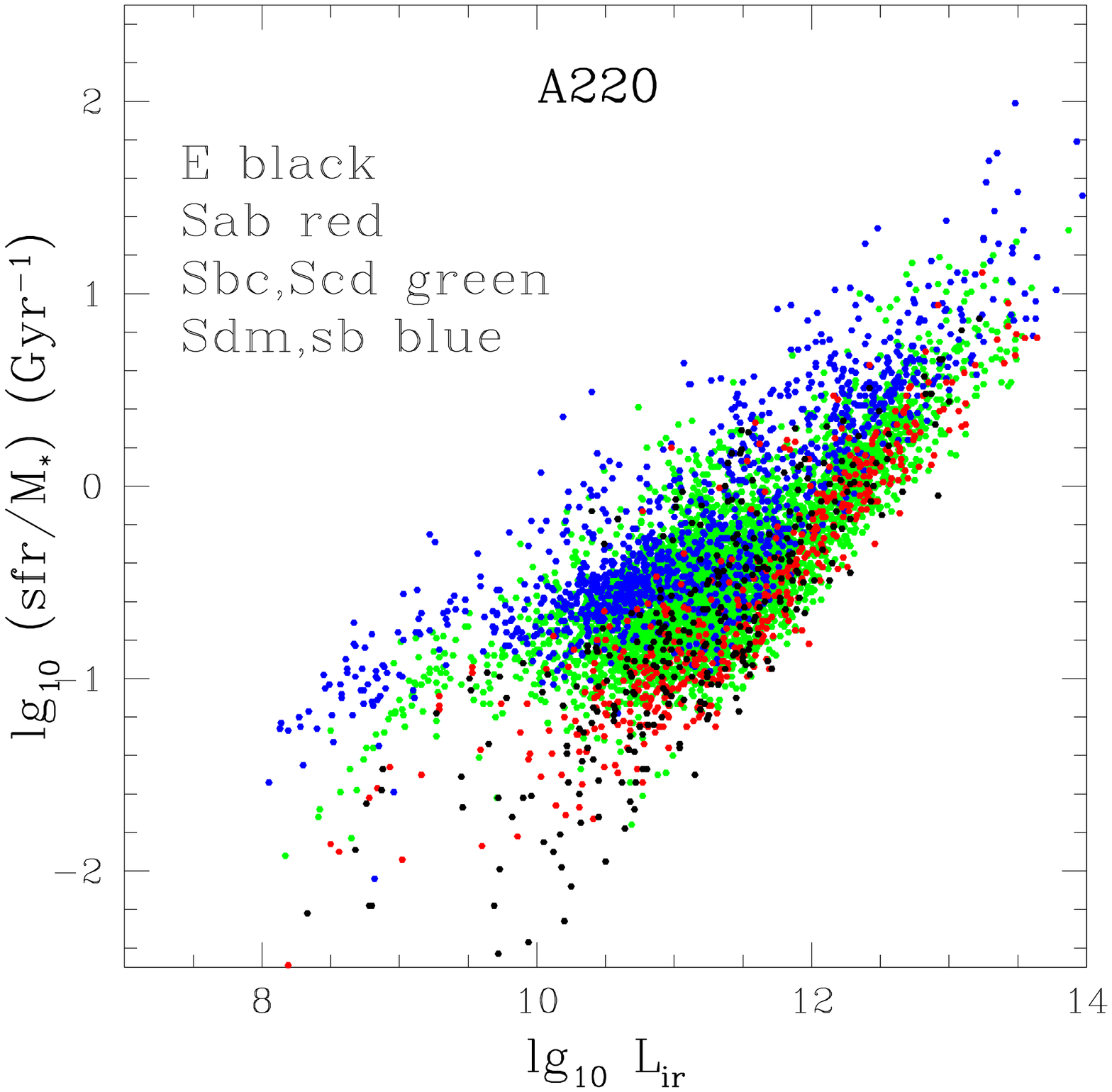,angle=0,width=7cm}
\caption{
LH: Specific star-formation rate, $\phi_*/M_*$, in $Gyr^{-1}$, versus stellar mass, $M_*$,
 for SWIRE galaxies with photometric redshifts determined from at least 6 photometric bands and with 
reduced $\chi^2 <$5 and with 24  $\mu$m detections, with infrared excess fitted by M82 template.
RH: Specific star-formation rate, $\phi_*/M_*$, in $Gyr^{-1}$, versus stellar mass, $M_*$,
for SWIRE galaxies with photometric redshifts determined from at least 6 photometric bands and with 
reduced $\chi^2 <$5 and with 24  $\mu$m detections, with infrared excess fitted by A220 template.
}
\end{figure*}

Figure 17R shows $log_{10}(L_{tor}/L_{opt})$ versus $L_{tor}$ for objects dominated by AGN dust tori at 8 $\mu$m, 
with different coloured symbols for different optical SED types.  $L_{tor}/L_{opt}$ can be interpreted as 
$f_{tor} k_{opt}$ where $f_{tor}$ is the covering factor of the torus and $k_{opt}$ is the bolometric correction that needs
to be applied to the optical luminosity to acount for emission shortward of the Lyman limit.  Fig 19 shows the distribution of $log_{10} (L_{tor}/L_{opt})$ for galaxies and quasars with $log_{10} L_{tor} > 11.5, z < 2$, which is an approximately volume-limited sample.   The distribution for QSOs is well-fitted by a 
Gaussian with mean -0.10, and standard deviation 0.26.  

\begin{figure*}
\epsfig{file=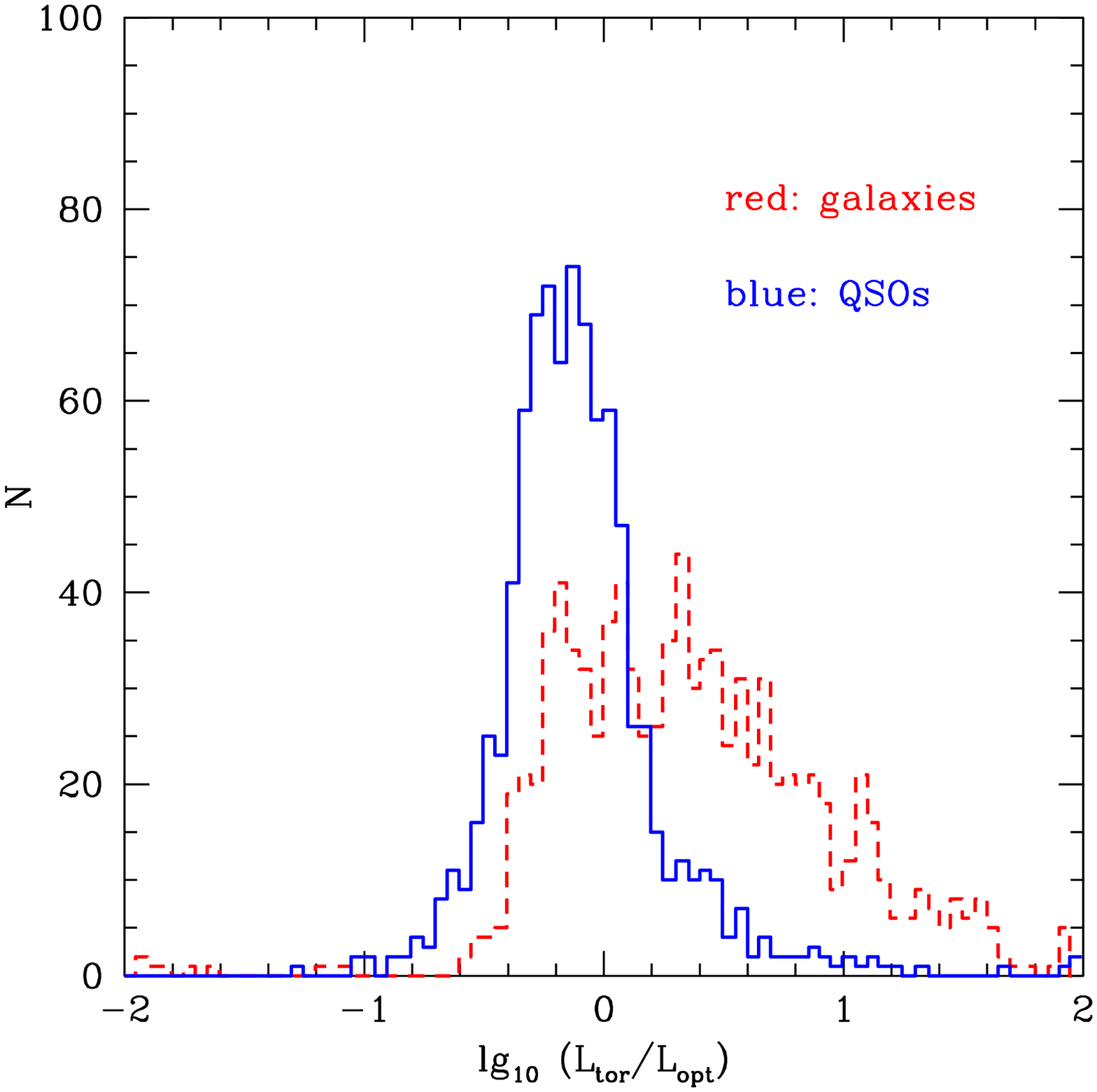,angle=0,width=14cm}
\caption{
Histogram of $log_{10}(L_{tor}/L_{opt})$   for QSOs (blue, solid) and galaxies (red, broken).
}
\end{figure*}

Assuming $k_{opt} \sim 2$, as implied by the composite QSO templates of Telfer et al (2002) and Trammell et al (2007),
we deduce that the mean value of the dust covering factor $f_{tor}$ for quasars is 0.40.   
The 2-$\sigma$ range would be 0.1-1.0 
We can use this mean value to infer the approximate luminosity of the underlying QSO in the galaxies
with AGN dust tori, since $L_{QSO} \sim 2.5 L_{tor}$, and hence deduce that most of the galaxies
with $L_{tor}/L_{opt} > 2.5$ should be Type 2 objects, since the implied QSO luminosity, if it was being viewed
face-on, would be sufficient to outshine the host galaxy.  There are 413 such galaxies in this distribution.
QSOs with $L_{tor}/L_{opt} > 2$ should also be Type 2 objects, although they must be presumed to be cases 
similar to SWIRE J104409.95+585224.8 (Polletta et al 2006), where the optical light is scattered 
light representing only a fraction of the total intrinsic optical output.  There are 84 of these.  These 497
Type 2 objects can be compared with the 796 QSOs with $L_{tor}/L_{opt}  < 2$, which are Type 1 objects,
yielding a Type 2 fraction of 0.41, in good agreement with the covering factor deduced above.

Most of the galaxies with  $log_{10} L_{tor} < 11.5$  in Fig 17R tend to lie at progressively lower values of 
$L_{tor}/L_{opt}$ than the QSOs, by an amount that increases
 towards lower $L_{tor}$, consistent with an increasing contribution of starlight to $L_{opt}$.

Figure 20 shows spectral energy distributions for examples of two populations of interest.  Fig 20L shows 
the SEDs of 5 galaxies with good quality photometric redshifts (at least 6 photometric bands, reduced $\chi^2 < 5$,
and 24 and 70 $\mu$m detections (in 2 cases also 160 $\mu$m), with the infrared template fitting requiring
a very luminous cool component ($> 10^{12} L_{\odot}$).  The luminosity in this cool component is clearly
substantially greater than the optical bolometric luminosity, suggesting that the optical depth of the interstellar
medium in these galaxies is $>$ 1 (Efstathiou and Rowan-Robinson 2003, Rowan-Robinson et al 2004).

Figure 20R shows the SEDs of 4 galaxies with elliptical galaxy template fits in the optical, spectroscopic redshifts 
and  $L_{ir} > L_{opt}$. These are galaxies that in optical surveys would be classified as red, early-type galaxies.
In the infrared two are fitted by cirrus components (objects 1 and 3), one by an Arp220 starburst (object 4) and one by a combination 
of an AGN dust torus and an M82 starburst (object 2).  Since highly extinguished starbursts like Arp220, which are probably
the product of a major merger, do look like ellipticals
in the optical because the optical light from the young stars is almost completely extinguished, object 4 is
consistent with being a highly obscured starburst.  For object 2, an elliptical  galaxy SED 
in the optical coupled
with an AGN dust torus in the mid infrared implies the presence of a Type 2 QSO, in which the QSO is hidden behind the dust 
torus. The infrared SED of this object (object 2) also shows evidence for strong star formation, so this also
appears to be a case of an obscured starburst.  
The two galaxies with  elliptical-like SEDs in the optical and with
strong cirrus components (objects 1 and 3) are harder to understand.  If the emission is from interstellar dust and the dust
were being illuminated solely by the old stellar population, then a high optical depth would be implied by
the fact that $L_{ir} > L_{opt}$, but the optical SED shows no evidence of extinction.  The implication is that they
are probably also obscured starbursts, perhaps with star formation extended through the galaxy to account for
the form of the infrared SED.  To see whether the morphology can help the interpretation, we show in
Fig 21  IRAC images of these 4 galaxies, colour-coded as blue for 3.6 $\mu$m and
red for 8 $\mu$m.  Objects 1 and 2 look elliptical and object 3 looks lenticular.  All three are strongly reddened in their outer parts,
indicating that the long wavelength radiation is coming from the outer parts of the galaxies.  This suggests that the 
infrared emission is associated with infalling gas and dust.  Object 4  appears more compact, consistent with
being similar to Arp 220.

Figure 22R shows the specific star formation rate versus stellar mass for 4135 SWIRE galaxies with $1.5 < z_{phot} < 2.5$, 24 $\mu$m detections and ir excesses in at least two bands, and reduced $\chi^2_{\nu} < 5$.
This can be contrasted with  Fig 15 of Reddy et al (2006), based on 200 galaxies identified by Lyman drop-out
and near-infrared excess techniques.   SWIRE is clearly a source of high redshift galaxies with infrared data 
which far exceeds previous optical-based samples.   Compared with the Reddy et al (2007) we note
that in the range $log_{10} M_*/M_{\odot} \sim 11-12)$, we find a far larger range of specific
star formation rate.  Because the depth of our optical surveys is typically r $\sim$ 23.5-25, we do not sample
the low-mass end of the galaxy distribution at these redshifts as well as Reddy et al (2006).  

Figure 22L shows the same plot for $z<0.5$.    We see that at the present epoch specific star formation rates $> 1 Gyr^{-1}$ are
rare, whereas they seem to have been common, at least amongst massive galaxies, at z $\sim$ 2.  At low redshifts we
see some evidence for 'down-sizing', in that elliptical galaxies, which tend to be of higher mass, have significantly 
lower specific star-formation rates than lower mass galaxies, which tend to have late-type spiral or starburst spectral energy distributions.

Figure 23L shows the star formation rate $\phi_*$ against redshift.  Although the distribution is subject to strong selection
effects, with the minimum detectable $\phi_*$ increasing steeply with redshift to z $\sim$ 2 and then decreasing
towards higher redshift because of the negative K-correction at 8 $\mu$m, the plot is consistent with a star-formation 
history in which the star formation rate increases from z = 0-1.5 and then remains steady from z =1.5-3.

We can use our radiative transfer models for cirrus, M82 and A220 starburst components to estimate the approximate dust mass
for each galaxy, using the recipe (Andreas Efstathiou, 2007, private communication):

$M_{dust}/M_{\odot} = k L_{ir}/L_{\odot}$, 

where k = $1.3x10^{-3}$ for cirrus, $1.1x10^{-4}$ for M82 and $4.4x10^{-4}$ for A220.

These dust masses have been given in the catalogue (section 5) and Fig 23R shows the dust mass, $M_{dust}$, versus
the stellar mass, $M_*$, for SWIRE galaxies, coded by optical SED type.  For most galaxies $ M_{dust}/M_*$ ranges from
$10^{-6}$ to $10^{-2}$, with the expected progression through this ratio with Hubble type.  Galaxies with exceptionally
high values of this ratio presumably have gas masses comparable to the stellar mass, assuming the usual gas-to-dust
ratios.  The dust masses will be uncertain because the uncertainty in photometric redshifts makes luminosities uncertain by 
a factor ranging from 0.06 dex at z = 1 to 0.20 dex at z = 0.2 (see start of this section).  There will also be a larger uncertainty
associated with ambiguity in the template fitting.  To estimate this we looked at all fits to the infrared excess where there is
an excess in at least two bands, excluding cases with an AGN dust torus component, and with reduced $\chi^2$ for
the infrared fit $<$ 5.  Relative to the best fit dust mass we find an rms uncertainty of $\pm$0.8 dex, or a factor of 6, so these dust
mass estimates have to be treated with caution.  This
uncertainty is substantially reduced if  70 or 160 $\mu$m data are available.
Data from HERSCHEL will allow these dust mass estimates to be greatly improved.

\section{Conclusions}

We have presented the SWIRE Photometric Redshift Catalogue, 1025119 redshifts of unprecedented reliability and good accuracy.
Our methodology is an extension of earlier work by Rowan-Robinson (2003), Rowan-Robinson et al (2004, 2005),
Babbedge et al (2004) and is based on fixed galaxy and QSO templates applied to data at 0.36-4.5 $\mu$m, and on a set of 4
infrared emission templates fitted to infrared excess data at 3.6-170 $\mu$m.  The galaxy templates are initially empirical, but 
have been given greater physical validity by fitting star-formation histories to them.  The code involves two passes through the
data, to try to optimize recognition of AGN dust tori.  A few carefully justified priors are used and are the key to supression of 
outliers.  Extinction, $A_V$, is allowed as a free parameter.  We have provided the full reduced $\chi^2_{\nu}(z)$ distribution 
for each source, so that the full error distribution can be used, and aliases investigated

We use a set of 5982 spectroscopic redshifts taken from the VVDS survey (LeFevre et al 2004), the ELAIS survey 
(Rowan-Robinson et al 2003), the SLOAN survey, NED, our own spectroscopic surveys (Smith et al 2007, Trichas et al
2007) and Swinbank et al (2007), to analyze the performance of our method as a function of the number of photometric
bands used in the solution and the reduced $\chi^2_{\nu}$.  For 7 photometric bands (5 optical + 3.6, 4.5 $\mu$m) the
rms value of $(z_{phot}-z_{spec})/(1+z_{spec})$ is 3.5$\%$, and the percentage of catastrophic outliers (defined as $>$ 
15$\%$ error in (1+z), is $\sim 1\%$.  These rms values are comparable with those achieved by the COMBO-17 collaboration
(Wolf et al 2004), and the outlier fraction is significantly better.  The inclusion of the 3.6 and 4.5 $\mu$m IRAC bands is 
crucial in supression of outliers.

We have shown the redshift distributions at 3.6 and 24 $\mu$m.  In individual fields structure in the redshift distribution
corresponds to real large-scale structure which can be seen in the spectroscopic redshift distribution, so these redshifts are
a powerful tool for large-scale structure studies.  

10$\%$ of sources in the SWIRE photometric redshift catalogue have z $>$2, and 5$\%$ have z$>$3, so this catalogue 
is a huge resource for high redshift galaxies.  The redshift calibration is less reliable at z $>$2 and high redshift sources often have
significant aliases, because the sources are detected in fewer bands.

We have shown the distribution of the mean extinction, $A_V$, as a function of redshift.  It shows a peak at z$\sim$0.5-1.5 and
then a decline to higher redshift, as expected from the star-formation history for galaxies.

A key parameter for understanding the evolutionary status of infrared galaxies is $\L_{ir}/L{opt}$, which can be analyzed
by optical and infrared template type.  For cirrus galaxies this ratio is a measure of the mean extinction in the interstellar medium of the
galaxy.  There appears to be a population of ultraluminous galaxies with cool dust and we have shown SEDs for some of the reliable examples.  For starbursts $\L_{ir}/L{opt}$ can be converted to the specific star-formation rate.  Although the very highest
values of this ratio tend to be associated with Arp220 starbursts, by no means all ultraluminous galaxies are.  There is a
population of galaxies with elliptical SEDs in the optical and with luminous starbursts, and we have shown SEDs for four of these 
with data in all Spitzer
bands and spectroscopic redshifts.  For 3 of them the IRAC colour-coded images show that the 8 $\mu$m emission
is coming from the outer regions of the galaxies, suggesting that the star-formation is associated with infalling gas and
dust.

\begin{figure*}
\epsfig{file=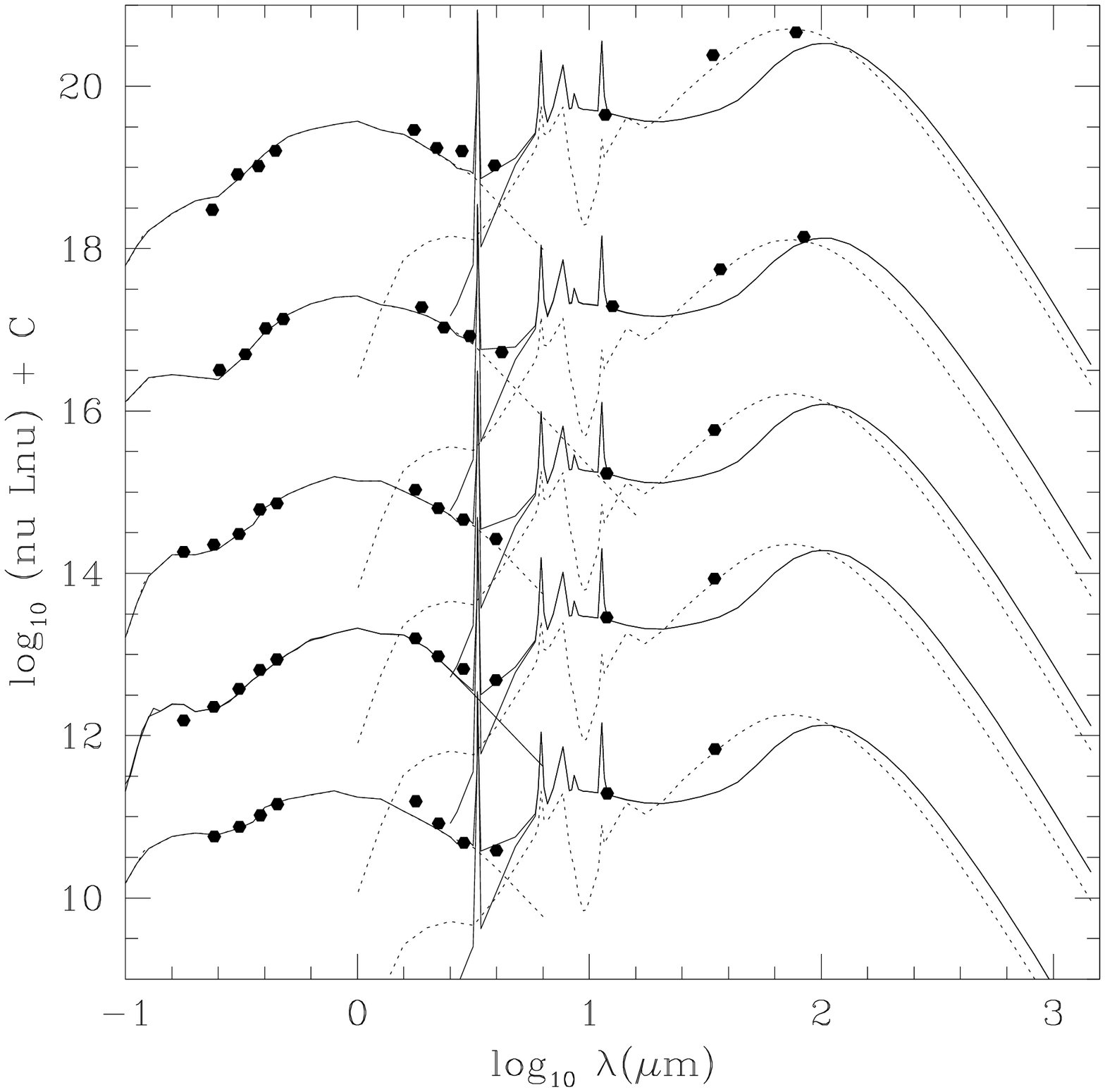,angle=0,width=7cm}
\epsfig{file=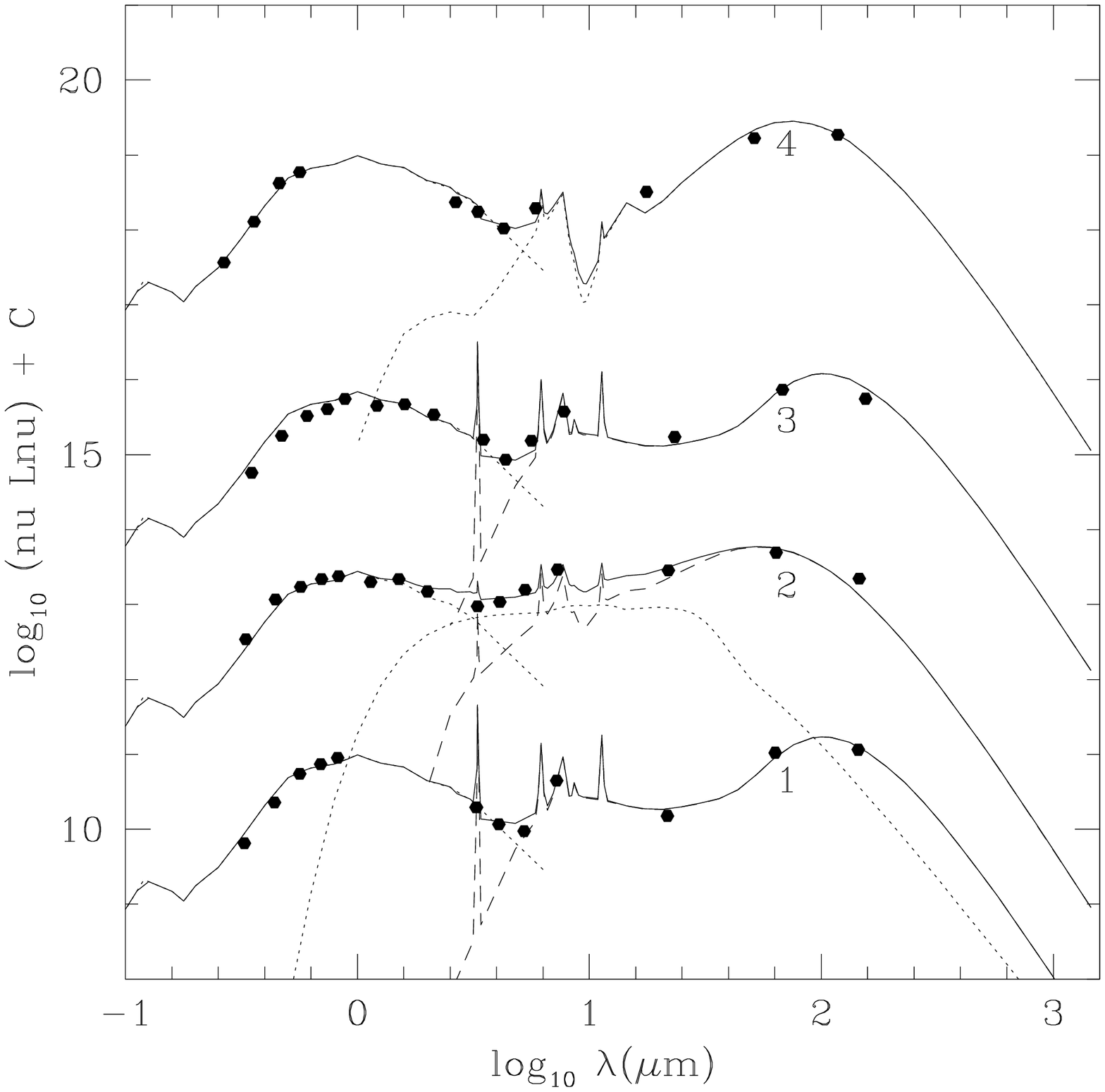,angle=0,width=7cm}
\caption{
LH: Spectral energy distributions of luminous cool galaxies with good photometric redshifts
($>$ 5 photometric bands, reduced $\chi^2 < 5$) and 24 and 70 $\mu$m detections.  In each case 
the infrared SEDs are fitted by cirrus and M82 starburst components, with the dominant luminosity 
coming from the cool cirrus component.
RH: Spectral energy distributions of luminous infrared galaxies with elliptical galaxy SEDs
in the optical, 24, 70 and 160 $\mu$m detections, and spectroscopic redshifts.
}
\end{figure*}

\begin{figure*}
\epsfig{file=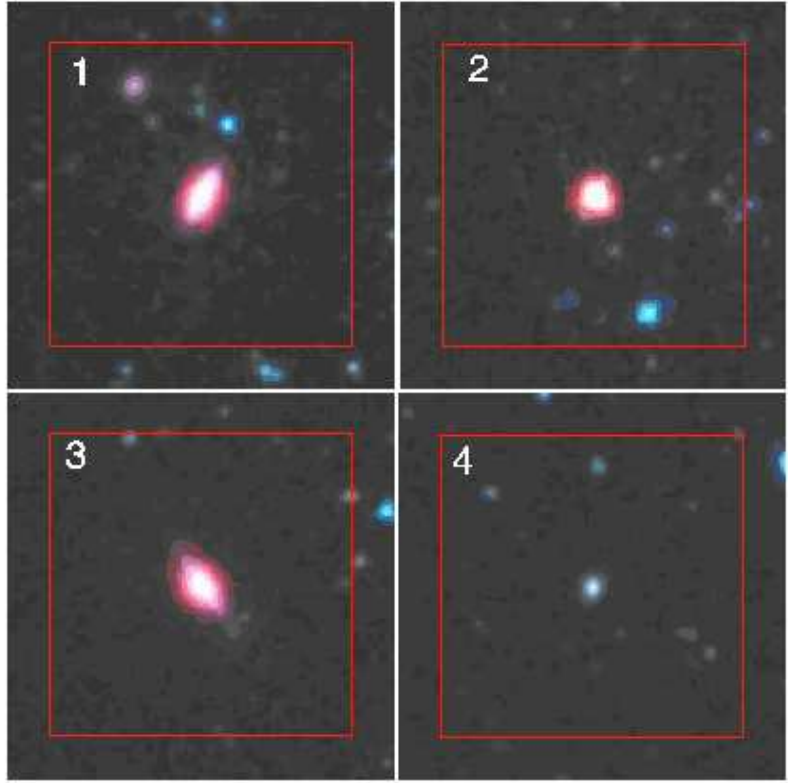,angle=0,width=15cm}
\caption{Postage stamps for the 4 elliptical galaxies of Fig 19R, colour-coded with blue=3.6$\mu$m,
red=8$\mu$m.  The size of each postage stamp is 1'.}
\end{figure*}

\begin{figure*}
\epsfig{file=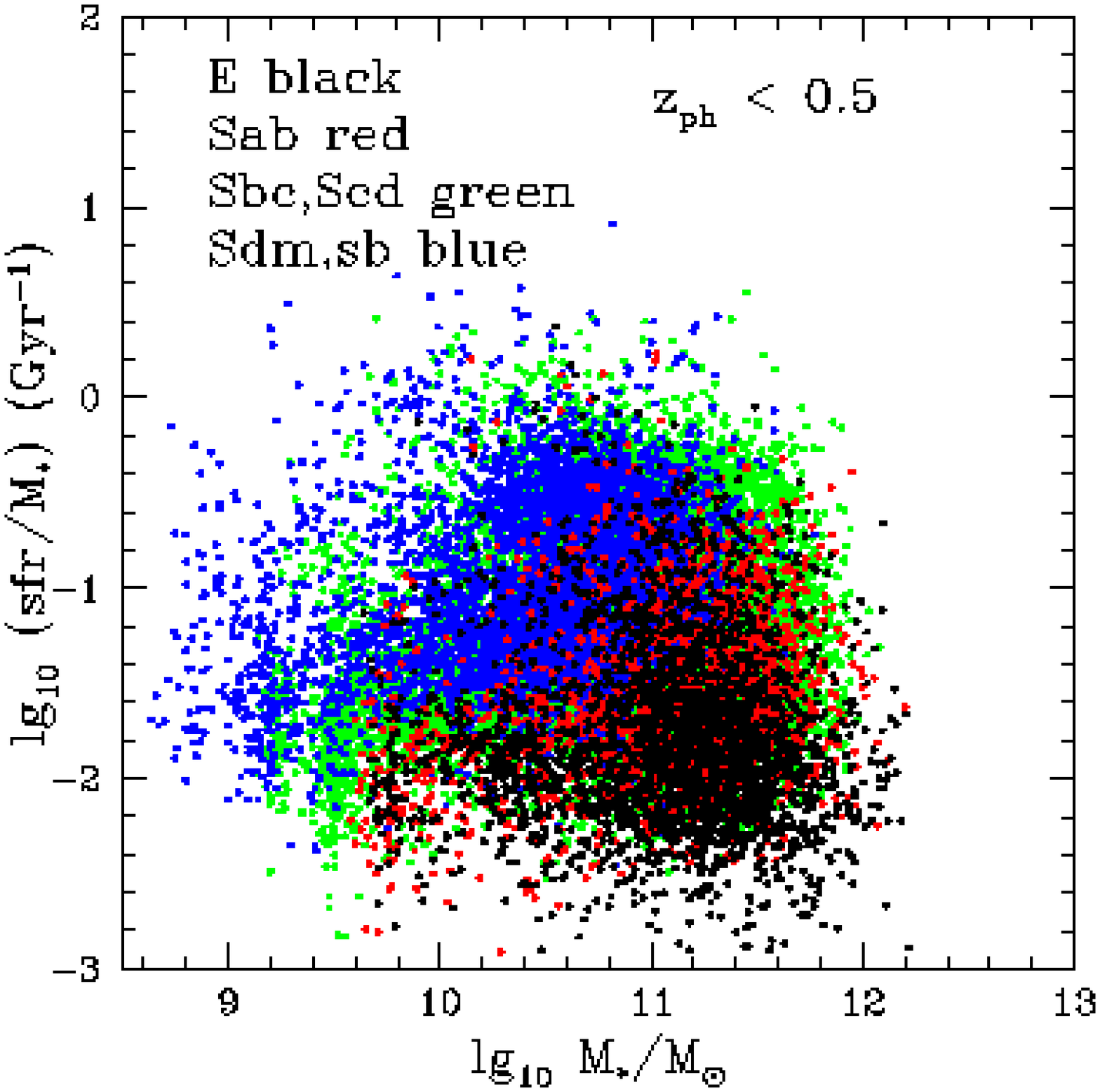,angle=0,width=7cm}
\epsfig{file=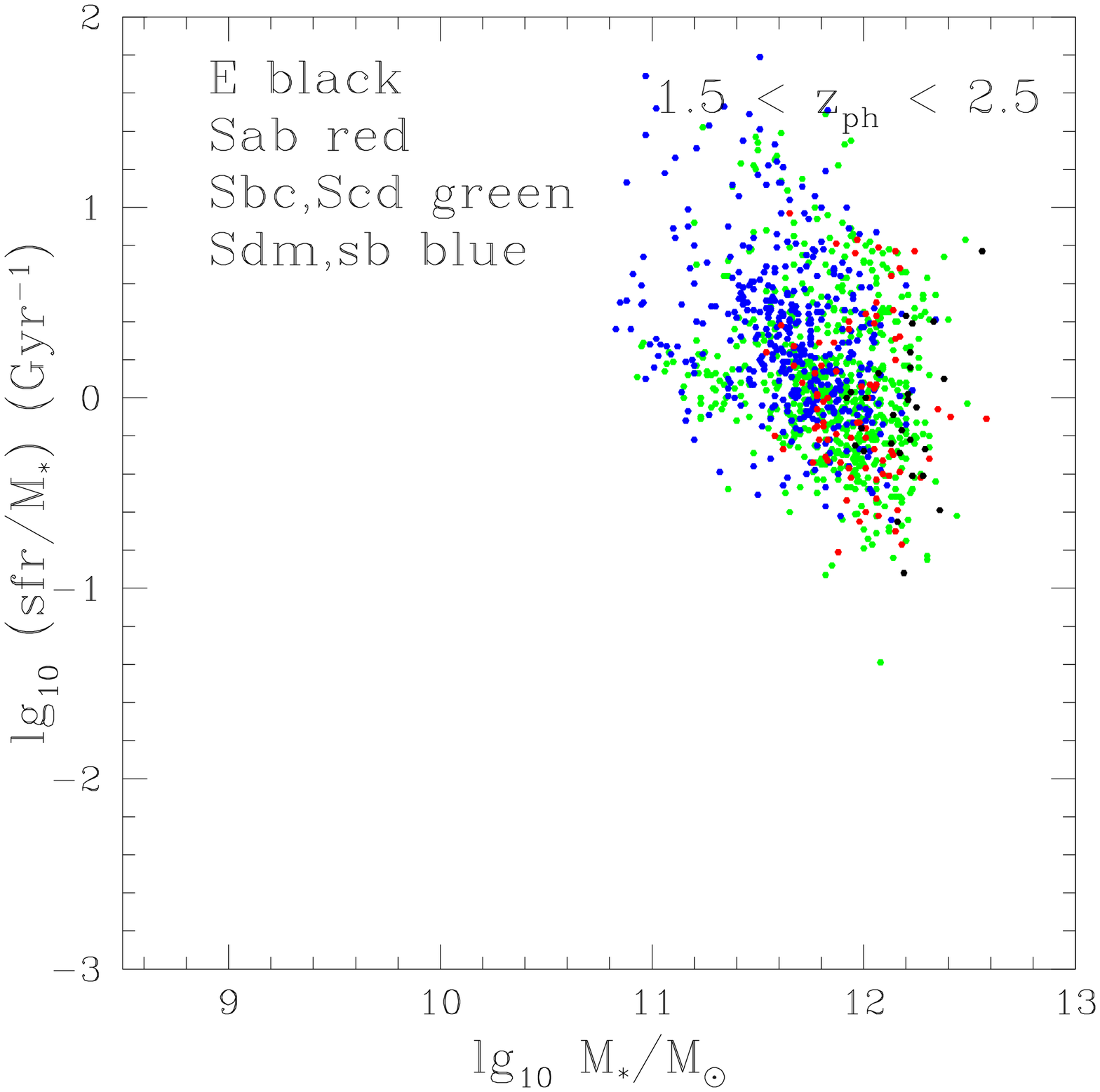,angle=0,width=7cm}
\caption{
The specific star-formation rate $\phi_*/M_*$, versus stellar mass, $M_*$, for galaxies with (L) z$<$0.5,
(R)$1.5 < z_{phot} < 2.5$, colour-coded by optical SED type.
}
\end{figure*}

\begin{figure*}
\epsfig{file=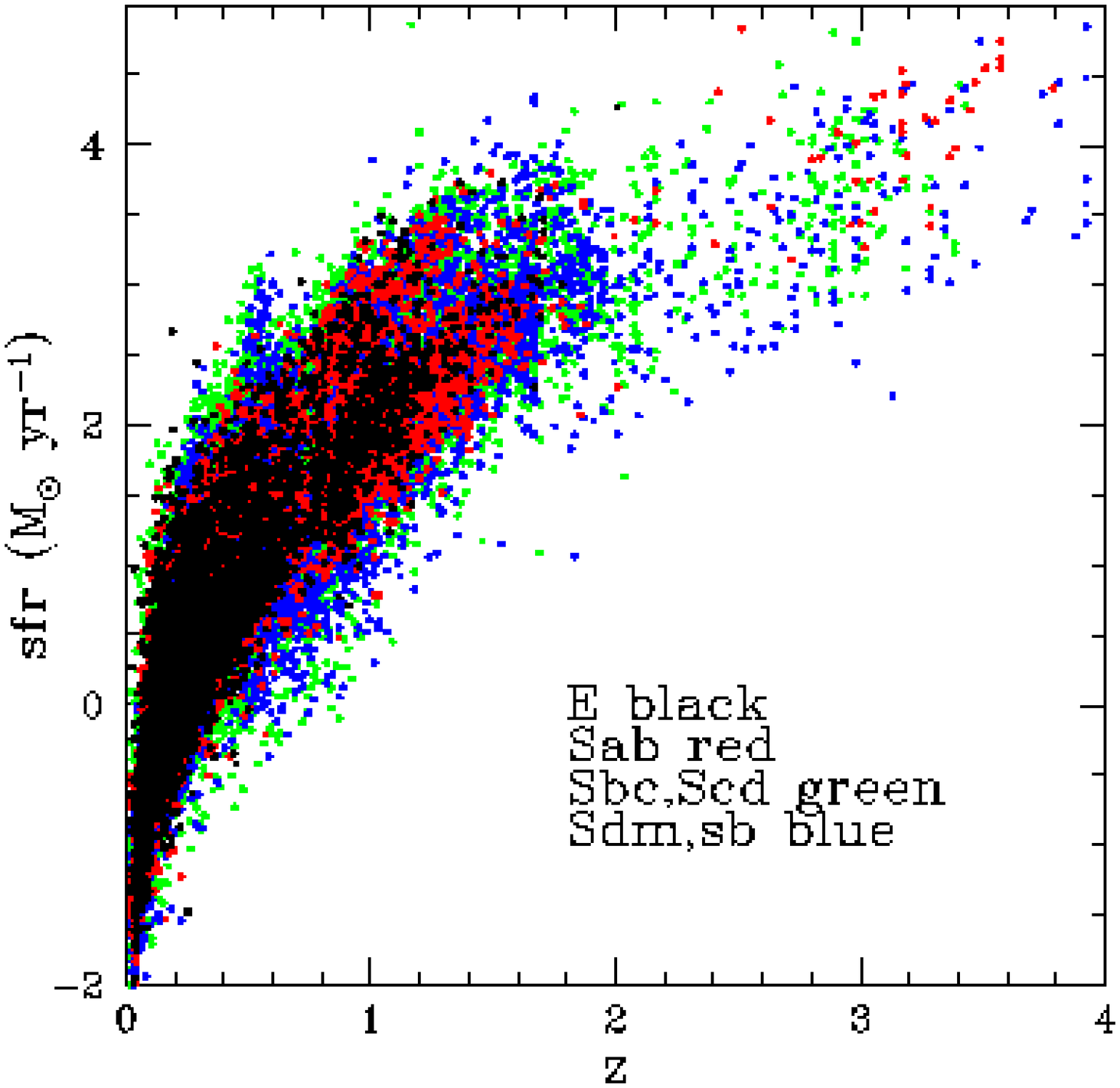,angle=0,width=7cm}
\epsfig{file=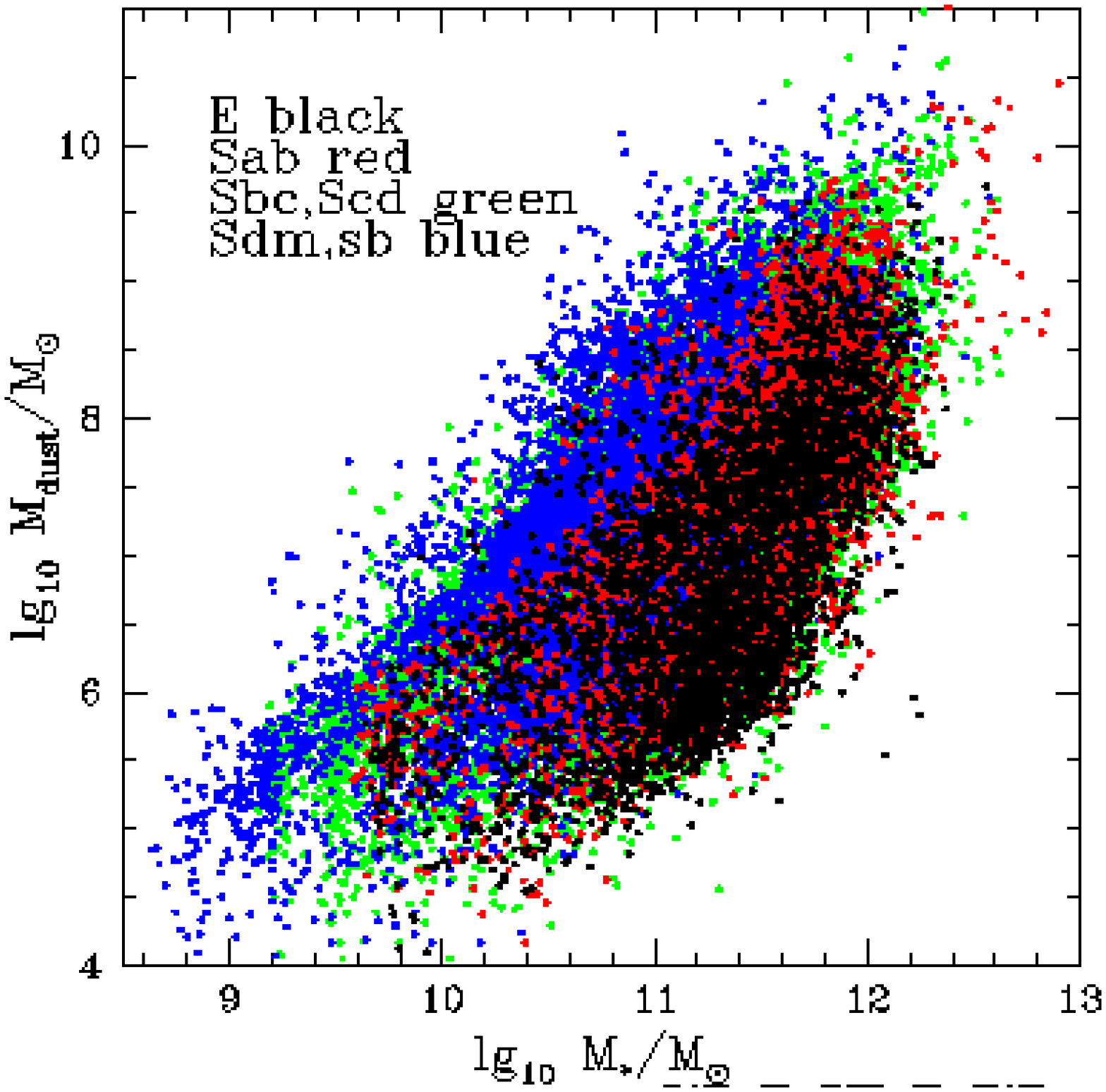,angle=0,width=7cm}
\caption{
L: The star-formation rate $\phi_*$, in $M_{\odot} yr^{-1}$, versus redshift, colour-coded by optical SED type.
R: The dust mass, $M_{dust}/M_{\odot}$ versus stellar mass, $M_*$, for SWIRE galaxies, colour-coded by optical SED type.
}
\end{figure*}

\section{Acknowledgements}
Analysis of the VVDS area was based on observations obtained with MegaPrime/MegaCam, a joint project of
CFHT and CEA/DAPNIA, at the Canada-France-Hawaii Telescope (CFHT), which is operated by the
National Research Council (NRC) of Canada, the Institut National des Sciences de l'Univers of the
Cntre national de la Reserche Scientifique (CNRS) of France, and the University of Hawaii.  The work is
based in part on data products produced at TERAPIX and the Canadian Astronomy Data Centre as part
of the CFHT Legacy Survey, a collaborative project of NRC and CNRS.

We are grateful fir financial support from NASA and from PPARC/STFC.

We thank Max Pettini for helpful comments.


\end{document}